%% file: example_paper.tex
\documentclass{article}

\usepackage{microtype}
\usepackage{graphicx}
\usepackage{subcaption}
\usepackage{booktabs} % for professional tables

\usepackage{hyperref}

\usepackage[accepted]{icml2026}

\usepackage{amsmath}
\usepackage{amssymb}
\usepackage{mathtools}
\usepackage{amsthm}

\usepackage{wrapfig}
\usepackage{mathtools}
\usepackage{amsmath, amssymb, amsthm}
\usepackage{bm}
\usepackage{enumitem}
\usepackage{multirow}
\usepackage{makecell}
\usepackage{siunitx}
\usepackage{adjustbox}
\usepackage{tabularx}
\usepackage{diagbox}
\usepackage[most]{tcolorbox}
\usepackage{anyfontsize}
\usepackage{url}            % simple URL typesetting
\usepackage{amsfonts}       % blackboard math symbols
\usepackage{nicefrac}       % compact symbols for 1/2, etc.
\usepackage{xcolor}         % colors
\usepackage{natbib, mathtools}
\usepackage{multirow}
\usepackage{graphicx}
\usepackage[table]{xcolor}
\usepackage{array}
\usepackage{siunitx}
\usepackage{ragged2e}
\usepackage{listings}

\usepackage[capitalize,noabbrev]{cleveref}

\theoremstyle{plain}
\newtheorem{theorem}{Theorem}[section]
\newtheorem{proposition}[theorem]{Proposition}
\newtheorem{lemma}[theorem]{Lemma}
\newtheorem{corollary}[theorem]{Corollary}
\theoremstyle{definition}
\newtheorem{definition}[theorem]{Definition}
\newtheorem{assumption}[theorem]{Assumption}
\theoremstyle{remark}
\newtheorem{remark}[theorem]{Remark}
\newcommand{\KL}{D_{\mathrm{KL}}}
\newcommand{\E}{\mathbb{E}}

\newcommand{\R}{\mathbb{R}}
\def\eqref#1{equation~\ref{#1}}
\def\Eqref#1{Equation~\ref{#1}}

\usepackage[textsize=tiny]{todonotes}

\icmltitlerunning{RLCracker: Evaluating the Worst-Case Vulnerability of LLM Watermarks with Adaptive RL Attacks}

\begin{document}

\twocolumn[
  \icmltitle{RLCracker: Evaluating the Worst-Case Vulnerability of \\LLM Watermarks with Adaptive RL Attacks}



  \icmlsetsymbol{equal}{*}
  \icmlsetsymbol{contact}{\dagger}

  \begin{icmlauthorlist}
    \icmlauthor{Hanbo Huang}{sjtu}
    \icmlauthor{Yiran Zhang}{sjtu}
    \icmlauthor{Hao Zheng}{sjtu}
    \icmlauthor{Xuan Gong}{sjtu}
    \icmlauthor{Yihan Li}{NUDT}
    \icmlauthor{Lin Liu}{NUDT}
    \icmlauthor{Zhuotao Liu}{qinghua}
    \icmlauthor{Shiyu Liang}{sjtu}
  \end{icmlauthorlist}

  \icmlaffiliation{sjtu}{Shanghai Jiao Tong University, Shanghai, China}
  \icmlaffiliation{NUDT}{National University of Defense Technology}
  \icmlaffiliation{qinghua}{Tsinghua University}
  \icmlsetsymbol{equal}{*}
  
  % \icmlaffiliation{mail}{hhuang417@sjtu.edu.cn}

  \icmlcorrespondingauthor{Shiyu Liang}{lsy18602808513@sjtu.edu.cn}

  \icmlkeywords{Machine Learning, ICML}

  \vskip 0.3in
]



\printAffiliationsAndNotice{}  % no special notice (required even if empty)

\begin{abstract}
{Large language model (LLM)} watermarking has shown promise in detecting AI-generated content and mitigating misuse, with prior work claiming robustness against paraphrasing and text editing. In this paper, we argue that existing evaluations are not sufficiently adversarial, obscuring critical vulnerabilities and overstating the security. To address this, we introduce the \textit{adaptive robustness radius}, a formal metric that quantifies the {worst-case} resilience of watermarks against adaptive adversaries. {By lifting the paraphrase space into a KL-divergence ball, we approximate this radius and} theoretically demonstrate that optimizing the attack context and model parameters can significantly reduce the {approximated} radius, making watermarks highly vulnerable to paraphrase attacks.
Leveraging this insight, we propose RLCracker, a reinforcement learning (RL)–based adaptive attack that erases watermarks while preserving semantic fidelity. RLCracker requires only \textit{limited} watermarked examples and \textit{zero} access to the detector. Despite weak supervision, it empowers a 3B model to achieve 98.5\% removal success {with minimal semantic shift} on 1,500-token Unigram-marked texts after training on only \textit{100} short samples. This performance dramatically exceeds 6.75\% by GPT-4o and generalizes across five model sizes over ten watermarking schemes. Our code is available in this \href{https://github.com/OTTO-OTO/RLCracker}{repository}.
% Our results confirm that adaptive attacks are broadly effective and pose a fundamental threat to current watermarking defenses.
\end{abstract}

\section{Introduction}
With the rapid advancement and increasing accessibility of large language models (LLMs), they are being widely applied across diverse applications, generating fluent, human-like content~\citep{yang2025qwen3}. However, this extensive adoption raises pressing concerns about misuse, ranging from misinformation and copyright violations to prompt injection and model theft~\citep{liu2024can, wei2023jailbroken, wang2024unveiling}. As a safeguard, \textit{text watermarking} has become a leading defense: by subtly embedding statistical signals into model outputs, watermarking allows reliable attribution while preserving the output quality~\citep{kirchenbauer2023watermarkKGW, liu2023semanticSIR, zhao2024permutePF}.

Most existing watermarking schemes follow a generate-and-detect paradigm. A detection algorithm scans the model outputs for hidden patterns that distinguish AI-generated text from human-written text~\citep{liu2023unforgeableUPV}. 
% Prior evaluations show robustness to naive paraphrasing~\citep{liu2023semanticSIR}, while translation can break a few standard watermarks unless strengthened designs are used~\citep{he2024canXSIR}. 
Prior evaluations show robustness to naive paraphrasing~\citep{liu2023semanticSIR} and to translation attacks under strengthened designs~\citep{he2024canXSIR}.
However, these evaluations focus on \textit{average-case} prompting under fixed, handcrafted instructions, leaving their \textit{worst-case robustness} under adaptive, 
high-capacity attacks largely unexplored.

To address this, researchers have investigated watermark removal attacks to more rigorously assess the watermark security. Nevertheless, existing approaches suffer from significant drawbacks, including ineffectiveness on challenging long-form text ($\geq$500 tokens)~\citep{krishna2023paraphrasing}, poor generalization ability~\citep{jovanovic2024watermark}, and excessive data requirements (i.e., up to 100k samples)~\citep{huang2024b4}. As LLMs continue to advance, there is an urgent need for both a principled metric and an efficient, generalizable, and data-efficient methodology for evaluating the worst-case robustness of watermarking schemes.

\begin{figure*}[t]
    \centering
    \includegraphics[width=0.93\linewidth]{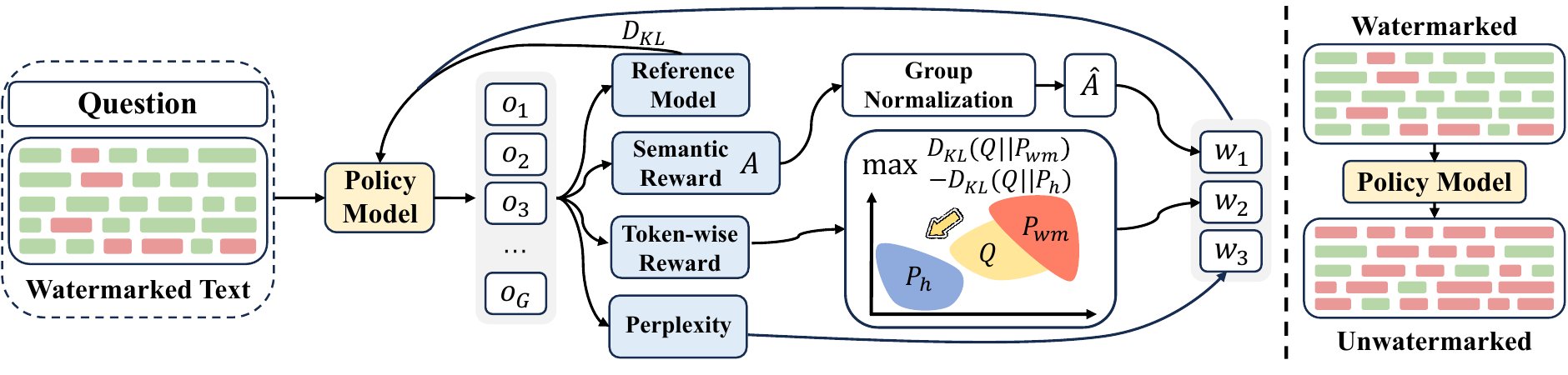}
    \vspace{-0.1cm}
    \caption{The RLCracker algorithm for watermark removal. The model is trained on question and watermarked-text pairs, jointly optimizing a semantic reward to preserve meaning and a token-wise KL reward to shift the model's output $Q$ from the watermark distribution ($P_{wm}$) and towards a human-written distribution ($P_{h}$), thereby effectively removing the watermark in texts.}
    \label{fig:workflow}
    \vspace{-0.4cm}
\end{figure*}

In this paper, we systematically investigate the vulnerability of LLM watermarking schemes by introducing the \textbf{adaptive robustness radius}, a semantic margin within which watermark detection remains reliable under adversarial paraphrasing. Inspired by certified robustness in classification~\citep{cohen2019certified, zhu2021certified}, this metric quantifies the minimum semantic shift required to erase a watermark. However, directly computing this radius is intractable: it requires optimizing over paraphrased outputs without access to the watermark key or detector. To address this, we relax the problem from instance-level attacks to distribution-level shifts. Specifically, we lift the paraphrase space into a KL-divergence ball centered on the watermark distribution and derive a distributional surrogate, the \textit{KL adaptive radius}. This formulation reveals that optimizing attack context and model parameters to steer the paraphrased distribution toward human-like text and away from the watermark distribution can significantly reduce the adaptive radius, enabling more effective watermark removal.

Leveraging this insight, we propose \textbf{RLCracker}, a reinforcement learning–based attacker that removes watermarks without accessing the detector. As shown in Figure~\ref{fig:workflow}, RLCracker jointly optimizes semantic fidelity and token-wise KL divergence to shift model outputs away from the watermark distribution and toward natural text. Despite using only \textbf{100 short watermarked samples}, RLCracker enables a compact 3B model to achieve a \textbf{98.5\% success rate} against the Unigram watermark on \textit{1,500-token texts}, 
surpassing GPT-4o by over 90 points. Extensive experiments across five model scales over ten distinct watermarking schemes validate our theoretical claims and the broad effectiveness 
of our attack, exposing critical vulnerabilities in current watermark defenses. Our contributions are as follows:
\begin{itemize}[leftmargin=*, itemsep=1pt, parsep=0pt, topsep=0pt]
    \item We introduce the adaptive robustness radius, a theoretical framework to formally quantify a watermark's worst-case robustness against paraphrase attacks. We prove that this radius can be systematically minimized by optimizing the attack context and model parameters. (Section~\ref{sec::KLRadius})
    \item We propose RLCracker, an efficient RL-based adaptive attack that removes watermarks while preserving high semantic fidelity, requiring only limited watermarked samples and zero access to watermark detectors. (Section~\ref{sec::RLCracker})
    \item We conduct extensive experiments across five model scales and ten distinct watermarking schemes. The results validate our theoretical claims and demonstrate the broad effectiveness of RLCracker, exposing critical vulnerabilities in current watermarking methods. (Section~\ref{sec::Experiments})
\end{itemize}

\section{Related Works}

\textbf{LLM Watermarking.}
Watermarking schemes for LLMs, which embed imperceptible but algorithmically detectable patterns in generated text, have become essential for content attribution, copyright protection, and misuse mitigation~\citep{wu2025survey,liu2024preventing,sander2024watermarking}.  
Existing methods can be broadly classified into two dominant paradigms. The first, \textit{logit-based} methods~\citep{hu2023unbiased,wu2023dipmark}, partition the vocabulary into "green" and "red" lists based on a secret hash key and bias the output logits of LLM toward greenlist tokens during generation. The detection is then based on the observation of a statistically significant frequency of these tokens~\citep{kirchenbauer2023watermarkKGW}. In contrast, the second paradigm, \textit{sampling-based} methods~\citep{christ2024undetectable, zhao2024permutePF}, embed watermarks by modifying the token selection process itself rather than applying a fixed logit bias, enabling finer-grained control. 
Recent advances in both paradigms have focused on minimizing detectable patterns, aiming to make the watermarked output indistinguishable from the natural text~\citep{kuditipudi2023robust}.

\textbf{LLM Watermark Removal Attacks.}
To assess the robustness of watermarking schemes, researchers have explored various watermark removal attacks. A common approach involves straightforward text manipulations using LLMs. For instance, GPT-4o and the specialized DIPPER~\citep{krishna2023paraphrasing} paraphraser have been used to remove watermarks through semantic rewriting~\citep{liu2023semanticSIR}. Similarly, SIRA~\citep{SIRA} performs watermark removal using training-free paraphrasing templates guided by semantically constrained reference texts. 
Other methods train large proxy models to approximate the watermarking process. WS~\citep{jovanovic2024watermark} mimics the KGW detector scoring mechanism to infer watermark embedding patterns, while $B^4$~\citep{huang2024b4} learns the distribution of watermarked text to guide a paraphraser away from watermark-likely tokens.
However, these methods suffer from key limitations. GPT, DIPPER, and SIRA exhibit sharp performance drops on longer texts ($\geq$500 tokens), failing to reveal the worst-case vulnerability of watermarking schemes. WS is tailored specifically to the KGW watermark and generalizes poorly, whereas $B^4$ requires impractically large datasets ($\sim$ 100k samples). These limitations underscore the urgent need for effective, generalizable, and practical attack strategies to rigorously evaluate the watermark security.
{ More detailed related works on LLM watermarking and removal attacks is provided in Appendix~\ref{append::relatedworks}.}

\section{Preliminaries}\label{sec::prelim}
In this section, we develop a theoretical framework grounded in widely adopted logit-based watermarking schemes. We introduce the \textit{adaptive robustness radius} to quantify the worst-case watermark robustness and formally characterize the optimization objective of adversaries.

\textbf{LLM Generation and Logit-based Watermark.}  
Let \( \mathcal{V} \) be a vocabulary and \( \mathcal{V}^* \) the set of all finite token sequences. {A paraphraser \(\pi\) with parameters \(\theta\) defines a distribution over 
{output} sequences \(\mathbf{X}' = (x'_1, x'_2, \dots)\) conditioned on an {input} \(\mathbf{X}\) 
and a context \(c\):
\(
\pi_\theta(\mathbf{X}' \mid \mathbf{X}, c)
= \prod_{t \ge 1} \pi_\theta(x'_t \mid x'_{<t}, \mathbf{X}, c),
\)
where its support is
\(
\mathrm{supp}(\pi_\theta(\cdot \mid \mathbf{X}, c))
:= \{ \mathbf{X}' \in \mathcal{V}^* : \pi_\theta(\mathbf{X}' \mid \mathbf{X}, c) > 0 \}.
\)}
Independently of the paraphraser, a watermark is embedded in a generated 
sequence \(\mathbf{X} = (x_1, x_2, \dots)\) using a secret key \(s\) and a target 
greenlist rate \(q \in (0,1)\), which define a keyed hash function 
\(H_s : \mathcal{V}^* \times \mathcal{V} \to [0,1]\). 
At each position \(t\), the \emph{greenlist} is
\(
\mathcal{G}_t := \{ x \in \mathcal{V} : H_s(x_{<t}, x) \le q \},
\)
and tokens in \(\mathcal{G}_t\) are softly upweighted during generation of \(\mathbf{X}\).
Given any sequence \(\mathbf{X}\), the detector reconstructs each \(\mathcal{G}_t\) and 
computes a score \(f(\mathbf{X}, s)\) from the token--greenlist pairs 
\(\{(x_t, \mathcal{G}_t)\}_{t \ge 1}\).
A watermark is detected if \(f(\mathbf{X}, s) > \delta\) for a fixed threshold \(\delta\); 
otherwise, the sequence is deemed unmarked.
%considered

\textbf{Threat Model: Semantics-Preserving Paraphrasing.}  
We consider an adversary aiming to erase the watermark while preserving semantic meaning. The adversary is defined by three components: (1) \textbf{\textit{Objective}}: Given a watermarked sequence \( \mathbf{X} \), generate a paraphrased output \( \mathbf{X}' \) that is semantically equivalent yet undetectable. (2) \textbf{\textit{Knowledge}}: {The adversary does not know the watermarking algorithm’s internal design,} the secret key \( s \), detector function \( f \), or threshold \( \delta \), but may query the generator to observe watermarked outputs~\citep{jovanovic2024watermark}. (3) \textbf{\textit{Capability}}: The adversary { can collect question-watermarked sample pairs}, select and tune a paraphrasing model \( \pi_\theta \), adjust parameters \( \theta \), and choose any conditioning context \( c \), while having no access to the detector.  We define a semantic distance function \( d(\mathbf{X}, \mathbf{X}') \), where smaller values indicate higher similarity. An attack is considered successful if (i) \( d(\mathbf{X}, \mathbf{X}')\) within a certain semantic distance and (ii) \( f(\mathbf{X}', s) \le \delta \).

\textbf{Limitations of Standard Robustness.}
Prior works assess watermark robustness by testing against fixed paraphrasers or manually crafted prompts~\citep{kirchenbauer2023reliability}, reporting average-case detection success. However, such evaluations fail to reflect worst-case watermark vulnerability. In practice, attackers can adapt both the model and the attack prompts to suppress watermark signals. To bridge this gap, we propose \textit{adaptive robustness}, a certified notion of integrity under adversarial control of both generation and paraphrasing, constrained by semantic drift.

\textbf{A Certified View of Watermark Integrity.}  
Our formulation draws inspiration from certified robustness in classification~\citep{cohen2019certified, zhu2021certified} and distributionally robust optimization (DRO)~\citep{sinha2018certifying, duchi2019distributionally}, where models are guaranteed to remain correct under bounded perturbations. Similarly, we define a semantic robustness radius within which watermark detection is verifiably preserved, even when the adversary adaptively chooses paraphrasing models and prompts. Unlike prior watermarking work focused on average-case robustness~\citep{liu2023semanticSIR}, our framework yields worst-case semantic guarantees under adaptive attacks.

\textbf{Adaptive Robustness Radius \( r^\star(\mathbf{X}) \).}  
Let \( \Theta \) be a constrained space of paraphrasing models \( \pi_\theta \), and \( \mathcal{C} \) a restricted prompt space. We define the \textit{adaptive robustness radius} as: $r^\star(\mathbf{X}) := \sup \{ \rho \ge 0 \mid \forall \theta \in \Theta,\; c \in \mathcal{C},\;  \forall  \mathbf{X}' \in \mathrm{supp}(\pi_\theta(\cdot\mid\mathbf{X}, c)),\; d(\mathbf{X}, \mathbf{X}') \le \rho \Rightarrow f(\mathbf{X}', s) > \delta\}$
% \begin{align*}
%     r^\star(\mathbf{X}) := \sup \left\{ \rho \ge 0 \mid \forall \theta \in \Theta,\; c \in \mathcal{C},\;  \forall  \mathbf{X}' \in \mathrm{supp}(\pi_\theta(\cdot\mid\mathbf{X}, c)),\; d(\mathbf{X}, \mathbf{X}') \le \rho \Rightarrow f(\mathbf{X}', s) > \delta \right\}.
% \end{align*}

% \resizebox{\linewidth}{!}{$
% \begin{aligned}
% r^\star(\mathbf{X}):= 
% \sup \Bigl\{ \rho \ge 0 \,\Bigm| &\,
%     \forall \theta \in \Theta,\; c \in \mathcal{C}, \forall \mathbf{X}' \in \mathrm{supp}(\pi_\theta(\cdot\mid\mathbf{X}, c)), \\
% & d(\mathbf{X}, \mathbf{X}') \le \rho \Rightarrow f(\mathbf{X}', s) > \delta
% \Bigr\}.
% \end{aligned}
% $}

This measures the worst-case semantic margin preserving the watermark under adaptive manipulation. It parallels certified prediction radii in adversarial NLP~\citep{zhu2021certified}, but applies to watermark detection rather than classification.

\textbf{Adversarial Optimization Objective.}
Given a watermark with hash key \(s\), the attacker seeks to generate the closest paraphrase that evades detection. This yields the equivalent form: $r^\star(\mathbf{X}) = \inf_{\theta \in \Theta,\; c \in \mathcal{C},\;\mathbf{X}'}  d(\mathbf{X}, \mathbf{X}') \quad \text{s. t.} \quad \mathbf{X}'\in\mathrm{supp}(\pi_\theta(\cdot\mid\mathbf{X},c)),\;f(\mathbf{X}', s) \le \delta$.
Accordingly, the adversary minimizes the required semantic deviation by directly tuning model parameters \(\theta\) and attack context \(c\), approaching \(r^\star(\mathbf{X})\).
This objective enables both empirical and certified evaluation of watermark robustness, analogous to robust training in adversarial learning~\citep{madry2018towards}, and facilitates benchmarking watermarking schemes via worst-case guarantees rather than average-case attack success.

\section{Methodology}

\subsection{A Computable Certificate via KL Divergence}
\label{sec::KLRadius}

\textbf{Direct instance-level computation of $r^\star(\mathbf{X})$ is intractable.}
The adaptive robustness radius $r^\star(\mathbf{X})$ measures how far an adversarial paraphrase $\mathbf{X}'$ can drift semantically from the original input $\mathbf{X}$ while remaining detectable. But identifying this worst-case $\mathbf{X}'$ requires a combinatorial search over the entire paraphrase space of $\mathbf{X}$, a discrete, irregular domain with no gradients and no exploitable structure. This makes instance-level certification fundamentally infeasible.

\textbf{From deterministic instance to local distribution: a new robustness lens.}
We model the watermarked input $\mathbf{X}$ as inducing a \emph{local distribution} $P_{\mathbf{X},c,\theta}$, formed by passing $\mathbf{X}$ through a lightweight paraphraser $\pi_\theta(\cdot\mid\mathbf{X},c)$ with context $c$ and parameters $\theta$ that generates meaning-preserving variants.
This shift is motivated by a key observation: naive paraphrases of $\mathbf{X}$ often preserve the watermark, suggesting that $\mathbf{X}$ carries a stochastic watermark signature rather than a fixed pattern. This distributional view captures far more about the watermark's behavior than any single paraphrase.

\textbf{A smooth geometric view of the adaptive radius.}
Rather than searching for a single elusive paraphrase, we evaluate how far an \emph{adversarial paraphrasing distribution} $Q$ moves from the local watermark distribution $P_{\mathbf{X},c,\theta}$ in KL space. Here $Q$ represents an attacker’s distribution over meaning-preserving rewrites that aim to reduce or erase the watermark signal. Measuring this distributional deviation yields the \emph{KL adaptive radius}, a geometric robustness certificate quantifying the maximum permissible drift under which all such adversarial paraphrase distributions remain detectable.

% \resizebox{\linewidth}{!}{$
% \begin{aligned}
% r_{\mathrm{KL}}(\mathbf{X}, c, \theta)
% :=&\sup \Bigl\{ \rho \ge 0 \ \Bigm|\ 
% \forall Q:\mathrm{D}_{\mathrm{KL}}(Q \| P_{\mathbf{X},c,\theta}) \le \rho \\
% &\qquad \Rightarrow \mathbb{E}_{\mathbf{X}''\sim Q}[f(\mathbf{X}'', s)] > \delta \Bigr\}.
% \end{aligned}
% $}
\begin{definition}[KL Adaptive Radius]
For a given {watermarked input} $\mathbf{X}$, context $c$ and paraphraser parameters $\theta$, the KL adaptive radius is $r_{\mathrm{KL}}(\mathbf{X}, c, \theta)
:=\sup \{ \rho \ge 0 \ |\ 
\forall Q:\mathrm{D}_{\mathrm{KL}}(Q \| P_{\mathbf{X},c,\theta}) \le \rho \Rightarrow \mathbb{E}_{\mathbf{X}''\sim Q}[f(\mathbf{X}'', s)] > \delta \}.$
\end{definition}
% The KL adaptive radius is the robustness notion for the relaxed distributional problem, not the original instance-level radius. It is a surrogate because it replaces worst-case search over individual paraphrases with a local paraphrasing distribution.
% The KL adaptive radius measures how much local distributional shift the watermark can withstand. It defines a safety region around the local watermark distribution: as long as an attack distribution stays within this region, its expected detection score remains above threshold. A small radius signals a fragile watermark; a large one indicates resilience to substantial distributional perturbations. 
The KL adaptive radius is not the original instance-level radius, but the robustness notion of a relaxed distributional problem. It replaces the intractable worst-case search over individual paraphrases with a local paraphrasing distribution around the watermarked text. Under this relaxation, the radius measures how much KL shift the watermark can tolerate: within this KL neighborhood, every attack distribution still has expected detection score above the threshold. Thus, a small radius indicates fragility, while a large radius indicates resilience to local distributional perturbations.

\textbf{Advantages over instance-level robustness definitions.}
Prior notions, including the $(\varepsilon,\delta)$-robustness~\citep{diaa2024optimizing} and the instance-level radius in Section~\ref{sec::prelim}, measure deviation of a \emph{single} paraphrase from 
$\mathbf{X}$, a brittle criterion in the discrete paraphrase space. 
By contrast, our definition compares \emph{distributions}: the attack 
distribution $Q$ and the local watermark distribution $P_{\mathbf{X},c,\theta}$ 
in KL space. This provides two key advantages. 
(1) \textit{More informative:} $P_{\mathbf{X},c,\theta}$ reveals how the 
watermark behaves across natural variations of $\mathbf{X}$, exposing structure 
invisible to any individual paraphrase. 
(2) \textit{Geometrically smoother:} KL divergence endows paraphrasing 
distributions with a differentiable geometry, enabling gradient-based reasoning 
and formal robustness certificates that the discrete instance space cannot 
support.

\textbf{A necessary condition for paraphrase-robust watermarking: detectability holds only within a KL neighborhood, and watermark removal requires a large KL deviation.}
The distributional relaxation enables us to express robustness as a concrete requirement on the adversarial distribution: the watermark remains detectable only if the attacker’s paraphrasing distribution stays sufficiently close, in KL divergence, to the local watermark distribution. Making this requirement quantitative requires a mild concentration assumption on the detector score, which allows us to derive an explicit lower bound on the KL adaptive radius. This assumption is natural: green-list watermarking schemes produce z-scores that are empirically tightly concentrated and, as shown in Appendix~\ref{appen::justification-assumption}, provably sub-Gaussian under mild sequential assumptions.

\begin{assumption}[Sub-Gaussian Detector Score]\label{assump:sg}
Under the local watermark distribution $P_{\mathbf{X},c,\theta}$ (i.e., $\mathbf{X}'' \sim P_{\mathbf{X},c,\theta}$), 
the detector score $f(\mathbf{X}'',s)$ concentrates around its mean $\mu(\mathbf{X},s,c,\theta)$ with sub-Gaussian tails governed by variance proxy $\sigma^2(\mathbf{X},s,c,\theta)$.
\end{assumption}

% \begin{theorem}[KL lower bound for watermark robustness]\label{thm:kl}
% Let $Q$ be any adversarial paraphrasing distribution. Under the sub-Gaussian assumption,
% $\mathbb{E}_{\mathbf{X}''\sim Q}[f(\mathbf{X}'',s)] 
% \;\ge\; 
% \mu(\mathbf{X},s,c,\theta) 
% -\sqrt{2\,\sigma^2(\mathbf{X},s,c,\theta)\,\mathrm{D}_{\mathrm{KL}}(Q\|P_{\mathbf{X},c,\theta}) }.$
% This implies that the KL adaptive radius is lower-bounded by a computable certificate $\rho^*$:
% $r_{\mathrm{KL}}(\mathbf{X},c,\theta) \ge \rho^*(\mathbf{X},s,c,\theta) := \frac{[\left(\mu(\mathbf{X},s,c,\theta)-\delta\right)_+]^2}{2\,\sigma^2(\mathbf{X},s,c,\theta)},$
% where $(x)_+ = \max\{x,0\}$. 
% \end{theorem}

\begin{theorem}[KL lower bound for watermark robustness]\label{thm:kl}
Let $Q$ be any adversarial paraphrasing distribution. Under the sub-Gaussian assumption,
$\mathbb{E}_{\mathbf{X}''\sim Q}[f(\mathbf{X}'',s)] 
\;\ge\; 
\mu(\mathbf{X},s,c,\theta) 
-\sqrt{2\,\sigma^2(\mathbf{X},s,c,\theta)\,\mathrm{D}_{\mathrm{KL}}(Q\|P_{\mathbf{X},c,\theta}) }.$

This implies that the KL adaptive radius is lower-bounded by a computable certificate $\rho^*$:
\begin{align*}
r_{\mathrm{KL}}(\mathbf{X},c,\theta)
\ge
\rho^*(\mathbf{X},s,c,\theta)
:=
\frac{
\left[\mu(\mathbf{X},s,c,\theta)-\delta\right]_+^2
}{
2\sigma^2(\mathbf{X},s,c,\theta)
},
\end{align*}
where $[x]_+=\max\{x,0\}$.
\end{theorem}

% \textbf{Remark 1: }
\begin{remark}
\textbf{A large KL shift is necessary to scrub the watermark.}
The theorem imposes a geometric constraint on the attacker: the expected detection score drops only when the adversarial distribution moves far away from the natural distribution of meaning-preserving rewrites $P_{\mathbf{X},c,\theta}$. Scrubbing the watermark therefore requires a \emph{substantial} distributional shift, not a single cleverly crafted paraphrase. Because the score decreases only at the rate $\sqrt{\mathrm{KL}(Q\|P)}$, any successful attack must push the KL divergence beyond the certified radius. This yields a natural and differentiable attack objective: maximize $\mathrm{KL}(Q\|P_{\mathbf{X},c,\theta})$ to leave the detectable region. The proof is given in Appendix~\ref{appen::theproof}.
\end{remark}

\begin{remark}
\textbf{The lower bound $\rho^*$ is computable and strongly tracks watermark removability.}
Our framework lower-bounds the intractable KL adaptive radius $r_{\mathrm{KL}}$ by a single computable certificate $\rho^*$, derived from detector statistics $(\mu,\sigma^2)$. Although these quantities are not publicly accessible, they are available to the watermark designer, enabling internal robustness assessment without executing adaptive attacks. Empirically, Appendix~\ref{appen::adaptiveRadiusEmpirical} provides evidence that $\rho^*$ exhibits a strong negative association with watermark removal rate across models and contexts, supporting its use as a practical and diagnostic robustness metric.

%Our framework highlights that watermark robustness is sensitive to the adversary’s search space. In particular, expanding the choices of context $c$ and parameters $\theta$ can reduce the minimal attainable adaptive KL radius. We further show in Appendix~\ref{appen::adaptiveRadiusEmpirical} that $r_{\mathrm{KL}}(\mathbf{X},c,\theta)$ is strongly negatively correlated with watermark removal rate across models and contexts, supporting its use as a practical robustness metric for watermark designers.

\end{remark}

\subsection{RLCracker: Crack Watermarks with RL Attack}
\label{sec::RLCracker}

% \textbf{From KL guidance to practical attack strategies.} A practical strategy for inducing large KL divergence under semantic similarity constraints. A practical attack strategy for inducing large KL divergence while preserving semantic similarity.

\textbf{A practical strategy for inducing large KL divergence under semantic similarity constraints.}
The KL analysis in Section~\ref{sec::KLRadius} shows that watermark removal 
requires increasing the divergence between the attack distribution $Q$ and the 
local watermark distribution $P_{\mathbf{X},c,\theta}$. In practical terms, a 
successful attack must place more probability mass on regions where the 
watermark signal is weak. 
\begin{wrapfigure}{r}{0.22\textwidth}
  \centering
  \vspace{-0.3cm}
    \includegraphics[width=\linewidth]{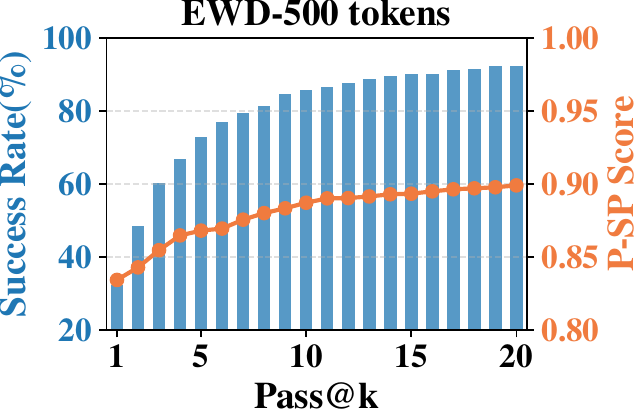}
    % \vspace{-0.3cm}
    \caption{Pass@20 on EWD.}
    \vspace{-0.3cm}
    \label{fig:pass20}
\end{wrapfigure}
Multi-sample attacks provide a simple mechanism for 
achieving this: by drawing $k$ candidates from $\pi_\theta$, the chance that 
\emph{at least one} sample naturally falls into a low-watermark region 
increases with $k$.

To validate this effect, we run a Pass@20 attack on the EWD watermark 
\cite{lu2024entropyEWD} using Qwen2.5-3B-Instruct. As shown in 
Figure~\ref{fig:pass20}, Pass@20 reaches an 89\% success rate at a semantic 
threshold of 0.7, far above the 32\% Pass@1 baseline. This sharp improvement 
shows that \emph{multi-sample} attacks are highly effective: with enough 
samples, at least one candidate typically lands in a low-score region that 
scrubs the watermark. It also suggests a natural goal: \emph{can we raise 
single-sample (Pass@1) success to approach multi-sample (Pass@k) performance?}

% \textbf{From multi-sample attacks to single-sample attacks.}
{\textbf{From best-of-$k$ selection to effective single-sample attacks.}
While best-of-$k$ selection via multi-sample attacks is effective, it is impractical in realistic black-box settings because the attacker cannot query the detector to identify the successful candidate.} This motivates an alternative: learn a policy whose \emph{single} output already behaves like the best-of-$k$ candidates, achieving high success without an oracle detector. Prior work by~\cite{yue2025does} shows that reinforcement learning can approximate multi-sample behavior by shaping a model’s output distribution toward high-reward regions. Building on this insight, we reformulate watermark removal as policy optimization: our aim is to train a model whose attack distribution $Q$ achieves Pass@k-level success in a single sample, without detector access.

% \resizebox{\linewidth}{!}{$
  % \begin{aligned}
  %   \mathcal{J}(\theta) \approx\; 
  %   \mathbb{E}_{\{o_i\}\sim\pi_{\theta_{\mathrm{old}}}}
  %   \frac{1}{G} \sum_{i=1}^G \frac{1}{|o_i|} \sum_{t=1}^{|o_i|} & 
  %     \Biggl[ \frac{\pi_\theta(o_{i,t}\mid q, o_{i,<t})}
  %     {\pi_{\theta_{\mathrm{old}}}(o_{i,t}\mid q, o_{i,<t})}
  %     \bigl(w_1 \hat{A}_i + w_2 \Delta r_{i,t} \bigr)  \\
  %   & - \beta\KL\bigl[\pi_\theta\|\pi_{\mathrm{ref}}\bigr]
  %   \Biggr] - w_3\,\mathrm{PPL}(\pi_{\theta}, \{o_i\}),
  % \end{aligned}$}
  
\textbf{Algorithm Design.}  
Contemporary LLM watermarking techniques often induce detectable shifts in output distributions~\citep{liu2024can}. We exploit this property to propose RLCracker, an adaptive RL-based attack that reshapes the adversarial output distribution \(Q\) to align with human text \(P_h\), while diverging from the watermark distribution \(P_{wm}\). 
Formally, the objective is defined as: $\max_{\theta \in \Theta}\KL(Q\|P_{wm})-\KL(Q\|P_h)$.
RLCracker employs token-wise optimization derived from GRPO~\citep{shao2024deepseekmath}, requiring only question--watermarked response pairs \((q, wr)\) to effectively learn watermark evasion. 
Concretely, the attack policy \(\pi_\theta\) is iteratively updated by sampling a set of outputs \(\{o_1, \dots, o_G\}\) from the previous policy \(\pi_{\theta_{\mathrm{old}}}\), maximizing the following training objective:
$\mathcal{J}(\theta) \approx 
    \mathbb{E}_{\{o_i\}\sim\pi_{\theta_{\mathrm{old}}}}
    \frac{1}{G} \sum_{i=1}^G \frac{1}{|o_i|} \sum_{t=1}^{|o_i|}  
      \bigl[ \frac{\pi_\theta(o_{i,t}\mid wr, o_{i,<t})}
      {\pi_{\theta_{\mathrm{old}}}(o_{i,t}\mid wr, o_{i,<t})}(w_1 \hat{A}_i + w_2 \Delta r_{i,t}) - \beta\KL[\pi_\theta\|\pi_{\mathrm{ref}}] \bigr] - w_3 \mathrm{PPL}(\pi_{\theta}, \{o_i\}),$
where $\Delta r_{i,t} = D_{\text{KL},i,t}(Q\|P_{wm}) - D_{\text{KL},i,t}(Q\|P_h)$ is the token-wise KL reward,  
\(\pi_{\mathrm{ref}}\) denotes the reference model, and \(w_1, w_2, w_3\) are the weights of the respective components.  
The remaining terms impose token-wise KL regularization and a perplexity penalty to preserve fluency.

\textbf{Rewards Design.} We employ two reward components to ensure semantic fidelity while evading watermark detection.  First, the \emph{semantic reward} (\(A\)) is computed from the P-SP score~\citep{wieting2021paraphrastic} between the generated text and the original text, scaled by a sigmoid transformation to amplify gradients and preserve semantic similarity. We normalize $A$ across the rollout group to obtain the semantic advantage 
$\hat{A}_i = \frac{A_i - \mathrm{mean}(A)}{\mathrm{std}(A)}$, which stabilizes 
learning and preserves semantic consistency.

Second, the \emph{token-wise KL reward} ($D_{\text{KL},i,t}$) steers $\pi_\theta$ 
toward human-like behavior while discouraging watermark-like patterns. Each 
token’s distribution is encouraged to align with a human-like reference 
(approximating $P_h$) and diverge from a watermark-induced reference 
(approximating $P_{wm}$). In practice, $P_{wm}$ is instantiated by passing the watermarked instance $wr$ 
through a lightweight reference model $\pi_{\mathrm{ref}}$ with a simple prompt, inspired by the weak rewriting behavior that typically preserves watermarks. $P_h$ is obtained by querying the same unwatermarked reference model $\pi_{\mathrm{ref}}$ with the question 
$q$ that was used to generate the watermark, yielding a human-like distribution.

Finally, we estimate the token-wise KL divergence using the following estimator~\citep{schulman2020kl}:
$D_{\text{KL},i,t}(Q\|P_{*})
\approx 
\frac{\pi_{\mathrm{ref}}(o_{i,t}\mid *,\, o_{i,<t})}
     {\pi_{\theta}(o_{i,t}\mid wr,\, o_{i,<t})}
-
\log 
\frac{\pi_{\mathrm{ref}}(o_{i,t}\mid *,\, o_{i,<t})}
     {\pi_{\theta}(o_{i,t}\mid wr,\, o_{i,<t})} - 1,$
where \( * \in \{h, wm\} \) indicates whether \(P_{*}\) corresponds to the 
human-like distribution \(P_{h}\) or the watermark-induced distribution \(P_{wm}\). 
Both are approximated using the reference model \(\pi_{\mathrm{ref}}\) evaluated 
with different inputs: 
\(
P_{h}(o_{i,t}) \approx \pi_{\mathrm{ref}}(o_{i,t}\mid q,\, o_{i,<t})\) and \(
P_{wm}(o_{i,t}) \approx \pi_{\mathrm{ref}}(o_{i,t}\mid wr,\, o_{i,<t}),
\)
where \(q\) is the question and \(wr\) is its corresponding watermarked response 
available to the attacker.

% \resizebox{\linewidth}{!}{$
% \begin{aligned}
% D_{\text{KL},i,t}(Q\|P_{*})
% \approx 
% \frac{\pi_{\mathrm{ref}}(o_{i,t}\mid *,\, o_{i,<t})}
%      {\pi_{\theta}(o_{i,t}\mid wr,\, o_{i,<t})}
% \;-\;
% \log 
% \frac{\pi_{\mathrm{ref}}(o_{i,t}\mid *,\, o_{i,<t})}
%      {\pi_{\theta}(o_{i,t}\mid wr,\, o_{i,<t})}
% \;-\; 1,
% \end{aligned}
% $}

\section{Experiments}
\label{sec::Experiments}

\subsection{Experimental setup}

% We begin by introducing our experimental setups. Details can be found in Appendix~\ref{appen::expDetails}.

We provide detailed experimental setups in Appendix~\ref{appen::expDetails}.

\textbf{Victim Models and Attackers.} We consider three victim models of varying sizes from two well-known model families for watermark text generation: Qwen2.5-1.5B-Instruct, LLaMA3.1-8B-Instruct~\citep{grattafiori2024llama}, and Qwen2.5-32B-Instruct~\citep{qwen2025qwen25technicalreport}. For attackers, we consider five models with different reasoning capabilities: Qwen3-0.6B, Qwen3-1.7B, Qwen3-4B, and Qwen3-8B~\citep{yang2025qwen3} as reasoning-capable, and Qwen2.5-3B-Instruct as non-reasoning.

% \textbf{Watermarking Schemes} We evaluate 10 distinct watermarking schemes across different generative strategies. Specifically, for green-list–based watermarking, we include EWD~\citep{lu2024entropyEWD}, KGW, KGW\_selfhash~\citep{kirchenbauer2023watermarkKGW}, UPV~\citep{liu2023unforgeableUPV},   SWEET~\citep{lee2023wroteSWEET}, Unigram~\citep{zhao2023provableUnigram},  SIR~\citep{liu2023semanticSIR}, and X-SIR~\citep{he2024canXSIR}. To reflect real-world advances, we also incorporate the sampling-based SynthID-Text~\citep{dathathri2024scalableSynthID} and the cryptographic PF-Watermark~\citep{zhao2024permutePF}. All the watermarks are implemented and detected using the well-adopted MarkLLM toolkit~\citep{pan2024markllm} under default settings.

\textbf{Watermarking Schemes.} We evaluate ten distinct watermarking schemes from both logit-based and sampling-based families. For logit-based watermarking, we include EWD~\citep{lu2024entropyEWD}, KGW, KGW\_selfhash~\citep{kirchenbauer2023watermarkKGW}, UPV~\citep{liu2023unforgeableUPV}, SWEET~\citep{lee2023wroteSWEET}, Unigram~\citep{zhao2023provableUnigram}, SIR~\citep{liu2023semanticSIR}, and X-SIR~\citep{he2024canXSIR}. To reflect recent advances, we also incorporate the sampling-based SynthID-Text~\citep{dathathri2024scalableSynthID} and cryptographic PF-Watermark~\citep{zhao2024permutePF}. All schemes are implemented and detected using the widely-used MarkLLM toolkit~\citep{pan2024markllm} under their default settings.

\textbf{Datasets.}
We consider four datasets: Reddit WritingPrompts~\citep{verma2024ghostbuster}, LFQA dataset~\citep{krishna2023paraphrasing}, MMW BookReport and FakeNews~\citep{piet2025markmywords}. For each watermarking scheme, we generate samples of 250, 500 and 1500 tokens. The training set comprises 100 prompt–watermarked response pairs generated from WritingPrompts. For evaluation, 250 and 500-token samples are created using 100 unique entries from each dataset. In the 1500-token setting, LFQA is excluded due to its short-text nature; instead, 400 samples are generated from BookReport, FakeNews, and WritingPrompts in a 1:1:2 ratio.

% \textbf{Attack Methods.} Under the black-box threat model, in addition to RLCracker, we evaluate five attack methods to expose vulnerabilities in watermarking. \textbf{Base} directly paraphrases watermarked samples via simple prompting. \textbf{Think} leverages the model’s reasoning ability to generate more diverse paraphrases. \textbf{SysP.} expands the prompt context via system prompts to increase attack success. We also employ SIRA~\citep{SIRA}, DIPPER~\citep{krishna2023paraphrasing}, and LLM Paraphraser~\citep{liu2023semanticSIR}, where we adopt GPT-4o-2024-08-06.

\textbf{Attack Methods.} Under the black-box threat model, in addition to RLCracker, we evaluate seven attack methods to expose vulnerabilities in watermarking. \textbf{Base} directly paraphrases watermarked samples via simple prompting. \textbf{Think} leverages the model's reasoning ability to generate more diverse paraphrases. \textbf{SysP.} expands the prompt context via system prompts to increase attack success. \textbf{Think+SysP.} combines these two factors by prompting the model to reason when rewriting under an adversarial system prompt. We also employ SIRA~\citep{SIRA}, DIPPER~\citep{krishna2023paraphrasing}, and LLM Paraphraser~\citep{liu2023semanticSIR}, where we adopt GPT-4o-2024-08-06.

\textbf{Implementation Details of RLCracker.} We apply RLCracker to strengthen watermark removal. The semantic reward employs a reparameterized sigmoid scaling, with 0.85 as the threshold separating positive from negative rewards, thereby promoting higher-quality rephrasings. Training on 100 samples of 500 tokens each, with a batch size of 48 and a group size \( G = 12 \), completes in approximately 1.5 hours for Qwen3-4B and 0.5 hours for Qwen3-0.6B using four NVIDIA A100 GPUs. In our experiments, we find that using dynamic $w_1$ with $w_2 = 0.9$, and $w_3 = 0.1$ yields the best performance across most watermarking schemes.

% Training is performed for 10 epochs with a batch size of 48, group size $G=12$, and learning rates of $2 \times 10^{-7}$ for Qwen3-4B and $1 \times 10^{-6}$ for the other models. In our experiments, we find that setting $w_1=12$, $w_2=0.9$, and $w_3=0.1$ achieves the best performance across most watermarks.

\textbf{Metric.} We use the Evasion Success Rate (ESR) to evaluate the robustness of watermarks. It is defined as the proportion of rephrased texts classified as unwatermarked while preserving high semantic similarity to the original (P-SP score $>0.7$ following prior work~\citep{jovanovic2024watermark}), relative to all rephrased texts.  We also employ perplexity and ChatGPT as a Judge to evaluate the quality of rephrased texts~\citep{SIRA}, with results in Appendix~\ref{appen::AttackResults}.%, with detailed numbers provided in Appendix~\ref{appen::AttackResults}.

% \newpage

\subsection{Main Results}
\label{sec:mainresults}

In this subsection, we evaluate the robustness of different watermarking schemes based on data generated by the Llama-3.1-8B-Instruct. Results are shown in Table~\ref{tab:wm-results}, with additional details of rephrase quality provided in Appendix~\ref{appen::AttackResults}.
% \newpage

\begin{table*}[t]
% \vspace{-0.4cm}
  \centering
  \caption{Evasion Success Rate (ESR, \%) across models and watermarking schemes. \textbf{Bold} indicates the \textbf{best} ESR per model. RLCracker is tested using the same prompt as the \textit{SysP.} method. Abbreviations: SWEET (SWE.), Unigram (Unig.), KGW\_selfhash (KG\_s). {More results for SIR, SynthID-Text and UPV watermarks are presented in Table~\ref{tab::MainresultDetailtab2},~\ref{tab::MainresultDetailtab3},~\ref{tab::MainresultDetailtab5} and ~\ref{tab::MainresultDetailtab6}} in Appendix~\ref{appen::AttackResults}.}
  % \vspace{-0.1cm}
  \label{tab:wm-results}
  \footnotesize
  \setlength{\tabcolsep}{3.2pt} % 默认是 6pt
  \renewcommand{\arraystretch}{1} % 数字越大行距越大，默认是 1.0
  \begin{tabular}{@{}llcccccccccccccc}
    \toprule
    & & \multicolumn{7}{c}{\textbf{500 tokens}} & \multicolumn{7}{c}{\textbf{1500 tokens}} \\
    \cmidrule(lr){3-9} \cmidrule(lr){10-16}
    \makecell[l]{\textbf{Models}} & \makecell[l]{\textbf{Methods}}
      & {\footnotesize\textbf{EWD}} & {\footnotesize\textbf{SWE.}} & {\footnotesize\textbf{XSIR}}
      & {\footnotesize\textbf{Unig.}} & {\footnotesize\textbf{KGW}} & {\footnotesize\textbf{KG\_s.}}  & {\footnotesize\textbf{PF}}
      & {\footnotesize\textbf{EWD}} & {\footnotesize\textbf{SWE.}} & {\footnotesize\textbf{XSIR}}
      & {\footnotesize\textbf{Unig.}} & {\footnotesize\textbf{KGW}} & {\footnotesize\textbf{KG\_s.}} & {\footnotesize\textbf{PF}} \\
    \midrule
    \multirow{6}{*}{Qwen3-0.6B}
      & Base        & 13.5 & 21.3 & 29.0 & 35.8 & 60.0 & 35.5 & 14.5 & 3.50 & 4.25 & 26.0 & 5.00  & 6.00 & 17.5 & 11.8 \\
      & SysP        & 36.5 & 44.8 & 52.5 & 42.5 & 52.3 & 63.3 & 37.3 & 5.75 & 12.5 & 39.3 & 6.25  & 15.0 & 37.0 & 22.0 \\
      & Think       & 28.5 & 35.3 & 37.3 & 34.3 & 45.3 & 66.3 & 27.3 & 10.8 & 10.3 & 32.0 & 6.00  & 15.8 & 33.0 & 22.5 \\
      & Think+SysP  & 48.2 & 58.8 & 50.7 & 37.5 & 34.5 & 69.2 & 53.8 & 14.8 & 16.2 & 38.2 & 5.20 & 19.8 & 47.8 & 30.0 \\
      & SIRA        & 0.00 & 0.00 & 0.00 & 0.00 & 24.8 & 0.25 & 0.00 & 0.00 & 0.00 & 0.00 & 0.25  & 0.00 & 0.00 & 0.00 \\
      & RLCracker & \textbf{91.5} & \textbf{92.5} & \textbf{86.5} & \textbf{67.0} & \textbf{80.8} & \textbf{85.3} & \textbf{76.8} & \textbf{63.3} & \textbf{61.5} & \textbf{60.0} & \textbf{87.5} & \textbf{48.5} & \textbf{66.8} & \textbf{62.0} \\
    \midrule
    
    \multirow{6}{*}{Qwen3-1.7B}
      & Base        & 34.3 & 42.8 & 43.0 & 34.0 & 62.3 & 56.8 & 36.5 & 4.25 & 6.75 & 30.3 & 5.00 & 9.25 & 32.5 & 19.3 \\
      & SysP        & 59.8 & 70.8 & 70.8 & 40.5 & 51.3 & 76.3 & 63.3 & 7.50 & 10.5 & 39.5 & 5.80 & 16.0 & 41.0 & 30.0 \\
      & Think       & 80.8 & 88.0 & 75.5 & 31.0 & 28.3 & 86.0 & 71.3 & 42.5 & 48.0 & 53.5 & 4.50 & 47.0 & 66.3 & 55.5 \\
      & Think+SysP  & 84.0 & 86.5 & 79.8 & 34.8 & 28.2 & 86.0 & 77.5 & 41.5 & 45.5 & 59.0 & 6.20 & 52.5 & 68.5 & 60.2 \\
      & SIRA        & 0.25 & 0.00 & 0.25 & 0.00 & 5.8 & 0.25 & 0.00 & 0.00 & 0.00 & 0.00 & 0.00 & 0.00 & 0.00 & 0.00 \\
      & RLCracker & \textbf{91.8} & \textbf{91.0} & \textbf{86.0} & \textbf{73.0} & \textbf{76.8} & \textbf{87.5} & \textbf{79.5} & \textbf{62.7} & \textbf{61.3} & \textbf{62.3} & \textbf{77.8} & \textbf{60.5} & \textbf{68.0} & \textbf{61.0} \\
    \midrule
    
    \multirow{6}{*}{Qwen3-4B}
      & Base        & 39.3 & 53.3 & 50.3 & 39.3 & 72.0 & 70.0 & 47.8 & 10.0 & 12.3 & 38.0 & 4.50 & 15.8 & 36.0 & 25.8 \\
      & SysP        & 54.3 & 65.8 & 66.8 & 43.8 & 68.8 & 79.3 & 63.0 & 12.5 & 15.0 & 45.0 & 5.50 & 19.3 & 40.8 & 34.5 \\
      & Think       & 76.0 & 83.5 & 75.3 & 39.5 & 51.7 & 85.3 & 71.8 & 40.8 & 45.0 & 62.0 & 4.75 & 52.3 & 78.0 & 59.0 \\
    & Think+SysP  & 78.8 & 88.0 & 79.2 & 38.8 & 47.5 & 88.2 & 81.2 & 38.2 & 46.2 & 63.2 & 6.50 & 54.2 & 79.5 & {63.5} \\
      & SIRA        & 0.00 & 0.00 & 0.25 & 0.75 & 12.3 & 0.25 & 0.00 & 0.00 & 0.75 & 0.00 & 1.25 & 0.00 & 0.00 & 0.00 \\
      & RLCracker & \textbf{93.3} & \textbf{95.8} & \textbf{85.5} & \textbf{73.5} & \textbf{89.0} & \textbf{88.8} & \textbf{82.3} & \textbf{64.0} & \textbf{65.3} & \textbf{66.3} & \textbf{63.8} & \textbf{57.3} & \textbf{82.3} & \textbf{64.3} \\
    \midrule

    \multirow{6}{*}{Qwen3-8B}
      & Base        & 72.0 & 70.8 & 77.0 & 40.5 & 74.5 & 81.3 & 68.3 & 33.5 & 30.0 & 48.8 & 5.25 & 46.8 & 58.8 & 44.3 \\
      & SysP        & 72.8 & 81.5 & 82.0 & 43.5 & 72.0 & 85.5 & 73.0 & 32.0 & 38.8 & 52.5 & 6.75 & 51.0 & 64.5 & 50.3 \\
      & Think       & 89.0 & 91.0 & 81.5 & 40.8 & 53.8 & 83.8 & 71.8 & 54.0 & 51.5 & 57.8 & 7.25 & 55.8 & 71.3 & 59.5 \\
    & Think+SysP  & 90.2 & 92.0 & 86.8 & 41.0 & 57.5 & 86.2 & 74.0 & 56.2 & 59.8 & 64.5 & 7.20 & 58.0 & 78.2 & 63.2 \\
      & SIRA        & 0.00 & 0.00 & 0.25 & 0.50 & 22.3 & 0.25 & 0.00 & 0.00 & 0.25 & 0.00 & 0.50 & 0.00 & 0.00 & 0.25 \\
      & {RLCracker}   & {\textbf{94.8}} & {\textbf{96.3}} & {\textbf{90.2}} & {\textbf{80.5}} & \textbf{91.5} & {\textbf{90.8}} & {\textbf{84.3}} & {\textbf{70.0}} & {\textbf{71.3}} & {\textbf{69.3}} & {\textbf{81.8}}  & \textbf{64.5} & {\textbf{84.3}} & {\textbf{73.3}} \\
    \midrule
    
    \multirow{4}{*}{\makecell[l]{\makecell{Qwen2.5-3B\\-Instruct}}}
      & Base        & 31.3 & 46.8 & 68.8 & 37.3 & 46.0 & 58.8 & 37.0 & 11.5 & 17.0 & 39.3 & 3.50 & 20.5 & 50.8 & 16.0 \\
      & SysP        & 75.5 & 83.0 & 78.5 & 43.5 & 31.8 & 84.5 & 74.0 & 27.3 & 35.5 & 45.3 & 5.50 & 44.8 & 56.8 & 42.8 \\
      & SIRA        & 0.25 & 0.25 & 0.25 & 0.75 & 2.25 & 0.75 & 0.00 & 0.00 & 0.00 & 0.00 & 0.00 & 0.50 & 0.50 & 0.00 \\
      & RLCracker & \textbf{93.3} & \textbf{95.3} & \textbf{89.8} & \textbf{78.5} & \textbf{97.5} & \textbf{89.8} & \textbf{81.8} & \textbf{73.0} & \textbf{71.5} & \textbf{66.5} & \textbf{98.5} & \textbf{58.0} & \textbf{78.0} & \textbf{77.8} \\
    \midrule

    \multirow{1}{*}{GPT-4o}
      & ---        & 71.0 & 81.5 & 77.3 & 49.8 & 53.8 & 86.8 & 78.0 & 10.3 & 20.5 & 50.3 & 6.75 & 49.3 & 61.8 & 49.3 \\
    \multirow{1}{*}{DIPPER}
      & ---        & 30.3 & 52.8 & 0.00 & 33.3 & 29.0 & 72.0 & 43.5 & 1.15 & 5.75 & 0.00 & 4.50 & 11.5  & 41.5 & 21.0 \\
      
    \bottomrule
  \end{tabular}
  \vspace{-0.3cm}
\end{table*}

\begin{figure}[t]
    \centering
    \vspace{0.1cm}
    \includegraphics[width=0.95\linewidth]{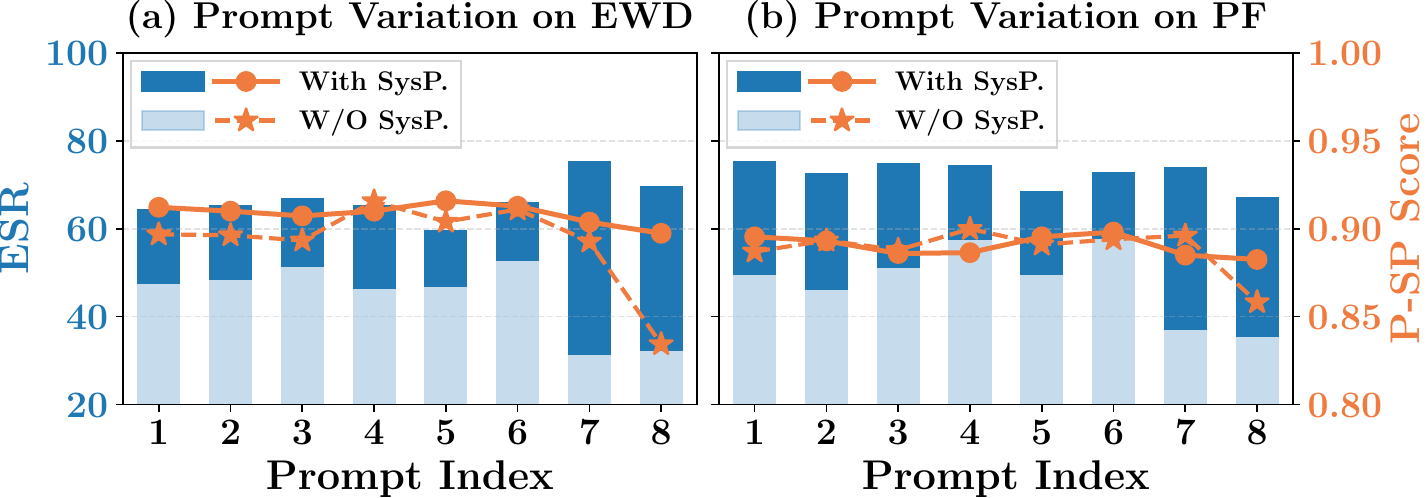}
    \caption{ESR and P-SP variation across user prompts (UserP.), with and without system prompt (SysP.).}
    \label{fig:SysP}
    \vspace{-0.5cm}
\end{figure}

\textbf{System prompts are an overlooked adversarial tool: simple in design but powerful in effect.}  While prior work has shown that user prompts can impact watermark evasion~\citep{kirchenbauer2023reliability,kirchenbauer2023watermarkKGW}, evaluations have largely focused on user input variation, overlooking the broader influence of system-level instructions. We confirm the known sensitivity to user prompts: Figure~\ref{fig:SysP} shows ESR on EWD varies from 31\% to 53\% across prompt indices. However, our key insight goes even further: by introducing a role-assignment system prompt (framing the model as a skilled assistant), we observe a consistent boost in ESR and a reduction in variance. The dark blue bars (with SysP.) outperform the light blue (without SysP.) across the board. Crucially, this boost comes without sacrificing semantic fidelity, as shown by improved P-SP scores. Table~\ref{tab:wm-results} further confirms that SysP consistently outperforms Base across all watermark schemes, model sizes, and token lengths. For example, on Qwen3-4B (500 tokens), ESR rises from 39.3\% to 54.3\% on EWD, and from 70.0\% to 79.3\% on KGW\_selfhash. This efficacy supports Theorem~\ref{thm:kl}, showing that richer attack contexts improve success. Despite the effectiveness, system prompt design remains largely overlooked in watermark robustness evaluations. More details are in Appendix~\ref{appen::Sysprompts}.

% \begin{wraptable}{r}{0.37\textwidth}
% \vspace{-0.72cm}
%   \centering
%   \footnotesize  % 使用更小字体
%   \setlength{\tabcolsep}{3.5pt}  % 缩小列间距，提高紧凑性
%   \caption{ESRs on long-texts (1500 tokens) under Base attack.}
%   \label{tab:robustness}
%   % \vspace{-0.2cm}
%   \begin{tabular}{@{}lcccc@{}}
%     \toprule
%     Model        & KGW   & SIR   & UPV   & Syn. \\
%     \midrule
%     Qwen3-0.6B   & 6.00  & 33.3  & 39.0  & 38.7 \\
%     Qwen3-1.7B   & 9.25  & 35.3  & 39.0  & 49.0 \\
%     Qwen3-4B      & 15.8  & 37.0  & 30.5  & 54.0 \\
%     Qwen3-8B     & 46.8  & 62.5  & 70.0  & 78.3 \\
%     GPT-4o  & 49.3  & 65.5  & 77.8  & 88.3 \\
%     \bottomrule
%   \end{tabular} 
%   \vspace{-0.4cm}
% \end{wraptable}
% \textbf{Paraphraser capability is critical yet not fully explored in evaluating watermark robustness.}  
% While prior work has focused on text length as a vulnerability axis~\citep{kirchenbauer2023reliability}, it largely ignores the adversarial role of model strength. 
% {Tables~\ref{tab:wm-results} and~\ref{tab:robustness} reveal a clear trend}: as paraphraser capability improves, ESR increases significantly across all {ten different} watermarking schemes. 
% These results highlight a fundamental limitation of current watermarking schemes: they do not scale with paraphraser ability and are easily broken by more powerful models. { Note that Table~\ref{tab:robustness} presents a separate ablation study with a different subset of watermarking schemes, so its results are not directly comparable to those in Table~\ref{tab:wm-results}.}

\textbf{The paraphraser’s reasoning ability is an underexplored adversarial factor in watermark robustness.}  
Beyond model size, we find that exploiting the paraphraser’s reasoning capability further degrades watermark resilience. When guided to perform intermediate reasoning before rewriting, the model produces semantically faithful yet structurally transformed outputs that evade detection. As shown in Table~\ref{tab:wm-results} (\textit{Think} rows), Qwen3-4B’s ESR increases from 39.3\% to 76.0\% on the 500-token EWD watermark, and from 36.0\% to 78.0\% on the 1500-token KGW\_selfhash watermark, gaining greater than 40 percentage points. Similar trends appear across models and watermark schemes, showing that watermark signals cannot survive structured paraphrasing enabled by reasoning.

\begin{figure*}[t]
    \centering
    % \vspace{-0.2cm}
    \includegraphics[width=0.95\textwidth]{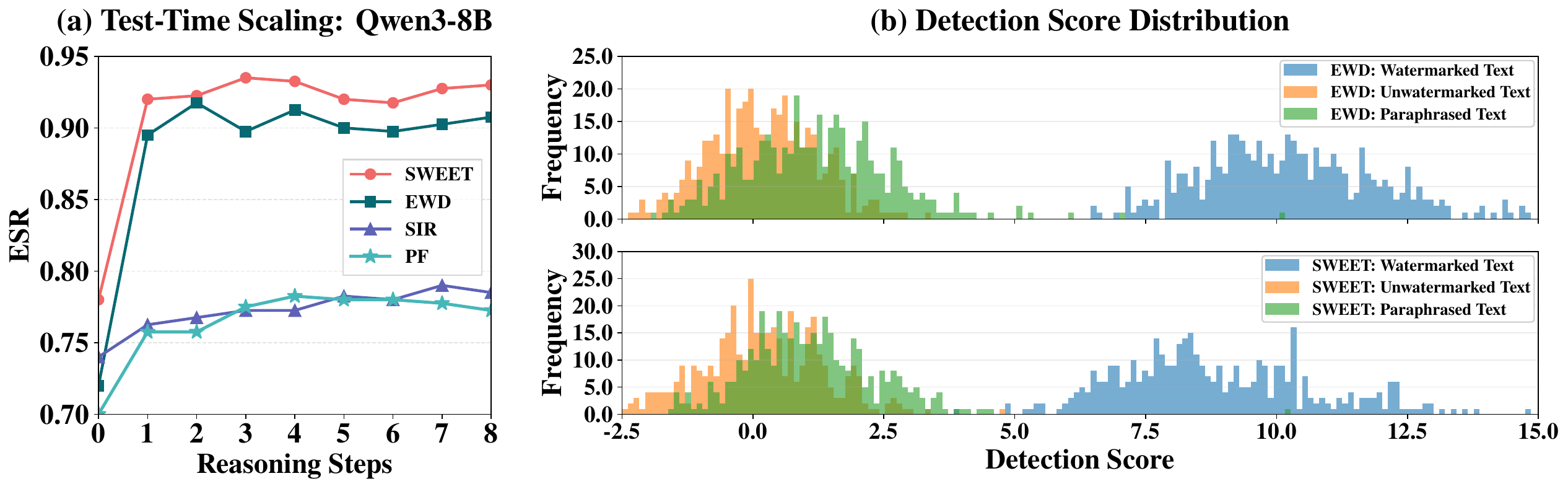}
    \vspace{-0.15cm}
    \caption{(a) shows the effectiveness of test-time scaling on watermark removal using Qwen3-8B; (b) illustrates the detection score distributions for EWD and SWEET watermarks across unwatermarked outputs, watermarked outputs, and paraphrased texts generated by Qwen3-4B trained with RLCracker. }
    \label{fig:Distribution_TestTime}
    \vspace{-0.2cm}
\end{figure*}

\begin{table*}[t]
    % \centerin
    \centering
    \setlength{\tabcolsep}{2.3pt} % 调整列间距
    \footnotesize
    \begin{minipage}{0.47\textwidth}
        \centering
        \caption{Cross-model watermark ESR (\%) of RLCracker.}
        \renewcommand{\arraystretch}{0.9} % Reduce row height
        \begin{tabular}{llcccccc}
        \toprule
        \multirow{2}{*}{\diagbox{\makecell{\textbf{Att.}}}{\makecell{\textbf{Gen.}}}}    &  
            & \multicolumn{3}{c}{\textbf{Qwen2.5-1.5B-Ins.}} 
            & \multicolumn{3}{c}{\textbf{Qwen2.5-32B-Ins.}} \\ 
        \cmidrule(lr){3-5} \cmidrule(lr){6-8}
          & \textbf{Meth.} & \textbf{EWD} & \textbf{SWE.} & \textbf{PF} & \textbf{EWD} & \textbf{SWE.} & \textbf{PF} \\ 
        \midrule
        \multirow{2}{*}{Qwen3-0.6B}        & Base      & 10.8 & 13.0 & 14.5 & 12.5 & 15.8 & 14.3 \\
                          & RLC. & {87.3} & {84.0} & {69.3} & {96.0} & {95.8} & {81.5} \\
        \midrule
        \multirow{2}{*}{\makecell[l]{Qwen2.5-3B\\-Instruct}} & Base      & 22.8 & 28.0 & 26.8 & 28.0 & 33.8 & 29.3 \\
                          & RLC. &  {86.0}  &  {84.8}  & {73.5} &  {97.3}  &  {98.8}  & {84.5} \\
        \midrule
        \multirow{2}{*}{Qwen3-4B}          & Base      & 21.0 & 22.8 & 36.0 & 42.3 & 48.3 & 43.8 \\
                          & RLC. & {84.0} & {83.0} & {73.5} & {98.0} & {98.0} & {84.5} \\
        \bottomrule
        \end{tabular}
        \label{table:crossmodel}
    \end{minipage}
    \hfill
    \begin{minipage}{0.26\textwidth}
        \centering
        \caption{ESR vs. Training size.}
        \renewcommand{\arraystretch}{0.97} % 行间距
        \begin{tabular}{@{}llccc@{}}
        \toprule
        & & \multicolumn{3}{c}{\textbf{Samples}} \\
        \cmidrule(lr){3-5}
         \textbf{Model}& \textbf{Tokens} & \textbf{50} & \textbf{100} & \textbf{200} \\
        \midrule
        \multirow{3}{*}{\makecell{Qwen3\\(0.6B)}} 
            & 250  & 82.5 & 87.8 & 89.3 \\
            & 500  & 86.5 & 91.5 & 92.0 \\
            & 1500 & 91.5 & 92.5 & 92.8 \\
        \midrule
        \multirow{3}{*}{\makecell{Qwen3\\(4B)}} 
            & 250  & 87.8 & 92.0 & 92.8 \\
            & 500  & 90.0 & 93.3 & 93.8 \\
            & 1500 & 92.3 & 95.3 & 95.3 \\
        \bottomrule
        \end{tabular}
        \label{tab:trainsamples}
    \end{minipage}
    \hfill
    \begin{minipage}{0.26\textwidth}
    
        \renewcommand{\arraystretch}{0.98} % 行间距
        \centering
        \caption{Performance of RLCracker trained on Single $|$ Mixed keys data}
        \begin{tabular}{@{}llccc@{}}
        \toprule
        \textbf{Model} & \textbf{Key} & \textbf{EWD} & \textbf{SWE.} & \textbf{PF} \\
        \midrule
        \multirow{2}{*}{\makecell{Qwen3\\(0.6B)}} & Single & 91.5 & 92.5 & 76.8 \\
                                      & Mixed  & {88.8} & {90.5} & {75.5} \\
        \midrule
        \multirow{2}{*}{\makecell{Qwen3\\(1.7B)}} & Single & 91.8 & 91.0 & 79.5 \\
                                      & Mixed  & {90.8} & {90.0} & {78.3} \\
        \midrule
        \multirow{2}{*}{\makecell{Qwen3\\(4B)}}   & Single & 93.3 & 95.8 & 82.3 \\
                                      & Mixed  & {92.3} & {94.3} & {80.5} \\
        \bottomrule
        \end{tabular}
        \label{tab:mixedkeytype}
    \end{minipage}
    \vspace{-0.4cm}
\end{table*}

\textbf{Reasoning depth is a compounding and underexamined threat to watermark robustness.}  
We extend our analysis by employing test-time scaling~\citep{muennighoff2025s1}, where the paraphraser engages in multiple iterations of reasoning before generating a paraphrase. Specifically, whenever the model autonomously determines that the current reasoning process has concluded, we enforce additional reasoning cycles based on the model's previous outputs, thereby increasing the number of reasoning steps. For the Qwen3-8B model, imposing eight reasoning steps on 500-token watermarked inputs significantly amplifies ESR. As shown in Figure~\ref{fig:Distribution_TestTime}(a), ESR rises consistently for SIR and PF, while EWD and SWEET remain stable near 90\%. These results show that deeper reasoning enables more fluent and targeted rewrites, revealing that current watermarking schemes are not robust to increasingly capable inference-time attackers. Full results are in Appendix~\ref{appen::reasoning}.

\begin{wraptable}{r}{0.53\linewidth} % r=右侧，宽度占0.55行宽
% \vspace{-0.38cm}
    \centering
    \setlength{\tabcolsep}{2pt} % 列间距
    \caption{Performance of Qwen2.5-3B-Ins. on 1500-tokens Unigram.}
    % \vspace{0.1cm}
    \footnotesize
    \begin{tabular}{@{}lccc@{}}
    \toprule
    \textbf{Method} & \textbf{ESR} & \textbf{Rem.(\%)} & \textbf{P-SP} \\
    \midrule
    Base & 3.50 & 20.5 & 0.85 \\ 
    SysP. & 5.50 & 25.0 & 0.82 \\ 
    SIRA & 0.50 & 88.0 & 0.47 \\ 
    RLCracker & 98.5 & 100. & 0.92 \\ 
    \bottomrule
    \end{tabular}
    \label{tab:semantic}
\end{wraptable}
\textbf{RLCracker outperforms baseline attacks by combining key adversarial factors.}  
Our earlier findings highlight two key drivers of watermark removal: system prompts and the reasoning capability of the paraphraser. RLCracker integrates both by using a fixed system prompt and reinforcement learning to enhance reasoning. As shown in Table~\ref{tab:wm-results}, it consistently achieves higher ESR across models. On the 1500-token Unigram benchmark, RLCracker 
reaches 87.5\% for Qwen3-0.6B and 98.5\% for Qwen2.5-3B-Instruct, far surpassing GPT-4o’s 6.75\%. In contrast, while SIRA~\citep{SIRA} reports high removal rates, it fails under ESR evaluation, achieving near-zero success on long-form inputs due to poor semantic preservation. 
As shown 
in Table~\ref{tab:semantic}, SIRA removes 88\% of watermarks (Rem.) but yields a low average P-SP score of just 0.47, indicating that its outputs are no longer semantically faithful to the original.

\textbf{RLCracker achieves distribution-level alignment with unwatermarked text.}  
Beyond high ESR scores, we further examine how RLCracker reshapes detection behavior. Specifically, we analyze detection score distributions for EWD and SWEET watermarks across unwatermarked, watermarked, and RLCracker-rephrased texts using 400 samples of 500-token inputs. As shown in Figure~\ref{fig:Distribution_TestTime}(b), RLCracker substantially lowers the detection scores of paraphrased outputs, shifting them away from watermarked distributions and toward unwatermarked ones. This illustrates that RLCracker’s effectiveness stems not just from erasing watermark signals, but from aligning its output distribution with clean, unmarked text. More results in Appendix~\ref{appen::RLCrackerResults}.

\subsection{Ablation Study}
\label{sec:ablationStudy}
In this subsection, we examine RLCracker’s generalization to OOD watermark texts and sensitivity to training data. ESR results are reported in Table~\ref{table:crossmodel}, \ref{tab:trainsamples}, and \ref{tab:mixedkeytype}.

% \textbf{Generalizability of RLCracker on OOD watermarked texts.}
\textbf{RLCracker exhibits robust generalization to OOD watermarked texts.} To evaluate the performance of RLCracker on out-of-distribution (OOD) data, we test models trained on Llama-3.1-8B-Instruct outputs against 500-tokens watermarked texts generated by Qwen2.5-1.5B-Instruct and Qwen2.5-32B-Instruct. As shown in Table~\ref{table:crossmodel}, RLCracker attains high removal success rates across domains, substantially outperforming the \textit{Base} method. For instance, a 0.6B model achieves 96\% ESR on 32B-generated EWD watermarks, compared to only 12.5\% with \textit{Base}. Moreover, ESR is generally higher when targeting outputs from larger models, which may be due to their higher-quality generations. These results demonstrate that RLCracker generalizes effectively across OOD data, achieving high ESR and robust performance in watermark removal. Details in Appendix~\ref{appen::oodgeneralization}.

\textbf{Small training sets suffice for RLCracker to achieve robust performance.} We evaluate RLCracker’s performance with respect to two training data factors: (1) token length per sample and (2) number of training samples. Experiments are conducted using Qwen3-0.6B and Qwen3-4B on the EWD watermark. As shown in Table~\ref{tab:trainsamples}, we observe that even with only 50 samples of 250 tokens, Qwen3-0.6B achieves an ESR of 82.5\%. ESR further improves to 91.5\% as the token length increases under a fixed sample size. While enlarging the amount of training samples also yields gains, performance plateaus beyond 100 samples. These results suggest that relatively small training sets already suffice for RLCracker to achieve strong watermark removal. More details are provided in Appendix~\ref{appen::trainsetLength}.

% \begin{wraptable}{r}{0.35\linewidth} % r=右侧，宽度占0.55行宽
% \vspace{-0.7cm}
%     \centering
%     \setlength{\tabcolsep}{3.8pt} % 列间距
%     \caption{Performance of RLCracker trained on Single $|$ Mixed keys data}
%     \footnotesize
%     \begin{tabular}{@{}l l ccc@{}}
%     \toprule
%     \textbf{Model} & \textbf{Key} & \textbf{EWD} & \textbf{SWE.} & \textbf{PF} \\
%     \midrule
%     \multirow{2}{*}{\makecell{Qwen3\\(0.6B)}} & Single & 91.5 & 92.5 & 76.8 \\
%                                   & Mixed  & {88.8} & {90.5} & {75.5} \\
%     \midrule
%     \multirow{2}{*}{\makecell{Qwen3\\(1.7B)}} & Single & 91.8 & 91.0 & 79.5 \\
%                                   & Mixed  & {90.8} & {90.0} & {78.3} \\
%     \midrule
%     \multirow{2}{*}{\makecell{Qwen3\\(4B)}}   & Single & 93.3 & 95.8 & 82.3 \\
%                                   & Mixed  & {92.3} & {94.3} & {80.5} \\
%     \bottomrule
%     \end{tabular}
%     \label{tab:mixedkeytype}
% \end{wraptable}
\textbf{RLCracker shows minor degradation when trained on mixed-key data.} In real-world settings, collected watermarked data may be generated with different hash keys~\citep{liu2024can}, resulting in mixed watermark patterns. To evaluate RLCracker in this scenario, we construct three mixed-key training sets for watermark EWD, PF and SWEET. Each set contains 100 samples (500 tokens on average), with each sample generated using a distinct hash key (one sample per key). We train Qwen3-0.6B, 1.7B, and 4B on these datasets and evaluate on single-key test data {generated using a key distinct from the training data}. 
{As shown in Table~\ref{tab:mixedkeytype}, mixed-key training incurs only a minor performance drop relative to in-domain training with the test key. The largest ESR reduction is about 3\%, observed for EWD on the 0.6B model. Overall, RLCracker retains strong watermark removal performance under mixed-key training.}
%As shown in Table~\ref{tab:mixedkeytype}, mixed-key training causes only a slight performance decline compared to in-domain training that uses the same key as the test set. The maximum ESR reduction is around 3\%, occurring for EWD with the 0.6B model. These results demonstrate that RLCracker maintains strong watermark removal performance even under mixed-key training. 
Additional details are provided in Appendix~\ref{appen::mixedKeyRobustness}.

\section{Conclusion} % and Broader Impact
% \textbf{Conclusion.} In this paper, we introduce the adaptive robustness radius to characterize the worst-case resilience of LLM watermarking schemes and show that its KL approximation can be minimized by optimizing the attack context and model parameters. We further develop {RLCracker}, a detector-free RL attack that efficiently exploits distributional differences between watermarked and unwatermarked text with only 100 short training samples. RLCracker outperforms baselines and generalizes to long-form text with detection scores close to unwatermarked outputs.

% \textbf{Broader Impact.} Our findings reveal a secondary use-case: RLCracker functions as an effective stress-test for watermark designers and enabling pre-deployment robustness checks. We hope these findings motivate more rigorous evaluation protocols and more robust watermarking designs.

{In this paper, we introduce the adaptive robustness radius to characterize the worst-case resilience of LLM watermarking schemes and show that its KL approximation can be minimized by optimizing attack context and model parameters. We develop {RLCracker}, a detector-free RL attack that exploits distributional differences between watermarked and unwatermarked text using only 100 short training samples. Experiments show that RLCracker outperforms baselines and generalizes to long-form text, enabling effective stress testing and pre-deployment robustness checks. We hope this work motivates more rigorous evaluation protocols and more robust watermarking designs.}

%In this paper, we introduce the adaptive robustness radius to characterize the worst-case resilience of LLM watermarking schemes and show that its KL approximation can be minimized by optimizing the attack context and model parameters. We further develop {RLCracker}, a detector-free RL attack that efficiently exploits distributional differences between watermarked and unwatermarked text with only 100 short training samples. Extensive results show that RLCracker outperforms baselines and generalizes to long-form texts, establishing it as an effective stress-test for watermark designers and enabling pre-deployment robustness checks. We hope these findings motivate more rigorous evaluation protocols and more robust watermarking designs.

\section*{Impact Statement}
This work aims to advance the study of robustness evaluation for LLM watermarking. The proposed methods are intended for research use and do not directly involve sensitive personal data, security-critical systems, or legal decision-making. However, techniques that improve watermark removal or stress-test watermarking schemes could be misused to evade provenance mechanisms. We therefore recommend that any release of attack implementations be accompanied by responsible access controls and that practitioners deploy watermarking only with complementary safeguards.

On the positive side, our findings suggest a constructive use-case: RLCracker serves as a diagnostic stress-test for watermark designers by exposing empirical distribution gaps and enabling pre-deployment checks of indistinguishability under low-distortion attacks. More broadly, we highlight the role of prompts and model reasoning in watermark removal, and we hope this work motivates more rigorous evaluation protocols and more robust watermarking designs.

\section*{Acknowledgments}

This work was supported by the National Key R\&D Program of China
(2025YFF0516900 \& 2025YFF0516904), NSFC U25B2039 and National Natural Science Foundation of China (No.~62306179);
NSFC (No.~12326608).


\bibliography{example_paper}
\bibliographystyle{icml2026}

%%%%%%%%%%%%%%%%%%%%%%%%%%%%%%%%%%%%%%%%%%%%%%%%%%%%%%%%%%%%%%%%%%%%%%%%%%%%%%%
%%%%%%%%%%%%%%%%%%%%%%%%%%%%%%%%%%%%%%%%%%%%%%%%%%%%%%%%%%%%%%%%%%%%%%%%%%%%%%%
% APPENDIX
%%%%%%%%%%%%%%%%%%%%%%%%%%%%%%%%%%%%%%%%%%%%%%%%%%%%%%%%%%%%%%%%%%%%%%%%%%%%%%%
%%%%%%%%%%%%%%%%%%%%%%%%%%%%%%%%%%%%%%%%%%%%%%%%%%%%%%%%%%%%%%%%%%%%%%%%%%%%%%%
\newpage
\input{Appendix}
% \appendix
% \onecolumn

% You can have as much text here as you want. The main body must be at most $8$
% pages long. For the final version, one more page can be added. If you want, you
% can use an appendix like this one.

% The $\mathtt{\backslash onecolumn}$ command above can be kept in place if you
% prefer a one-column appendix, or can be removed if you prefer a two-column
% appendix.  Apart from this possible change, the style (font size, spacing,
% margins, page numbering, etc.) should be kept the same as the main body.
%%%%%%%%%%%%%%%%%%%%%%%%%%%%%%%%%%%%%%%%%%%%%%%%%%%%%%%%%%%%%%%%%%%%%%%%%%%%%%%
%%%%%%%%%%%%%%%%%%%%%%%%%%%%%%%%%%%%%%%%%%%%%%%%%%%%%%%%%%%%%%%%%%%%%%%%%%%%%%%

\end{document}

%% file: Appendix.tex
\appendix
\onecolumn

% \section{Use of LLMs}
% During the preparation of this manuscript, we employed GPT-4o\footnote {\url{https://chatgpt.com/}} to assist with language refinement, including improving the clarity, coherence, and academic tone of the writing. However, we emphasize that LLMs were not involved in the development of key components of our work. Specifically, they were not used for designing research ideas, writing critical code, formulating the experimental methodology, analyzing results, or drafting the related works section. All core contributions and technical decisions were made independently by the authors.

\section{Related Works}
\label{append::relatedworks}
\subsection{LLM Watermarking.}
Watermarking schemes for LLMs, which embed imperceptible but algorithmically detectable patterns in generated text, have become essential for content attribution, copyright protection, and misuse mitigation~\citep{wu2025survey,liu2024preventing,sander2024watermarking}.  
Existing methods can be broadly classified into two dominant paradigms.

The first class of methods, referred to as \textit{logit-based} or \textit{KGW Family} approaches~\citep{hu2023unbiased, wu2023dipmark, takezawa2023necessary, kirchenbauer2023watermarkKGW, liu2023unforgeableUPV, zhao2023provableUnigram,ren2023robust}, involves partitioning the vocabulary into two distinct categories: "green" and "red" lists, with the partitioning determined by a secret hash key~\cite{lu2024entropyEWD,kirchenbauer2023reliability}. During the text generation process, these methods modify the model's logits to bias the output towards tokens from the green list, thereby embedding a statistical signature into the generated text without significantly compromising its overall quality. The watermark is detectable by analyzing the frequency of greenlist tokens in the output, with a predefined threshold used to assess the likelihood that a given passage is watermarked~\citep{kirchenbauer2023watermarkKGW}. 
The primary distinction among methods within the KGW Family lies in the technique used for partitioning the greenlist and redlist tokens. For example, foundational methods~\citep{kirchenbauer2023watermarkKGW} partition tokens based on prior context tokens, while other approaches in this class refine the watermarking technique by incorporating additional mechanisms, such as entropy-based adjustments~\citep{lu2024entropyEWD}. These refinements aim to enhance the robustness of the watermark while further reducing its detectability. The detection process remains focused on the statistical analysis of token distributions, utilizing the frequency of greenlist tokens as a key metric for watermark identification.

In contrast, the second paradigm, referred to as \textit{sampling-based} or \textit{Christ Family} methods~\citep{christ2024undetectable, kuditipudi2023robust}, deviates from logit manipulation by focusing on guiding the sampling process to preferentially select certain tokens during generation. Rather than applying a fixed logit bias, these methods modify the token selection mechanism itself to embed the watermarks. By adapting strategies such as top-k sampling and temperature scaling, Christ-style methods increase the likelihood of selecting watermarked tokens. This approach can produce outputs that are statistically indistinguishable from unwatermarked text and are generally more robust to paraphrasing or other post-processing transformations~\citep{zhao2024permutePF}.
Recent advances in both paradigms have focused on minimizing detectable patterns, aiming to make the watermarked output indistinguishable from the natural text~\citep{kuditipudi2023robust}.

\subsection{LLM Watermark Removal Attacks.}
To evaluate the robustness of watermarking schemes, researchers have explored various watermark removal attacks.

A common approach involves text manipulation using traditional NLP techniques, such as word deletion, substitution, translation~\citep{he2024canXSIR}, or insertion~\citep{kirchenbauer2023watermarkKGW}. However, with advancements in watermarking, these techniques have become less effective~\citep{kirchenbauer2023reliability}.

Some researchers have extended these attacks by employing large language models (LLMs) in a strict black-box setting, where the attacker’s knowledge is limited to the watermarked text. For example, \citet{krishna2023paraphrasing} proposed DIPPER, a paraphrase generation model fine-tuned on an aligned paragraph dataset using T5-XXL~\citep{raffel2020exploring}. This model, along with other approaches like GPT Paraphraser~\citep{liu2023semanticSIR}, removes watermarks through semantic rewriting.

Other approaches focus on bypassing watermark detectors by modifying specific tokens. \cite{rastogi2024revisiting} achieves watermark removal by predicting and replacing low-confidence tokens. Similarly, SIRA~\citep{SIRA} uses training-free paraphrasing templates guided by semantically constrained reference texts, while watermark smoothing attack~\citep{chang2024watermark} guides token selection during rewriting using a reference model to facilitate watermark removal.

Further methods train proxy models to approximate the watermarking process. WS~\citep{jovanovic2024watermark} mimics the KGW detector to infer watermark patterns, while $B^4$~\citep{huang2024b4} learns the distribution of watermarked text to guide a paraphraser away from watermark-prone tokens.

Other research has framed watermark removal as a preference optimization problem. By constructing positive and negative sample pairs, DPO with reinforcement learning (RL) enhances watermark removal capabilities~\citep{diaa2024optimizing}. Additionally, the random walk attack~\citep{zhang2023watermarks} perturbs the watermarked text iteratively using multiple models, relying on detector feedback as the termination condition.

While these methods effectively remove watermarks, they have notable limitations. For instance, GPT, DIPPER, and SIRA exhibit performance drops on longer texts ($\geq$500 tokens) and fail to reveal worst-case vulnerabilities of watermarking schemes. WS is tailored to the KGW watermark and requires large datasets (up to 30,000 samples) for score fitting, limiting generalizability. $B^4$ demands impractically large datasets (up to 100,000 samples), and the iterative nature of the random walk attack introduces significant computational and time overhead. These limitations underscore the urgent need for effective, generalizable, and practical attack strategies to rigorously evaluate the watermark security.

\section{Proof of Theorem~\ref{thm:kl}}
\label{appen::theproof}

% For clarity, we restate the main theorem and its components. Let $P_{s,c,\bm\theta}$ be the watermarked output distribution on $\mathcal{V}^\ast$. The mean detector score is $\mu(s,c,\bm\theta) := \E_{P_{s,c,\bm\theta}}[f(\mathbf X,s)]$.

For clarity, we restate the main theorem and its components. Let $P_{\mathbf{X},c,\theta}$ be the local watermark distribution on $\mathcal{V}^\ast$ (i.e., $\mathbf{X}' \sim P_{\mathbf{X},c,\theta}$). The mean detector score is $\mu(\mathbf{X},s,c,\theta) := \mathbb{E}_{P_{\mathbf{X},c,\theta}}[f(\mathbf{X}',s)]$.

% \newtheorem*{thm1restated}{Theorem 1 (Restated)}
% \begin{thm1restated}[KL lower bound for watermark robustness]
% Let the centered score $f(\mathbf X,s)-\mu(s,c,\bm\theta)$ be $\sigma^2(s,c,\bm\theta)$-sub-Gaussian under $P_{s,c,\bm\theta}$. For any attack distribution $Q$ with $\KL(Q\|P_{s,c,\bm\theta}) < \infty$, the expected score is bounded by:
% \begin{equation}
% \E_Q[f(\mathbf X,s)] \ge \mu(s,c,\bm\theta) - \sqrt{2\,\sigma^2(s,c,\bm\theta)\,\KL(Q\|P_{s,c,\bm\theta})}.
% \label{eq:dv-lower}
% \end{equation}
% This implies the KL adaptive radius $r_{\mathrm{KL}}(s,c,\bm\theta)$ is lower-bounded by the certificate:
% \[
% r_{\mathrm{KL}}(s,c,\bm\theta) \ge \rho^\star(s,c,\bm\theta) := \frac{\left(\mu(s,c,\bm\theta)-\delta\right)_+^2}{2\,\sigma^2(s,c,\bm\theta)}.
% \]
% Furthermore, for $\mathcal C\subseteq\mathcal C'$ and $\bm\Theta\subseteq\bm\Theta'$, the worst-case radius $R_{\mathrm{KL}}(s;\mathcal C,\bm\Theta)$ is monotonically non-increasing, i.e., $R_{\mathrm{KL}}(s;\mathcal C',\bm\Theta') \le R_{\mathrm{KL}}(s;\mathcal C,\bm\Theta)$.
% \end{thm1restated}

% \newtheorem*{thm1restated}{Theorem 4.3}
\begin{theorem}[KL lower bound for watermark robustness]
Let the centered score $f(\mathbf{X}',s) - \mu(\mathbf{X},s,c,\theta)$ be $\sigma^2(\mathbf{X},s,c,\theta)$-sub-Gaussian under $P_{\mathbf{X},c,\theta}$. For any attack distribution $Q$ with $\mathrm{KL}(Q \| P_{\mathbf{X},c,\theta}) < \infty$, the expected score is bounded by: 
\begin{equation}
\mathbb{E}_{\mathbf{X}''\sim Q}[f(\mathbf{X}'',s)] 
\;\ge\; 
\mu(\mathbf{X},s,c,\theta) 
-\sqrt{2\,\sigma^2(\mathbf{X},s,c,\theta)\,\mathrm{D}_{\mathrm{KL}}(Q\|P_{\mathbf{X},c,\theta}) }.
\label{eq:dv-lower}
\end{equation}
This implies that the KL adaptive radius is lower-bounded by a computable certificate $\rho^*$:
\[
r_{\mathrm{KL}}(\mathbf{X},c,\theta) \ge \rho^*(\mathbf{X},s,c,\theta) := \frac{[\left(\mu(\mathbf{X},s,c,\theta)-\delta\right)_+]^2}{2\,\sigma^2(\mathbf{X},s,c,\theta)},
\]
where $(x)_+ = \max\{x,0\}$. 
\end{theorem}

The proof relies on the following change-of-measure inequality, which is a direct consequence of the Donsker-Varadhan variational representation of KL divergence.

\begin{lemma}[{Donsker-Varadhan Inequality~\citep{donsker1975variational}}\label{lem:dv}]

Let $P, Q$ be probability measures with $Q \ll P$. For any measurable function $g:\mathcal V^\ast\!\to\R$ and any $\lambda>0$:
\[
\E_Q[g] \le \frac{1}{\lambda}\Big(\KL(Q\|P)+\log \E_P[e^{\lambda g}]\Big).
\]
\end{lemma}

\begin{proof}[Proof]
% Let $P = P_{s,c,\bm\theta}$ and $\mu = \mu(s,c,\bm\theta)$. We first establish the bound in Eq.~{\Eqref{eq:dv-lower}}. Applying Lemma~\ref{lem:dv} with $g = -(f-\mu)$ and replacing $\lambda$ with $-\lambda$ for $\lambda > 0$ yields:
% \[
% \E_Q[f-\mu] \ge -\frac{1}{\lambda}\Big(\KL(Q\|P) + \log \E_P[e^{-\lambda (f-\mu)}]\Big).
% \]
% By the sub-Gaussian assumption, $\log \E_P[e^{-\lambda (f-\mu)}] \le \frac{1}{2}\sigma^2\lambda^2$. Substituting this into the inequality gives:
% \[
% \E_Q[f-\mu] \ge -\frac{\KL(Q\|P)}{\lambda} - \frac{\sigma^2\lambda}{2}.
% \]
% The right-hand side is maximized over $\lambda>0$ at $\lambda^\star = \sqrt{2\,\KL(Q\|P) / \sigma^2}$, yielding the tightest lower bound:
% \[
% \E_Q[f(\mathbf X,s)] - \mu \ge -\sqrt{2\,\sigma^2\,\KL(Q\|P)},
% \]
% which proves Eq.~{\Eqref{eq:dv-lower}}.

Let $P = P_{\mathbf{X},c,\theta}$ and $\mu = \mu(\mathbf{X},s,c,\theta)$. We first establish the bound in {\Eqref{eq:dv-lower}}. Applying Lemma~\ref{lem:dv} with $g = -(f - \mu)$ and replacing $\lambda$ with $-\lambda$ for $\lambda > 0$ yields:
\[
\mathbb{E}_Q[f - \mu] \ge -\frac{1}{\lambda}\Big(\mathrm{KL}(Q \| P) + \log \mathbb{E}_P[e^{-\lambda (f - \mu)}]\Big).
\]
By the sub-Gaussian assumption, $\log \mathbb{E}_P[e^{-\lambda (f - \mu)}] \le \frac{1}{2}\sigma^2(\mathbf{X},s,c,\theta) \lambda^2$. Substituting this into the inequality gives:
\[
\mathbb{E}_Q[f - \mu] \ge -\frac{\mathrm{KL}(Q \| P)}{\lambda} - \frac{\sigma^2(\mathbf{X},s,c,\theta) \lambda}{2}.
\]
The right-hand side is maximized over $\lambda > 0$ at $\lambda^\star = \sqrt{2\,\mathrm{KL}(Q \| P) / \sigma^2(\mathbf{X},s,c,\theta)}$, yielding the tightest lower bound:
\[
\mathbb{E}_Q[f(\mathbf{X},s)] - \mu \ge -\sqrt{2\,\sigma^2(\mathbf{X},s,c,\theta)\,\mathrm{KL}(Q \| P)},
\]
which proves {\Eqref{eq:dv-lower}}.

Next, to show $r_{\mathrm{KL}} \ge \rho^\star$, assume $\mu > \delta$ (otherwise $\rho^\star = 0$ and the bound is trivial). For any distribution $Q$ such that $\mathrm{KL}(Q \| P) < \rho^\star = \frac{(\mu - \delta)^2}{2\sigma^2(\mathbf{X},s,c,\theta)}$, {\Eqref{eq:dv-lower}} implies:
\[
\mathbb{E}_Q[f(\mathbf{X},s)] > \mu - \sqrt{2 \sigma^2(\mathbf{X},s,c,\theta) \cdot \frac{(\mu - \delta)^2}{2\sigma^2(\mathbf{X},s,c,\theta)}} = \mu - (\mu - \delta) = \delta.
\]
By the definition of $r_{\mathrm{KL}}$, this establishes that $r_{\mathrm{KL}}(\mathbf{X},c,\theta) \ge \rho^\star(\mathbf{X},s,c,\theta)$.

{Let $R_{\mathrm{KL}}(\mathbf{X},\mathcal{C},\Theta) 
= \min_{c\in\mathcal{C},\,\theta\in\Theta} r_{\mathrm{KL}}(\mathbf{X},c,\theta)$ denote the worst-case KL adaptive radius over context set $\mathcal{C}$ and parameter set $\Theta$. } The monotonicity of $R_{\mathrm{KL}}$ follows directly from its definition as an infimum. Since $\mathcal{C} \times \Theta \subseteq \mathcal{C}' \times \Theta'$, the infimum over the larger set cannot be greater than the infimum over the smaller set.

Finally, we note two conditions for a strict decrease in the certified radius. (i) If an adversary finds a strategy $(c', \theta') \in \mathcal{C}' \times \Theta'$ where $\mu(\mathbf{X},s,c',\theta') \le \delta$, then $\rho^\star(\mathbf{X},s,c',\theta') = 0$, causing $R_{\mathrm{KL}}(\mathbf{X}; \mathcal{C}', \Theta') = 0$. (ii) If the sub-Gaussian bound is tight (i.e., the score is truly Gaussian), one can construct an exponentially tilted distribution $Q$ that achieves the bound in {\Eqref{eq:dv-lower}} with equality. This implies $r_{\mathrm{KL}}(\mathbf{X},c',\theta') = \rho^\star(\mathbf{X},s,c',\theta')$, and if this value is smaller than the previous worst-case radius, the radius strictly decreases.

\end{proof}

\subsection{Justification of Assumption~\ref{assump:sg}}\label{appen::justification-assumption}
In this subsection, we justify Assumption~\ref{assump:sg} by demonstrating that the detected 
z-score is both empirically concentrated and theoretically sub-Gaussian. 
Figure~\ref{fig:Distribution_TestTime} (b) shows that the detected z-scores of watermarked text (blue) exhibit 
tight concentration and light-tailed behavior. This is expected: the green-list detector sums many 
bounded token-level indicators, each contributing only a small centered increment, so the overall 
score behaves like a normalized sum of bounded differences, even when $\mathbf{X}'$ is produced by a 
complex sequential generator. Such sums are known to satisfy sub-Gaussian tail bounds under very 
mild conditions. The next proposition formalizes this intuition and establishes that the normalized 
green-list z-score is provably sub-Gaussian for \emph{any} sequential text-generation process.

\begin{proposition}[Sub-Gaussianity of the normalized detector score]
\label{prop:subg-seq}
Fix the key $s$ and  sequence length $n$. Let $\mathbf{X}' = (X'_1,\dots,X'_n)$ be any sequentially
generated text under the local watermark distribution $P_{\mathbf{X},c,\theta}$;
that is, $X'_i$ is adapted to the filtration 
$\mathcal{F}_i := \sigma(X'_1,\dots,X'_i)$ with no independence or Markov assumptions.
For each position $i$, let 
$Y_i := \mathbf{1}\{X'_i \in \mathcal{G}_i(s)\} \in \{0,1\}, $
and define the variance term
\(
V := \sum_{i=1}^n p_i(1-p_i), \, p_i := \mathbb{E}[Y_i].
\)
Define the normalized z-score as
\[
f(\mathbf{X}',s):= \frac{1}{\sqrt{V}}\sum_{i=1}^n (Y_i-p_i).
\]
Then $f(\mathbf{X}',s) - \mathbb{E}[f(\mathbf{X}',s)]$ is sub-Gaussian with variance proxy $n^2/(4V)$. In particular,
for all $t \ge 0$,
\[
\mathbb{P}(| f(\mathbf{X}',s) - \mathbb{E}[f(\mathbf{X}',s)] | \ge t)
\;\le\;
2\exp\!\Big(-\frac{2Vt^2}{n^2}\Big).
\]
\end{proposition}
\begin{remark}[Insight]
Proposition~\ref{prop:subg-seq} follows from a simple and general fact:
any bounded random variable is sub-Gaussian, and any linear combination
of (possibly dependent) bounded variables remains sub-Gaussian with a variance
proxy determined by the total range of the sum rather than by independence.
Since each summand $(Y_i - p_i)$ lies in $[-1,1]$, the aggregate fluctuation
$\sum_{i=1}^n (Y_i - p_i)$ is itself a bounded random variable supported on
an interval of length at most $n$, which directly yields the sub-Gaussian
parameter $n^2/4$ used in Proposition~\ref{prop:subg-seq}. Normalizing by
$\sqrt{V}$ produces the variance proxy $n^2/(4V)$ for the centered z-score.
\end{remark}

\begin{remark}[Illustration]
A simple example highlights the meaning of the variance proxy.
Suppose a single Bernoulli variable $Z\sim\mathrm{Bernoulli}(1/2)$ is repeated
across all positions, so $Y_i := Z$ for all $i$. Then
\[
\sum_{i=1}^n (Y_i - p_i) = \pm \frac{n}{2},
\qquad
V = \frac{n}{4},
\qquad
 f(\mathbf{X}',s) - \mathbb{E}[f(\mathbf{X}',s)] = \pm \sqrt{n}.
\]
Thus, for any fixed sequence length $n$, the centered z-score takes only two
values in the interval $[-\sqrt{n},\sqrt{n}]$ and is therefore sub-Gaussian
with variance proxy $\Theta(n)$, exactly the form captured by
$n^2/(4V)=n$ in Proposition~\ref{prop:subg-seq}. This example illustrates that
the normalized detector score remains sub-Gaussian even under perfect
dependence across positions, though its scale naturally grows with $n$.
\end{remark}

\begin{proof}
If $V = 0$, then $p_i \in \{0,1\}$ for all $i$, so each $Y_i$ is almost surely constant and
\[
\sum_{i=1}^n (Y_i - p_i) = 0 \quad \text{a.s.}
\]
Hence $f(\mathbf{X}',s) \equiv 0$, and the stated tail bound holds trivially.  
Thus, in the remainder of the proof we assume $V > 0$.

Define
\[
S := \sum_{i=1}^n (Y_i - p_i).
\]
By linearity of expectation and the definition $p_i = \mathbb{E}[Y_i]$, we have
\[
\mathbb{E}[S] = \sum_{i=1}^n \mathbb{E}[Y_i - p_i]
= \sum_{i=1}^n \big(\mathbb{E}[Y_i] - p_i\big) = 0,
\]
so $S$ is a centered random variable. Next, we bound the range of $S$. Since $Y_i \in \{0,1\}$, we have
\[
Y_i - p_i \in \{-p_i,\, 1 - p_i\} \subseteq [-1,1]
\quad\text{for each } i.
\]
Therefore
\[
S = \sum_{i=1}^n (Y_i - p_i)
\in \Big[-\sum_{i=1}^n p_i,\; \sum_{i=1}^n (1 - p_i)\Big]
\quad\text{almost surely},
\]
and the length of this interval is
\[
\Big(\sum_{i=1}^n (1 - p_i)\Big) - \Big(-\sum_{i=1}^n p_i\Big)
= \sum_{i=1}^n \big[(1-p_i) + p_i\big]
= n.
\]
In particular, $S$ is a centered random variable supported on an interval of length at most $n$.

By Hoeffding's lemma~\citep{hoeffding1963probability}, for a single bounded random variable, if a random variable
$Z$ satisfies $\mathbb{E}[Z]=0$ and $Z \in [a,b]$ almost surely, then for all
$\lambda \in \mathbb{R}$,
\[
\mathbb{E}\big[ e^{\lambda Z} \big]
\;\le\;
\exp\!\Big( \frac{\lambda^2 (b-a)^2}{8} \Big).
\]
Applying this to $Z := S$ with $b-a \le n$ yields
\[
\mathbb{E}\big[ e^{\lambda S} \big]
\;\le\;
\exp\!\Big( \frac{\lambda^2 n^2}{8} \Big)
\quad\text{for all } \lambda \in \mathbb{R}.
\]
Thus $S$ is sub-Gaussian with variance proxy $n^2/4$, in the sense that
\[
\mathbb{P}(|S| \ge u) \;\le\; 2 \exp\!\Big( -\frac{2u^2}{n^2} \Big)
\quad\text{for all } u \ge 0.
\]

By definition, we have 
\[
f(\mathbf{X}',s) - \mathbb{E}[f(\mathbf{X}',s)]
= \frac{1}{\sqrt{V}} \sum_{i=1}^n (Y_i - p_i)
= \frac{S}{\sqrt{V}}.
\]
Therefore, for any $t \ge 0$,
\[
\mathbb{P}\big( |f(\mathbf{X}',s) - \mathbb{E}[f(\mathbf{X}',s)] | \ge t \big)
= \mathbb{P}\big( |S| \ge t \sqrt{V} \big)
\;\le\;
2 \exp\!\Big( -\frac{2 (t^2 V)}{n^2} \Big)
= 2 \exp\!\Big( -\frac{2 V t^2}{n^2} \Big).
\]
This shows that $f(\mathbf{X}',s) - \mathbb{E}[f(\mathbf{X}',s)]$ is sub-Gaussian with variance proxy $n^2/(4V)$ and establishes the claimed tail bound.
\end{proof}

\section{Experimental Setup and Configuration}
\label{appen::expDetails}

\subsection{Watermark algorithm setting}
\label{appen::algorithm_settings}

In this subsection, we detail the hyperparameter settings used for the watermarking algorithm evaluated in Section~\ref{sec::Experiments}. To ensure consistency and reproducibility, we adopt the default hyperparameter configuration provided by the publicly available and widely used MarkLLM toolkit~\citep{pan2024markllm}\footnote{https://github.com/THU-BPM/MarkLLM}. This toolkit is recognized in the community for its robustness and ease of integration, and has been frequently employed in prior watermarking studies.
% \begin{lstlisting}[language=json,caption={Hyperparameters for the KGW.}, captionpos=b]
% {
%     "algorithm\_name": "KGW",
%     "gamma": 0.5,
%     "delta": 2.0,
%     "hash\_key": 15485863,
%     "prefix\_length": 4,
%     "z\_threshold": 4.0,
%     "f\_scheme": "time",
%     "window\_scheme": "left"
% }
% \end{lstlisting}
\begin{tcolorbox}[
    title=Hyperparameters for the KGW watermark,
    colframe=gray, 
    colback=gray!15,
    coltitle=gray,
    fonttitle=\bfseries\color{white},
    rounded corners,
    enhanced,
    left=15pt, right=6pt, top=6pt, bottom=6pt,
    boxrule=1pt,
    arc=6pt,
    width=\linewidth
]
"algorithm\_name": "KGW",\\
"gamma": 0.5,\\
"delta": 2.0,\\
"hash\_key": 15485863,\\
"prefix\_length": 4,\\
"z\_threshold": 4.0,\\
"f\_scheme": "time",\\
"window\_scheme": "left"
\end{tcolorbox}

\begin{tcolorbox}[
    title=Hyperparameters for the KGW\_selfhash watermark,
    colframe=gray, 
    colback=gray!15,
    coltitle=gray,
    fonttitle=\bfseries\color{white},
    rounded corners,
    enhanced,
    left=15pt, right=6pt, top=6pt, bottom=6pt,
    boxrule=1pt,
    arc=6pt,
    width=\linewidth
]
"algorithm\_name": "KGW",\\
    "gamma": 0.25,\\
    "delta": 2.0,\\
    "hash\_key": 15485863,\\
    "prefix\_length": 4,\\
    "z\_threshold": 4.0,\\
    "f\_scheme": "min",\\
    "window\_scheme": "self"
\end{tcolorbox}

\begin{tcolorbox}[
    title=Hyperparameters for the EWD watermark,
    colframe=gray, 
    colback=gray!15,
    coltitle=gray,
    fonttitle=\bfseries\color{white},
    rounded corners,
    enhanced,
    left=15pt, right=6pt, top=6pt, bottom=6pt,
    boxrule=1pt,
    arc=6pt,
    width=\linewidth
]
"algorithm\_name": "EWD",\\
"gamma": 0.5,\\
"delta": 2.0,\\
"hash\_key": 15485863,\\
"prefix\_length": 1,\\
"z\_threshold": 4.0
\end{tcolorbox}

\begin{tcolorbox}[
    title=Hyperparameters for the SIR watermark,
    colframe=gray, 
    colback=gray!15,
    coltitle=gray,
    fonttitle=\bfseries\color{white},
    rounded corners,
    enhanced,
    left=15pt, right=6pt, top=6pt, bottom=6pt,
    boxrule=1pt,
    arc=6pt,
    width=\linewidth
]
    "algorithm\_name": "SIR",\\
    "delta": 1.0,\\
    "chunk\_length": 10,\\
    "scale\_dimension": 300,\\
    "z\_threshold": 0.2,\\
    "transform\_model\_input\_dim": 1024,\\
    "transform\_model\_name": "watermark/sir/model/transform\_model\_cbert.pth",\\
    "embedding\_model\_path": "watermark/sir/model/compositional-bert-large-uncased/",\\
    "mapping\_name": "watermark/sir/mapping/300\_mapping\_128256.json"
\end{tcolorbox}

\begin{tcolorbox}[
    title=Hyperparameters for the Unigram watermark,
    colframe=gray, 
    colback=gray!15,
    coltitle=gray,
    fonttitle=\bfseries\color{white},
    rounded corners,
    enhanced,
    left=15pt, right=6pt, top=6pt, bottom=6pt,
    boxrule=1pt,
    arc=6pt,
    width=\linewidth
]
    "algorithm\_name": "Unigram",\\
    "gamma": 0.5,\\
    "delta": 2.0,\\
    "hash\_key": 15485863,\\
    "z\_threshold": 4.0
\end{tcolorbox}

\begin{tcolorbox}[
    title=Hyperparameters for the PF watermark,
    colframe=gray, 
    colback=gray!15,
    coltitle=gray,
    fonttitle=\bfseries\color{white},
    rounded corners,
    enhanced,
    left=15pt, right=6pt, top=6pt, bottom=6pt,
    boxrule=1pt,
    arc=6pt,
    width=\linewidth
]
    "algorithm\_name": "PF",\\
    "ngram": 8,\\
    "seed": 0,\\
    "seeding": "hash",\\
    "salt\_key": 35317,\\
    "payload": 0,\\
    "max\_seq\_len": 2048
\end{tcolorbox}

\begin{tcolorbox}[
    title=Hyperparameters for the SWEET watermark,
    colframe=gray, 
    colback=gray!15,
    coltitle=gray,
    fonttitle=\bfseries\color{white},
    rounded corners,
    enhanced,
    left=15pt, right=6pt, top=6pt, bottom=6pt,
    boxrule=1pt,
    arc=6pt,
    width=\linewidth
]
"algorithm\_name": "SWEET",\\
    "gamma": 0.5,\\
    "delta": 2.0,\\
    "hash\_key": 15485863,\\
    "z\_threshold": 4.0,\\
    "prefix\_length": 1,\\
    "entropy\_threshold": 0.9
\end{tcolorbox}

\begin{tcolorbox}[
    title=Hyperparameters for the SynthID-Text watermark,
    colframe=gray, 
    colback=gray!15,
    coltitle=gray,
    fonttitle=\bfseries\color{white},
    rounded corners,
    enhanced,
    left=15pt, right=6pt, top=6pt, bottom=6pt,
    boxrule=1pt,
    arc=6pt,
    width=\linewidth
]
"algorithm\_name": "SynthID",\\
    "ngram\_len": 5,\\
    "keys": [
        654, 400, 836, 123, 340, 443, 597, 160, 57, 29,
        590, 639, 13, 715, 468, 990, 966, 226, 324, 585,
        118, 504, 421, 521, 129, 669, 732, 225, 90, 960
    ],\\
    "sampling\_table\_size": 65536,\\
    "sampling\_table\_seed": 0,\\
    "watermark\_mode": "non-distortionary",\\
    "num\_leaves": 2,\\
    "context\_history\_size": 1024,\\
    "detector\_type": "mean",\\
    "threshold": 0.52
\end{tcolorbox}

\begin{tcolorbox}[
    title=Hyperparameters for the UPV watermark,
    colframe=gray, 
    colback=gray!15,
    coltitle=gray,
    fonttitle=\bfseries\color{white},
    rounded corners,
    enhanced,
    left=15pt, right=6pt, top=6pt, bottom=6pt,
    boxrule=1pt,
    arc=6pt,
    width=\linewidth
]
"algorithm\_name": "UPV",\\
    "gamma": 0.5,\\
    "delta": 2.0,\\
    "z\_threshold": 4.0,\\
    "prefix\_length": 1,\\
    "bit\_number": 16,\\
    "sigma": 0.01,\\
    "default\_top\_k": 20,\\
    "generator\_model\_name": "watermark/upv/model/generator\_model\_b16\_p1.pt",\\
    "detector\_model\_name": "watermark/upv/model/detector\_model\_b16\_p1\_z4.pt",\\
    "detect\_mode": "network"
\end{tcolorbox}

\begin{tcolorbox}[
    title=Hyperparameters for the XSIR watermark,
    colframe=gray, 
    colback=gray!15,
    coltitle=gray,
    fonttitle=\bfseries\color{white},
    rounded corners,
    enhanced,
    left=15pt, right=6pt, top=6pt, bottom=6pt,
    boxrule=1pt,
    arc=6pt,
    width=\linewidth
]
    "algorithm\_name": "XSIR",\\
    "delta": 1.0,\\
    "chunk\_length": 10,\\
    "scale\_dimension": 300,\\
    "z\_threshold": 0.2,\\
    "transform\_model\_input\_dim": 768,\\
    "dictionary": "watermark/xsir/dictionary/dictionary.txt",\\
    "transform\_model\_name": "watermark/xsir/model/transform\_model\_x-sbert\_10K.pth",\\
    "embedding\_model\_path": "watermark/xsir/model/paraphrase-multilingual-mpnet-base-v2",\\
    "mapping\_name": "watermark/xsir/mapping/300\_mapping\_llama\_Ins.json"
\end{tcolorbox}

\subsection{Baselines.}

We consider seven distinct watermark removal (attack) methods. Their implementation details are described below:

\textbf{Base.} The Base method performs direct paraphrasing of watermarked samples using simple prompting. Specifically, we wrap the watermarked text using the following template:

\begin{tcolorbox}[
    title=Prompt template used in the Base method,
    colframe=gray, 
    colback=gray!15,
    coltitle=gray,
    fonttitle=\bfseries\color{white},
    rounded corners,
    enhanced,
    left=6pt, right=6pt, top=6pt, bottom=6pt,
    boxrule=1pt,
    arc=6pt,
    width=\linewidth
]
\detokenize{
###Target Text: 
  }{ [watermarked text]} 
  
\detokenize{
###Instruction: Rewrite the target text above using different words but keeping the same meaning and similar length. 
  }
  
\detokenize{
###Your Response:
  }
\end{tcolorbox}

\textbf{Think.} The Think method is built on the Base method by enabling the model's reasoning capability to produce more diverse and semantically rich paraphrases. We use the same wrapping prompt as in the Base method.

\textbf{SysP.} The SysP method further enriches the prompting context by applying a system prompt in addition to the base prompt. Specifically, we use the standard chat template and prepend the following system message:

\begin{quote}
\textit{You are an AI assistant skilled in rewriting prompts in diverse and effective ways. You can provide well-structured and detailed rewordings that maintain the original meaning while improving clarity and variety.}
\end{quote}

\textbf{SIRA.} In our implementation, we strictly adopt the hyperparameter settings from the original SIRA paper~\citep{SIRA}. In particular, the self‑information masking threshold is set to $\varepsilon= 0.30$, meaning that tokens whose self‑information exceeds the 30th percentile are masked during the attack. All algorithm implementations and watermarking applications are based on the official SIRA code\footnote{https://github.com/Allencheng97/Self-information-Rewrite-Attack}. For each watermarking scheme, we utilize the \textbf{same model} both to compute self‑information and to conduct the attack. This alignment ensures that our robustness evaluation is conducted under a threat model where the adversary uses the same model architecture for attack and estimation, but is still constrained in practical terms (e.g. limited resources).

\textbf{DIPPER.} We adopt the DIPPER~\citep{krishna2023paraphrasing} paraphrase model as one of our baseline methods. We adopt the same settings following~\citep{SIRA}, setting lexical diversity = 60 and order diversity = 40. These parameters respectively control the extent of vocabulary variation and the degree of reordering of sentences or content segments. Such settings allow us to test watermarking schemes under more challenging rewriting attacks while still preserving text coherence and meaning.

\textbf{LLM Paraphraser.} In this setting, we utilize GPT-4o-2024-08-06 as the paraphrasing model, with its performance assessed under the prompt configuration of our Base method.

\textbf{Implementation Details.} All methods—Base, Think, SysP, and SIRA—are executed using the vLLM inference engine to accelerate generation. We set the sampling parameters as follows: temperature = 0.7, top\_p = 0.95, and top\_k = 20.

\textbf{Think + SysP.} Additionally, we evaluate a combined method that integrates both the system prompt and reasoning capability (i.e., Think + SysP). Results, presented in Tables~\ref{tab::MainresultDetailtab1}, \ref{tab::MainresultDetailtab2}, \ref{tab::MainresultDetailtab3}, \ref{tab::MainresultDetailtab4}, \ref{tab::MainresultDetailtab5}, \ref{tab::MainresultDetailtab6}, show that combining these two techniques further improves attack performance. This aligns with our theoretical finding that enriching the attack context reduces the adaptive robustness radius of watermarking schemes.

\subsection{Implementation Details of RLCracker.}

Recall that our {RLCracker} method adopts a token-wise optimization framework inspired by GRPO~\citep{shao2024deepseekmath}, requiring only question--watermarked response pairs \((q, wr)\) to effectively learn watermark evasion policies. The RLCracker is implemented based on the GRPO component from the Hugging Face TRL library~\citep{vonwerra2022trl}, where we directly adopt the implementation of the KL divergence term $\KL[\pi_\theta|\pi_{ref}]$. Following the KL divergence setting in OpenR1 project~\citep{openr1}, we use $\beta = 0.04$ to control the distributional shift between the policy model and the reference model.

The attack policy \(\pi_\theta\) is updated iteratively by sampling outputs \(\{o_1, \dots, o_G\}\) from the previous policy \(\pi_{\theta_{\text{old}}}\) and maximizing the following training objective:
\begin{align*}
    \mathcal{J}(\theta) \approx\; 
    \mathbb{E}_{\{o_i\}\sim\pi_{\theta_{\mathrm{old}}}}
    \frac{1}{G} \sum_{i=1}^G \frac{1}{|o_i|} \sum_{t=1}^{|o_i|} & 
      \Biggl[ \frac{\pi_\theta(o_{i,t}\mid q, o_{i,<t})}
      {\pi_{\theta_{\mathrm{old}}}(o_{i,t}\mid q, o_{i,<t})}
      \bigl(w_1 \hat{A}_i + w_2 \Delta r_{i,t} \bigr)  \\
    & \quad - \beta\KL\bigl[\pi_\theta\|\pi_{\mathrm{ref}}\bigr]
    \Biggr] - w_3\,\mathrm{PPL}(\pi_{\theta}, \{o_i\}),
\end{align*}
where $\Delta r_{i,t} = D_{\text{KL},i,t}(Q\|P_{wm}) - D_{\text{KL},i,t}(Q\|P_h)$,  
\(\pi_{\mathrm{ref}}\) denotes the reference policy approximating human-like distributions, and \(w_1, w_2, w_3\) are the weights of the respective components.  
The remaining terms impose token-wise KL regularization and a perplexity penalty to preserve fluency. Here, we use the original state of the policy model as our reference model \(\pi_{\mathrm{ref}}\). {Notably, to better fit $P_{wm}$, we use the Qwen3-0.6B model as the reference for the KL rewards during the training of Qwen3-8B.}

\textbf{Reward Components.}
We incorporate two reward signals to encourage semantic preservation and effective watermark evasion:

\begin{itemize}
    \item \textbf{Semantic Reward.} The semantic reward \(A_i \in [-1, 1]\) is computed based on the P-SP score~\citep{wieting2021paraphrastic} between the generated output and the original watermarked response. To enhance gradient flow and emphasize semantic fidelity, we apply a sigmoid-based scaling with a threshold of 0.85. This maps P-SP scores in the range [0.7, 1.0] to reward values between -1 and 1. Specifically,
    \[
    A_i = \frac{2}{1 + e^{-x}} - 1, \quad \text{where } x = \log\left(\frac{0.975}{0.025}\right) \cdot \text{semantic\_score}.
    \]
    The advantage term is normalized as \(\hat{A}_i = \frac{A_i - \text{mean}(A)}{\text{std}(A)}\).

    \item \textbf{Token-wise KL Reward.} This reward encourages each token’s distribution under $\pi_\theta$ to align with the human-like reference while diverging from watermark-induced patterns. The token-wise KL term is estimated using the standard PPO-style estimator~\citep{schulman2020kl}:
\[
D_{\text{KL},i,t}(Q\|P_{*})
\approx 
\frac{\pi_{\mathrm{ref}}(o_{i,t}\mid *,\, o_{i,<t})}
     {\pi_{\theta}(o_{i,t}\mid wr,\, o_{i,<t})}
\;-\;
\log 
\frac{\pi_{\mathrm{ref}}(o_{i,t}\mid *,\, o_{i,<t})}
     {\pi_{\theta}(o_{i,t}\mid wr,\, o_{i,<t})}
\;-\; 1,
\]
where $* \in \{h, wm\}$ specifies whether the reference distribution corresponds to the human-like distribution $P_h$ or the watermark-induced distribution $P_{wm}$. In practice, we approximate $P_{wm}$ by passing the watermarked instance $wr$ through the reference model $\pi_{\mathrm{ref}}$ with a simple paraphrasing prompt, and we approximate $P_h$ by querying the same reference model with the original question $q$. During training, we use the \emph{same} instruction prompt as the \emph{Base method} to ensure consistency with prior watermarking setups.

\end{itemize}

\textbf{Reward Balancing and Dynamic Scaling.}
We introduce three coefficients \(w_1\), \(w_2\), and \(w_3\) to weight the contributions of the semantic advantage, KL reward, and perplexity penalty, respectively. To better balance semantic fidelity and watermark evasion effectiveness, we adopt a dynamic scaling strategy for \(w_1\):
$w_1 = \max\left(w_1' \cdot (1 - \text{mean}(A_i)), 1\right),$
which increases the emphasis on semantic alignment when P-SP scores are low, and encourages distributional exploration when semantic fidelity is already high.

\textbf{Hyperparameters.}
For each watermarking scheme, we train a dedicated RLCracker model. The training data consists of 100 watermarked samples, each containing 500 tokens. We train for 10 epochs using a batch size of 48 and a group size of \( G = 12 \), with a cosine learning rate scheduler. For the Qwen3-4B model, we set the learning rate to \( 2 \times 10^{-7} \), while for the remaining three models, we use a learning rate of \( 1 \times 10^{-6} \). Training takes approximately 1.5 hours for Qwen3-4B and 0.5 hours for Qwen3-0.6B when using four NVIDIA A100 GPUs. 

\begin{figure}[ht]
    \centering
    \includegraphics[width=1\textwidth]{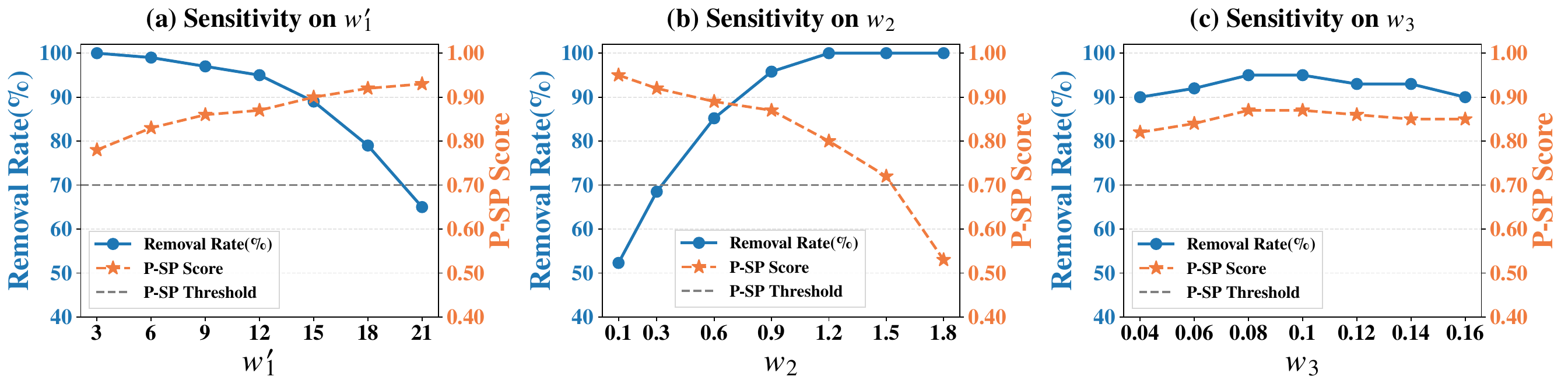}
    \caption{Sensitivity of RLCracker to Weight Settings under Qwen‑3‑4B for EWD Watermark}
    \label{fig:W_Sensitivity}
    \vspace{-0.3cm}
\end{figure}

We find that {RLCracker} is quite sensitive to the values of the weights \(w_{1}'\), \(w_{2}\), and \(w_{3}\): changes in these weights lead to significant variation in performance. As shown in Figure~\ref{fig:W_Sensitivity}, we trained Qwen‑3‑4B to remove the EWD watermark, and recorded how both the removal rate and the PSP score vary when each of the three weights is changed. Here, the removal rate is defined as the proportion of watermark that is removed, regardless of semantic fidelity. When testing \(w_{1}'\), we fix \(w_{2}=0.9\) and \(w_{3}=0.1\); when testing \(w_{2}\), we fix \(w_{1}'=12\) and \(w_{3}=0.1\); when testing \(w_{3}\), we fix \(w_{1}'=12\) and \(w_{2}=0.9\).
From these experiments, we observe that setting \((w_{1}' = 12,\; w_{2} = 0.9,\; w_{3} = 0.1)\) yields strong performance: the model achieves a high semantic similarity (P‑SP score of 0.87) while also maintaining a removal rate of 95\%, resulting in an ESR of 93.5\%. To facilitate reproducibility, we list the specific weight configurations used for each watermarking scheme and attack model in Table~\ref{tab::hyperparamsRLC}.

\begin{table}[ht]
\centering
\caption{Hyperparameters used for {RLCracker} across models and watermarking schemes.}
\label{tab::hyperparamsRLC}
\resizebox{\textwidth}{!}{
\begin{tabular}{lcccccccccccc}
\toprule
\textbf{Model} & \textbf{Weight} & EWD & PF & SWEET & Unigram & KGW & KGW\_self & SIR & XSIR & UPV & SynthID \\
\midrule

\multirow{3}{*}{Qwen3-0.6B} 
  & $w_1'$ & 12 & 12 & 12 & 5  & 6 & 8  & 12 & 12 & 12 & 12 \\
  & $w_2$  & 0.9 & 0.9 & 0.9 & 0.08 & 0.1 & 0.5 & 0.9 & 0.9 & 0.9 & 0.9 \\
  & $w_3$  & 0.1 & 0.1 & 0.1 & 0.1  & 0.1 & 0.1 & 0.1 & 0.1 & 0.1 & 0.1 \\
\midrule

\multirow{3}{*}{Qwen3-1.7B} 
  & $w_1'$ & 12 & 12 & 12 & 5  & 6 & 8  & 12 & 12 & 12 & 12 \\
  & $w_2$  & 0.9 & 0.9 & 0.9 & 0.08 & 0.1 & 0.5 & 0.9 & 0.9 & 0.9 & 0.9 \\
  & $w_3$  & 0.1 & 0.1 & 0.1 & 0.1  & 0.1 & 0.1 & 0.1 & 0.1 & 0.1 & 0.1 \\
\midrule

\multirow{3}{*}{Qwen3-4B} 
  & $w_1'$ & 12 & 12 & 12 & 5  & 6 & 8  & 12 & 12 & 12 & 12 \\
  & $w_2$  & 0.9 & 0.9 & 0.9 & 0.08 & 0.1 & 0.5 & 0.9 & 0.9 & 0.9 & 0.9 \\
  & $w_3$  & 0.1 & 0.1 & 0.1 & 0.1  & 0.1 & 0.1 & 0.1 & 0.1 & 0.1 & 0.1 \\
\midrule

\multirow{3}{*}{Qwen3-8B} 
  & $w_1'$ & 12 & 12 & 12 & 5  & 6 & 8  & 12 & 12 & 12 & 12 \\
  & $w_2$  & 0.9 & 0.9 & 0.9 & 0.08 & 0.1 & 0.5 & 0.9 & 0.9 & 0.9 & 0.9 \\
  & $w_3$  & 0.1 & 0.1 & 0.1 & 0.1  & 0.1 & 0.1 & 0.1 & 0.1 & 0.1 & 0.1 \\
\midrule

\multirow{3}{*}{\makecell{Qwen2.5-3B\\-Instruct}} 
  & $w_1'$ & 0 & 0 & 0 & 0  & 0 & 0  & 0 & 0 & 0 & 0 \\
  & $w_2$  & 0.9 & 0.6 & 0.8 & 0.03 & 0.1 & 0.4 & 0.7 & 0.7 & 0.8 & 0.5 \\
  & $w_3$  & 0.1 & 0.1 & 0.1 & 0.1  & 0.1 & 0.1 & 0.1 & 0.1 & 0.1 & 0.1 \\
\bottomrule
\end{tabular}
}
\end{table}

\section{Extended Experimental Results and Analysis}
\label{appen::expDetailsadnAna}

\subsection{Robustness of Watermarking Schemes across Models and Methods.}
\label{appen::AttackResults}

We present detailed watermark‐evasion results for various methods across different models and watermarking schemes, evaluated on two lengths of watermarked texts (500 and 1500 tokens). We compare the following methods: Base, SysP, Think, Think+SysP, SIRA, and our proposed RLCracker. Each table reports, for each method and watermark scheme, the evasion success rate (ESR), the Removal Rate (Rem.), the proportion of watermark removed regardless of semantic similarity, and the average P‑SP score after rephrasing. We also use perplexity (PPL.) and ChatGPT as a judge (GPTScore) to evaluate the quality of rephrased texts. Specifically, we use Llama3.1-8B-Instruct as the base model for measuring perplexity. For ChatGPT as a judge, we use GPT-4o-2024-08-06. The prompt template is as follows:

\begin{tcolorbox}[
    title=Prompt template for ChatGPT as a Judge,
    colframe=gray, 
    colback=gray!15,
    coltitle=gray,
    fonttitle=\bfseries\color{white},
    rounded corners,
    enhanced,
    left=6pt, right=6pt, top=6pt, bottom=6pt,
    boxrule=1pt,
    arc=6pt,
    width=\linewidth
]
\detokenize{
        You are an impartial evaluator. You will be provided with an original text and a paraphrased version of that text. 
        
        Your task is to assess the paraphrased text based on the following criteria:
        
        1. The degree to which the paraphrased text maintains the meaning of the original text.
        
        2. The fluency, coherence, and clarity of the paraphrased text.
        
        3. The extent to which the style and tone of the paraphrased text match those of the original text.

        Please provide an objective evaluation and rate the paraphrased text on a scale of 1 to 10:
        
        - A rating of 1 indicates that the paraphrased text completely alters the meaning of the original text.
        
        - A rating of 10 indicates that the paraphrased text preserves the exact meaning of the original text, maintaining fluency, clarity, and style.

        Your response should follow this strict format and return the score only: 
  }.
  
Here’s the original text: { [watermarked text]} 

Here’s the paraphrased text: { [rephrased text]}
\end{tcolorbox}

\textbf{Notably, the performance of RLCracker is evaluated using the same prompt as the SysP. method with the system prompt.}

As shown in Tables~\ref{tab::MainresultDetailtab1}, \ref{tab::MainresultDetailtab2}, \ref{tab::MainresultDetailtab3}, \ref{tab::MainresultDetailtab4}, \ref{tab::MainresultDetailtab5}, \ref{tab::MainresultDetailtab6}, under the Base method, smaller models ($\leq$ 4B parameters) consistently exhibit low ESR. For example, Qwen3‑0.6B attains approximately 13.5\% ESR on 500‑token EWD texts, but only about 3.5\% for 1500‑token EWD, with a removal rate of ~6.2\%. These observations align with prior work~\citep{kirchenbauer2023reliability}. At the same time, enriching the attack context via SysP, Think, or their combination (Think+SysP) yields substantial improvements in ESR, in agreement with Theorem~\ref{thm:kl}. {Moreover, as shown in Tables~\ref{tab::MainresultDetailtabQuality1}, \ref{tab::MainresultDetailtabQuality2}, \ref{tab::MainresultDetailtabQuality3}, \ref{tab::MainresultDetailtabQuality4}, \ref{tab::MainresultDetailtabQuality5}, \ref{tab::MainresultDetailtabQuality6}, all four methods maintain high semantic similarity, as evidenced by PPL remaining close to that of the original watermarked text, and the GPT score consistently exceeding 8. }However, the extent of improvement varies across watermark schemes, and no single method proves to be uniformly optimal.

{We also observe that SIRA~\citep{SIRA} maintains low ESR across different schemes. Although its removal rate is relatively high, its average P-SP scores remain low, which limits its overall effectiveness. Furthermore, its PPL, which exceeds that of the original text, and its GPT score below 5 indicate that the generated text quality is suboptimal.}
In contrast, RLCracker consistently achieves strong performance across all models, delivering high removal rates while preserving high average P‑SP. When using the same prompt, RLCracker outperforms all baselines, and in some settings even surpasses GPT‑4o in ESR, demonstrating its superior effectiveness.

\begin{table}[ht]
    \centering
    \caption{Evasion Success Rate (ESR, \%) across models and watermarking schemes (500 tokens).}
    \footnotesize
    \label{tab::MainresultDetailtab1}
    \setlength{\tabcolsep}{2.8pt} % 默认是 6pt
    \begin{tabular}{llcccccccccccc}
        \toprule
        \multirow{2}{*}{\textbf{Model}} & \multirow{2}{*}{\textbf{Method}} & \multicolumn{3}{c}{\textbf{EWD}} & \multicolumn{3}{c}{\textbf{SWEET}} & \multicolumn{3}{c}{\textbf{XSIR}} & \multicolumn{3}{c}{\textbf{Unigram}} \\
        \cmidrule(lr){3-5} \cmidrule(lr){6-8} \cmidrule(lr){9-11} \cmidrule(lr){12-14}
        & & ESR & Rem. & P-SP & ESR & Rem. & P-SP & ESR & Rem. & P-SP & ESR & Rem. & P-SP \\
        \midrule
        
        \multirow{6}{*}{Qwen3-0.6B} 
            & Base           & 13.5 & 14.8 & 0.94 & 21.2 & 22.2 & 0.94 & 29.0 & 31.2 & 0.92 & 35.8 & 64.8 & 0.78 \\
            & SysP.          & 36.5 & 39.0 & 0.91 & 44.8 & 46.8 & 0.93 & 52.5 & 55.8 & 0.90 & 42.5 & 82.2 & 0.73 \\
            & Think          & 28.5 & 33.0 & 0.91 & 35.3 & 40.6 & 0.92 & 37.2 & 40.8 & 0.91 & 34.2 & 76.8 & 0.71 \\
            & Think+SysP.    & 48.2 & 53.2 & 0.89 & 58.8 & 62.5 & 0.91 & 50.7 & 55.0 & 0.89 & 37.5 & 88.8 & 0.67 \\
            & SIRA             & 0.0 & 40.2 & 0.25 & 0.0 & 47.8 & 0.16 & 0.0 & 0.0 & 0.31 & 0.0 & 60.2 & 0.46 \\
            & RLCracker      & 91.5 & 98.5 & 0.83 & 92.5 & 99.0 & 0.82 & 86.5 & 94.5 & 0.83 & 67.0 & 71.5 & 0.86 \\
        
        \midrule
        \multirow{6}{*}{Qwen3-1.7B} 
            & Base           & 34.2 & 35.0 & 0.94 & 42.8 & 44.0 & 0.94 & 43.0 & 45.0 & 0.93 & 34.0 & 69.0 & 0.76 \\
            & SysP.          & 59.8 & 61.0 & 0.92 & 70.8 & 72.2 & 0.92 & 70.8 & 73.0 & 0.89 & 40.5 & 81.5 & 0.73 \\
            & Think          & 80.8 & 86.2 & 0.85 & 88.0 & 93.8 & 0.85 & 75.5 & 83.2 & 0.83 & 31.0 & 96.8 & 0.64 \\
            & Think+SysP.    & 84.0 & 89.0 & 0.85 & 86.5 & 92.0 & 0.85 & 79.8 & 87.2 & 0.83 & 34.8 & 97.8 & 0.65 \\
            & SIRA             & 0.2 & 86.0 & 0.28 & 0.0 & 91.2 & 0.19 & 0.2 & 85.5 & 0.33 & 0.0 & 93.5 & 0.39 \\
            & RLCracker      & 91.8 & 97.8 & 0.83 & 91.0 & 97.2 & 0.86 & 86.0 & 96.5 & 0.81 & 73.0 & 76.5 & 0.81 \\
        
        \midrule
        \multirow{6}{*}{Qwen3-4B}
            & Base           & 39.2 & 41.2 & 0.94 & 53.2 & 55.0 & 0.94 & 50.2 & 52.2 & 0.94 & 39.2 & 54.5 & 0.84 \\
            & SysP.          & 54.2 & 54.8 & 0.94 & 65.8 & 67.5 & 0.94 & 66.8 & 69.8 & 0.92 & 43.8 & 72.2 & 0.79 \\
            & Think          & 76.0 & 80.0 & 0.87 & 83.5 & 88.5 & 0.88 & 75.2 & 80.5 & 0.86 & 39.5 & 82.0 & 0.72 \\
            & Think+SysP.    & 78.8 & 83.0 & 0.87 & 88.0 & 91.0 & 0.88 & 79.2 & 85.0 & 0.85 & 38.8 & 85.0 & 0.72 \\
            & SIRA             & 0.0 & 78.5 & 0.28 & 0.0 & 83.0 & 0.18 & 0.2 & 79.8 & 0.33 & 0.8 & 89.2 & 0.41 \\
            & RLCracker      & 93.3 & 95.8 & 0.87 & 95.8 & 98.8 & 0.85 & 85.5 & 93.5 & 0.85 & 73.5 & 75.2 & 0.87 \\
        
        \midrule
        \multirow{6}{*}{Qwen3-8B}
            & Base           & 72.0 & 73.5 & 0.91 & 70.8 & 81.8 & 0.92 & 77.0 & 79.8 & 0.90 & 40.5 & 63.7 & 0.82 \\
            & SysP.          & 72.8 & 73.8 & 0.92 & 81.5 & 83.0 & 0.92 & 82.0 & 83.8 & 0.89 & 43.5 & 71.5 & 0.80 \\
            & Think          & 89.5 & 94.0 & 0.85 & 92.0 & 95.8 & 0.86 & 81.5 & 87.8 & 0.83 & 40.8 & 82.2 & 0.73 \\
            & Think+SysP.    & 90.2 & 95.2 & 0.85 & 92.0 & 96.5 & 0.85 & 86.8 & 93.8 & 0.82 & 41.0 & 87.8 & 0.72 \\
            & SIRA             & 0.0 & 80.2 & 0.29 & 0.0 & 84.8 & 0.19 & 0.2 & 81.0 & 0.34 & 0.5 & 86.0 & 0.45 \\
            & {RLCracker}   & {{94.8}} & {{95.5}} & {{0.88}} & {{96.3}} & {{98.0}} & {{0.88}} & {{90.2}} & {{92.8}} & {{0.88}} & {{80.5}} & {{88.3}} & {{0.80}} \\

        \midrule
        \multirow{4}{*}{\makecell{Qwen2.5-3B\\-Instruct}} 
            & Base             & 31.2 & 42.0 & 0.89 & 46.8 & 53.5 & 0.90 & 68.8 & 71.3 & 0.90 & 37.2 & 58.5 & 0.81 \\
            & SysP.            & 75.5 & 78.5 & 0.90 & 83.0 & 88.8 & 0.90 & 78.5 & 81.0 & 0.89 & 43.5 & 72.0 & 0.80 \\
            & SIRA             & 0.2 & 76.8 & 0.28 & 0.2 & 90.8 & 0.19 & 0.2 & 81.5 & 0.33 & 0.8 & 96.5 & 0.40 \\
            & RLCracker      & 93.3 & 98.0 & 0.85 & 95.3 & 99.0 & 0.85 & 89.8 & 92.2 & 0.84 & 78.5 & 90.2 & 0.81 \\

        \midrule
        \multirow{1}{*}{GPT-4o}
          & ---             & 71.0 & 72.2 & 0.93 & 81.5 & 83.2 & 0.93 & 77.2 & 79.2 & 0.90 & 49.8 & 89.8 & 0.73 \\
        \multirow{1}{*}{DIPPER}
          & ---        & 30.3 & 31.5 & 0.82 & 52.8 & 56.3 & 0.83 & 0.0 & 0.0 & 0.83 & 33.3 & 78.8 & 0.65 \\
        \bottomrule
    \end{tabular}
\end{table}

\subsection{Impact of Prompting Strategies}
\label{appen::Sysprompts}
We evaluate the impact of different prompting strategies using eight distinct prompts. Each prompt is assigned an index, as shown in the corresponding table. As shown in Section~\ref{sec::Experiments}, the choice of prompt leads to substantial variation in model ESR, underscoring the importance of considering prompt diversity when assessing watermark robustness. Only by validating vulnerability under various prompts can we obtain a complete picture of a watermarking scheme’s robustness.

The eight user prompts are as follows (index and content):

\begin{enumerate}
  \item \detokenize{
###Instruction: Paraphrase the above texts. ###Your Response:
  }
  \item \detokenize{
###Instruction: You are an expert copy‑editor. Please rewrite the target text above in your own voice and paraphrase all sentences. Ensure that the final output contains the same information as the original text and has roughly the same length. Do not leave out any important details when rewriting in your own voice. ###Your Response:
  }
  \item \detokenize{
###Instruction: As an expert copy‑editor, please rewrite the target text above in your own voice while ensuring that the final output contains the same information as the original text and has roughly the same length. Please paraphrase all sentences and do not omit any crucial details. Additionally, please take care to provide any relevant information about public figures, organizations, or other entities mentioned in the text to avoid any potential misunderstandings or biases. ###Your Response:
  }
  \item \detokenize{
###Instruction: Rewrite the target text above using different words but keeping the same meaning and similar length. ###Your Response:
  }
  \item \detokenize{
###Instruction: Paraphrase the target text above without changing its meaning or length. ###Your Response:
  }
  \item \detokenize{
###Instruction: Restate the target text above using different wording. Keep the meaning and text length nearly the same. ###Your Response:
  }
  \item \detokenize{
###Instruction: Generate a version of the target text above that means the same and is about the same length. ###Your Response:
  }
  \item \detokenize{
###Instruction: Write a similar target text to the one above, keeping both the meaning and the number of words roughly the same. ###Your Response:
  }
\end{enumerate}

To further enrich the attack context, we incorporate a system prompt alongside the user prompt to guide the model’s performance from a system perspective. Specifically, we employ the same system prompt used in the SysP method, which is as follows:

\begin{quote}
 \textit{You are an AI assistant skilled in rewriting prompts in diverse and effective ways. You can provide well-structured and detailed rewordings that maintain the original meaning while improving clarity and variety.}
 \end{quote}

\subsection{Reasoning and Test-Time Scaling}
\label{appen::reasoning}

In Section~\ref{sec:mainresults}, we empirically validate the effectiveness of test-time scaling, previously shown to be theoretically beneficial in mathematical reasoning tasks~\citep{muennighoff2025s1}, for watermark removal in natural language text. Specifically, we evaluate on test sets constructed from watermarked data (500-token sequences) generated using SWEET, EWD, SIR, and PF watermarking schemes.

\begin{figure}[ht]
    \centering
    \includegraphics[width=0.98\textwidth]{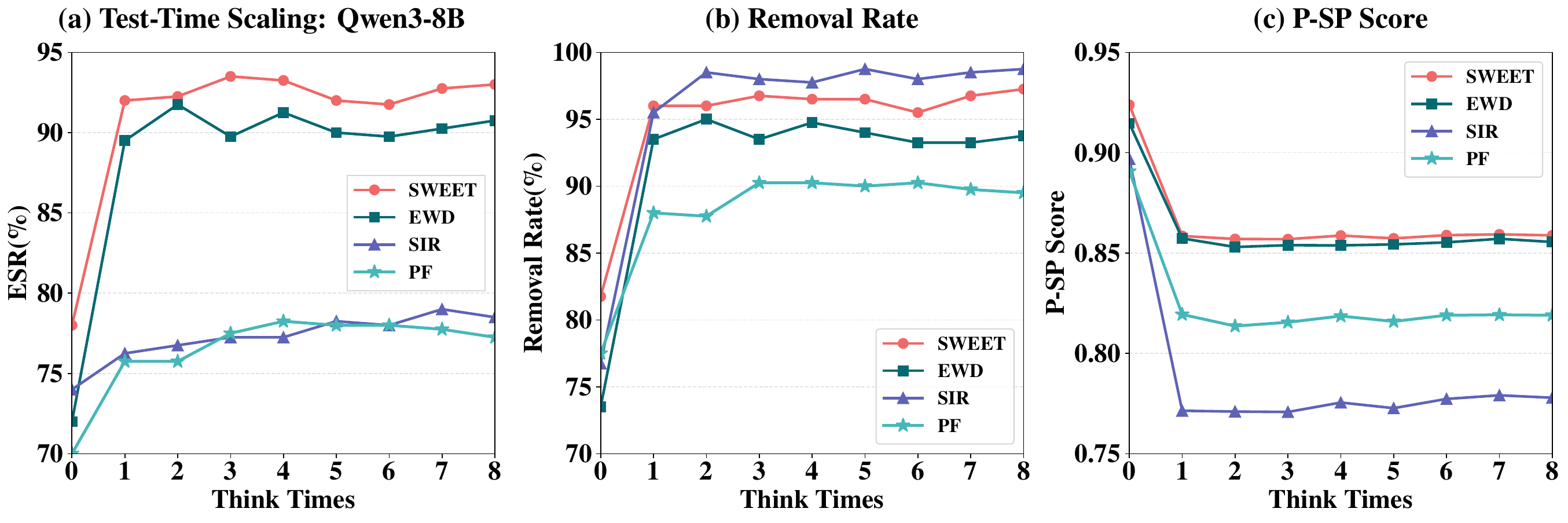}
    \caption{(a) shows test-time scaling on Qwen3-8B,  (b) shows remove rate on Qwen3-8B,  (c) shows p-sp score on Qwen3-8B. }
    \label{fig:TestTime_RemoveRate_Psp}
    \vspace{-0.3cm}
\end{figure}

In each trial, we truncate the reasoning text generated by the Qwen3-8B model just before the final output and append the following prompt: \textit{"Wait, I should rethink and reformulate it using different terms and structure, same meaning and same length."} This prompt encourages the model to continue reasoning in a deeper and more varied manner, effectively simulating the test-time scaling mechanism. We conduct all generations using the vLLM framework with the following decoding settings: temperature = 0.7, top-$p$ = 0.95, and top-$k$ = 20. The results are presented in Figure~\ref{fig:TestTime_RemoveRate_Psp}.

We observe that as the number of rethinking steps increases, both the ESR and the removal rate improve across all four watermarking schemes. However, the average P-SP score shows a sharp drop once rethinking is initiated, and only fluctuates mildly in subsequent steps. This drop is expected: the P-SP score positively correlates with watermark preservation. A higher P-SP score indicates higher semantic similarity and potentially lexical overlap with the original text, which in turn implies that the watermark is more likely to be preserved. As the model begins to rethink, the watermarked tokens are often rephrased or replaced, leading to reduced semantic similarity and hence a decrease in the P-SP score.

The consistent rise in ESR during this process demonstrates the efficacy of test-time scaling in weakening watermark robustness. This finding aligns well with our Theorem~\ref{thm:kl}, which suggests that richer generation contexts can degrade watermark stability.

\subsection{Detection Score Distribution}
\label{appen::RLCrackerResults}
The detection scores shown in Figure~\ref{fig:Distribution_TestTime}(b) indicate that RLCracker effectively pushes the distribution of rephrased texts toward that of unwatermarked outputs, and away from the watermarked distribution. To obtain the unwatermarked distribution, we use LLaMA3.1-8B-Instruct to generate outputs on the same test set without watermarking, and compute their detection scores. For the rephrased text distribution, we use detection scores of paraphrased outputs generated by Qwen3-4B trained with RLCracker. The clear separation between the rephrased and watermarked distributions further demonstrates the effectiveness of RLCracker in removing watermarks, highlighting a fundamental vulnerability in current watermarking schemes.

\subsection{RLCracker Generalization to Out-of-Distribution Settings}
\label{appen::oodgeneralization}
In Section~\ref{sec:ablationStudy}, we empirically evaluate whether models trained on data generated by LLaMA3.1-8B can effectively remove watermark patterns embedded in texts produced by other models under the same key \(s\). Specifically, we consider three watermarking schemes (EWD, SWEET, and PF) and generate 400 test samples of 500 tokens each using Qwen2.5-1.5B-Instruct and Qwen2.5-32B-Instruct, following the watermarking settings described in Appendix~C. We then assess model performance under both the base method and RLCracker, using the following paraphrasing prompt:
\detokenize{
###Instruction: Generate a version of the target text above that means the same and is about the same length. 
###Your Response:
}
The detailed results are provided in Table~\ref{tab::oodTabel1} and Table~\ref{tab::oodTabel2}.

Our findings show that RLCracker generalizes effectively to out-of-distribution (OOD) data, consistently achieving high Effective Semantic Rewriting (ESR) scores and strong watermark removal performance. The ESR achieved by RLCracker is generally higher when targeting outputs from larger models, which may be attributed to their higher quality and more coherent generations. These results further confirm both the effectiveness and the strong generalization ability of RLCracker, while also highlighting a fundamental vulnerability of current watermarking schemes.

\begin{table}[ht]
    \centering
    \caption{Performance of RLCracker on Qwen2.5-1.5B-Instruct generated data (500 tokens).}
    \footnotesize
    \label{tab::oodTabel1}
    \setlength{\tabcolsep}{5pt} % 默认是 6pt
    \begin{tabular}{llccccccccc}
        \toprule
        \multirow{2}{*}{\textbf{Model}} & \multirow{2}{*}{\textbf{Method}} & \multicolumn{3}{c}{\textbf{EWD}} & \multicolumn{3}{c}{\textbf{SWEET}} & \multicolumn{3}{c}{\textbf{PF}} \\
        \cmidrule(lr){3-5} \cmidrule(lr){6-8} \cmidrule(lr){9-11}
        & & ESR & Rem. & P-SP & ESR & Rem. & P-SP & ESR & Rem. & P-SP \\
        \midrule
        
        \multirow{2}{*}{Qwen3-0.6B} 
            & Base           & 10.8 & 16.0 & 0.89 & 13.0 & 20.0 & 0.89 & 14.5 & 21.3 & 0.84 \\
            & RLCracker      & 87.3 & 98.5 & 0.81 & 84.0 & 99.0 & 0.79 & 69.3 & 96.0 & 0.75 \\

        \midrule
        \multirow{2}{*}{Qwen3-1.7B} 
            & Base           &  22.3 & 28.0 & 0.89 & 25.3 & 31.5 & 0.90 & 29.8 & 37.5 & 0.86 \\
            & RLCracker      & 82.3 & 99.3 & 0.78 & 74.3 & 99.5 & 0.76 & 65.3 & 92.0 & 0.74 \\
        
        \midrule
        \multirow{2}{*}{\makecell{Qwen2.5-3B\\-Instruct}}
            & Base           & 22.8 & 34.0 & 0.86 & 28.0 & 39.8 & 0.86 & 26.8 & 40.0 & 0.92 \\
            & RLCracker      & 86.0 & 99.5 & 0.80 & 84.8 & 98.8 & 0.79 & 73.5 & 94.5 & 0.76 \\
        
        \midrule
        \multirow{2}{*}{Qwen3-4B}
            & Base           & 21.0 & 26.5 & 0.90 & 22.8 & 36.8 & 0.88 & 36.0 & 45.3 & 0.86 \\
            & RLCracker             & 84.0 & 96.5 & 0.81 & 83.0 & 96.8 & 0.81 & 73.5 & 93.8 & 0.77 \\

        \bottomrule
    \end{tabular}
\end{table}

\begin{table}[ht]
    \centering
    \caption{Performance of RLCracker on Qwen2.5-32B-Instruct generated data (500 tokens).}
    \footnotesize
    \label{tab::oodTabel2}
    \setlength{\tabcolsep}{5pt} % 默认是 6pt
    \begin{tabular}{llccccccccc}
        \toprule
        \multirow{2}{*}{\textbf{Model}} & \multirow{2}{*}{\textbf{Method}} & \multicolumn{3}{c}{\textbf{EWD}} & \multicolumn{3}{c}{\textbf{SWEET}} & \multicolumn{3}{c}{\textbf{PF}} \\
        \cmidrule(lr){3-5} \cmidrule(lr){6-8} \cmidrule(lr){9-11}
        & & ESR & Rem. & P-SP & ESR & Rem. & P-SP & ESR & Rem. & P-SP \\
        \midrule
        
        \multirow{2}{*}{Qwen3-0.6B} 
            & Base           & 12.5 & 13.3 & 0.94 & 15.8 & 16.5 & 0.95 & 14.3 & 20.8 & 0.88 \\
            & RLCracker      & 96.0 & 99.8 & 0.84 & 95.8 & 98.8 & 0.85 & 81.5 & 94.5 & 0.79 \\

        \midrule
        \multirow{2}{*}{Qwen3-1.7B} 
            & Base           & 36.0 & 36.8 & 0.94 & 41.5 & 42.3 & 0.94 & 34.0 & 40.0 & 0.88 \\
            & RLCracker      & 95.8 & 99.8 & 0.85 & 94.8 & 98.8 & 0.84 & 80.3 & 97.3 & 0.77 \\
        
        \midrule
        \multirow{2}{*}{\makecell{Qwen2.5-3B\\-Instruct}}
            & Base           & 28.0 & 28.8 & 0.93 & 33.8 & 35.3 & 0.93 & 29.3 & 36.0 & 0.88 \\
            & RLCracker      & 97.3 & 100. & 0.86 & 98.8 & 99.5 & 0.87 & 84.5 & 96.3 & 0.81 \\
        
        \midrule
        \multirow{2}{*}{Qwen3-4B}
            & Base           & 42.3 & 43.3 & 0.94 & 48.3 & 52.8 & 0.94 & 43.8 & 53.3 & 0.88 \\
            & RLCracker           & 98.0 & 98.8 & 0.88 & 98 & 99.8 & 0.87 & 84.5 & 96.0 & 0.83 \\

        \bottomrule
    \end{tabular}
\end{table}

\subsection{Impact of Training Set Length}
\label{appen::trainsetLength}
In Section~\ref{sec:ablationStudy}, we conduct an empirical study to examine how RLCracker’s effectiveness depends on two key training factors: the token length per sample and the total number of training samples. We train the model using data generated by Llama3.1-8B-Instruct with EWD watermark. We train the model for 10 epochs, using the learning rate of $1\times10^{-6}$ for Qwen3-0.6B and $2\times10^{-7}$ for Qwen3-4B. We test the model performance on 500-token EWD test dataset. Our results, detailed in Table~\ref{tab::trainsetScales}, indicate that Qwen3-0.6B attains an ESR of 82.5\% even when trained on as few as 50 samples of 250 tokens each. Increasing the token length under a fixed sample size further boosts ESR to 91.5\%. Expanding the number of training samples also improves performance, although the gains diminish once the sample size exceeds 100. These findings suggest that RLCracker can achieve strong watermark removal even with relatively modest training resources.

\begin{table}[ht]
    \centering
    \caption{Performance of RLCracker on EWD watermark under different scales of training sets.}
    \footnotesize
    \label{tab::trainsetScales}
    \setlength{\tabcolsep}{5pt} % 默认是 6pt
    \begin{tabular}{llccccccccc}
        \toprule
        \multirow{2}{*}{\textbf{Model}} & \multirow{2}{*}{\textbf{Tokens}} & \multicolumn{3}{c}{\textbf{50 Samples}} & \multicolumn{3}{c}{\textbf{100 Samples}} & \multicolumn{3}{c}{\textbf{200 Samples}} \\
        \cmidrule(lr){3-5} \cmidrule(lr){6-8} \cmidrule(lr){9-11}
        & & ESR & Rem. & P-SP & ESR & Rem. & P-SP & ESR & Rem. & P-SP \\
        \midrule
        
        \multirow{3}{*}{Qwen3-0.6B} 
        & 250 & 82.5 & 90.2 & 0.87 & 87.8 & 92.8 & 0.86 & 89.3 & 93.5 & 0.86 \\ 
        & 500 & 86.5 & 93.3 & 0.85 & 91.5 & 98.5 & 0.83 & 92.0 & 98.3 & 0.83 \\ 
        & 1500 & 91.5 & 98.3 & 0.83 & 92.5 & 98.5 & 0.83 & 92.8 & 99.3 & 0.82 \\ 
        \midrule
        \multirow{3}{*}{Qwen3-4B}
        & 250 & 87.8 & 94.5 & 0.86 & 92.0 & 95.3 & 0.87 & 92.8 & 98.3 & 0.87 \\ 
        & 500 & 90.0 & 96.3 & 0.87 & 93.3 & 95.8 & 0.87 & 93.8 & 99.8 & 0.87 \\ 
        & 1500 & 92.3 & 97.0 & 0.87 & 95.3 & 98.0 & 0.86 & 95.3 & 100. & 0.86 \\ 

        \bottomrule
    \end{tabular}
\end{table}

\subsection{RLCracker Effectiveness on Mixed-Key Training Data}
\label{appen::mixedKeyRobustness}

\textbf{Training data.}
To mimic realistic collection pipelines where watermarked text may be produced under different hash keys~\citep{liu2024can}, we construct mixed-key training sets for watermark EWD, PF, and SWEET. We construct the training datasets from the Reddit WritingPrompts~\citep{verma2024ghostbuster}. 
% Watermarks are applied with MarkLLM under the key assignments in Table~\ref{tab:mixedkey_keys}. Key$_1$ is the default key from the MarkLLM toolkit~\citep{pan2024markllm}; the remaining keys are randomly generated integers.
% For each watermark, we generate a training set of 100 samples with an average length of 500 tokens by mixing four distinct hash keys (25 samples per key) and the data used under different keys do not overlap.
{During the dataset construction process, we randomly generate 100 distinct 8-digit hash keys. Each hash key is used to generate a unique question-watermarked response pair, which forms the basis of our training data. Subsequently, we use this reconstructed mixed-key training set to train the model, with} all other watermarking hyperparameters and text-generation settings following the configuration used in the single-key experiments.
% All other watermarking hyperparameters and text-generation settings follow the single-key experiments.

\begin{table}[ht]
    \centering
    \caption{ESR, Removal Rate and P-SP of RLCracker trained on Single $|$ Mixed keys data}
    \footnotesize
    \label{tab:mixedkeytype_rem_psp}
    \setlength{\tabcolsep}{6pt}
    \begin{tabular}{llccccccccc}
        \toprule
        \multirow{2}{*}{\textbf{Model}} & \multirow{2}{*}{\textbf{Method}} & \multicolumn{3}{c}{\textbf{EWD}} & \multicolumn{3}{c}{\textbf{SWEET}} & \multicolumn{3}{c}{\textbf{PF}} \\
        \cmidrule(lr){3-5} \cmidrule(lr){6-8} \cmidrule(lr){9-11}
        & & \textbf{ESR} & \textbf{Rem.} & \textbf{P-SP} & \textbf{ESR} & \textbf{Rem.} & \textbf{P-SP} & \textbf{ESR} & \textbf{Rem.} & \textbf{P-SP} \\
        \midrule

        \multirow{2}{*}{Qwen3-0.6B}
            & Single & 91.5 & 98.5 & 0.83 & 92.5 & 99.0 & 0.82 & 76.8 & 94.8 & 0.78 \\
            & Mixed  & {88.8} & {91.3} & {0.87} & {90.5} & {96.5} & {0.88} & {75.5} & {88.3} & {0.81} \\
        \midrule

        \multirow{2}{*}{Qwen3-1.7B}
            & Single & 91.8 & 97.8 & 0.83 & 91.0 & 97.2 & 0.86 & 79.5 & 95.8 & 0.80 \\
            & Mixed  & {90.8} & {98.5} & {0.83} & {90.0} & {96.5} & {0.86} & {78.3} & {94.3} & {0.81} \\
        \midrule

        \multirow{2}{*}{Qwen3-4B}
            & Single & 93.3 & 95.8 & 0.87 & 95.8 & 98.8 & 0.85 & 82.3 & 95.0 & 0.82 \\
            & Mixed  & {92.3} & {95.5} & {0.86} & {94.3} & {97.5} & {0.87} & {80.5} & {89.5} & {0.85} \\
        \bottomrule
    \end{tabular}
\end{table}

\textbf{Evaluation.}
We evaluate models trained on the mixed-key datasets against single-key test sets, where the test key is the default MarkLLM key shown in Section~\ref{appen::algorithm_settings}. 
For each model (Qwen3-0.6B/1.7B/4B) and watermark (EWD, SWEET, PF), we report ESR, Rem. (removal rate), and P-SP score. 
As summarized in Table~\ref{tab:mixedkeytype_rem_psp}, mixed-key training yields only a slight degradation relative to in-domain single-key training, with the largest ESR drop being $\sim$2\% for PF on the 4B model, indicating that RLCracker remains robust under realistic mixed-key conditions.

\subsection{Impact of P-SP score Threshold on ESR}
\label{appen::pspThreshold}
\begin{figure}[ht]
    \centering
    \includegraphics[width=0.98\textwidth]{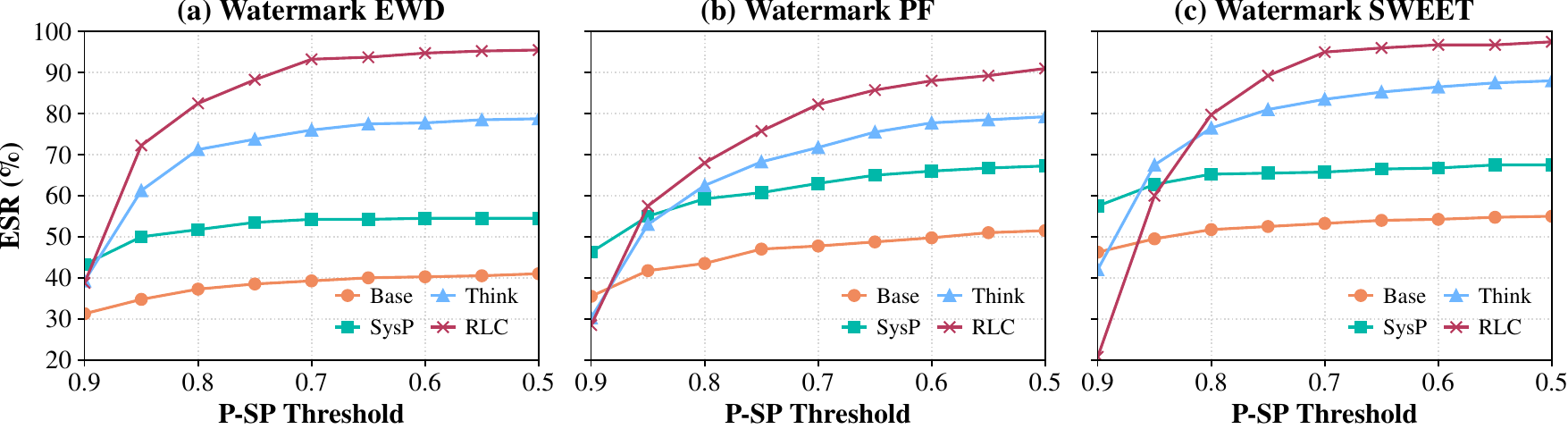}
    \caption{Variation of ESR with P-SP score threshold across three watermarks on 500-token texts}
    \label{fig:pspThresholdVar}
    \vspace{-0.3cm}
\end{figure}

{We analyzed the changes in ESR across four methods (Base, SysP., Think, and RLCracker) using the Qwen3-4B model as the P-SP score threshold varies. As depicted in Figure~\ref{fig:pspThresholdVar}, ESR increases for all methods as the P-SP score threshold decreases. The Base method effectively preserves the semantics of the text but struggles to remove the watermark adequately. In contrast, RLCracker demonstrates the highest ESR, successfully eliminating the watermark while maintaining semantic integrity, with ESR levels stabilizing above 90\%.}

\subsection{Potential Application of Adaptive Robustness Radius.}
\label{appen::adaptiveRadiusEmpirical}

For designers of green-list based watermarks, who have full knowledge of the details of their algorithms and detection hyperparameter settings, our adaptive robustness radius $\rho^*$ offers an estimate of their watermark's resistance to removal attacks. We empirically explore this potential application by validating the relationship between the radius and removal rate (Rem). Specifically, we calculate the radius for four watermarking schemes, EWD, SWEET, KGW, and KGW\_self, on 1500-token texts across different models and attack contexts. As shown in Table~\ref{tab::adaptiveRadiusEmpirical}, we observe a strong negative correlation between the radius and removal rate, with Pearson and Spearman coefficients of -0.77 and below -0.9, respectively. These results indicate that our robustness radius provides a useful metric for assessing watermark robustness against removal attacks.

\begin{table}[ht]
    \centering
    \caption{Correlation between removal rate (Rem.) and KL adaptive radius lower bound $\rho^*$.}
    \footnotesize
    \label{tab::adaptiveRadiusEmpirical}
    \setlength{\tabcolsep}{6pt}
    \begin{tabular}{llccccccccc}
        \toprule
        \multirow{2}{*}{\textbf{Model}} & \multirow{2}{*}{\textbf{Method}} & \multicolumn{2}{c}{\textbf{EWD}} & \multicolumn{2}{c}{\textbf{SWEET}} & \multicolumn{2}{c}{\textbf{KGW}} & \multicolumn{2}{c}{\textbf{KGW\_self}} \\
        \cmidrule(lr){3-4} \cmidrule(lr){5-6} \cmidrule(lr){7-8} \cmidrule(lr){9-10}
        & & \textbf{Rem.} & \textbf{Radius} & \textbf{Rem.} & \textbf{Radius} & \textbf{Rem.} & \textbf{Radius} & \textbf{Rem.} & \textbf{Radius} \\
        \midrule

        \multirow{2}{*}{Qwen3-0.6B}
            & Base & 6.20 & 1.35 & 10.2 & 0.97 & 33.8 & 0.21 & 23.5 & 0.32 \\
            & SysP.  & 13.0 & 0.58 & 25.2 & 0.33 & 55.8 & 0.01 & 47.2 & 0.04 \\
        \midrule

        \multirow{2}{*}{Qwen3-1.7B}
            & Base & 8.50 & 0.95 & 14.5 & 0.64 & 41.8 & 0.11 & 37.5 & 0.14 \\
            & SysP. & 15.5 & 0.58 & 23.0 & 0.41 & 55.2 & 0.02 & 49.0 & 0.05 \\
        \midrule

        \multirow{2}{*}{Qwen3-4B}
            & Base & 19.8 & 0.57 & 27.8 & 0.36 & 42.5 & 0.10 & 41.8 & 0.09 \\
            & SysP. & 24.8 & 0.41 & 32.5 & 0.25 & 57.5 & 0.00 & 47 & 0.02 \\
        \midrule

        \multirow{2}{*}{Qwen3-8B}
            & Base & 51.5 & 0.04 & 59.2 & 0.00 & 85.0 & 0.00 & 72.0 & 0.00 \\
            & SysP.  & 48.8 & 0.02 & 64.5 & 0.00 & 90.0 & 0.00 & 75.2 & 0.00 \\

        \midrule
        \multicolumn{2}{c}{{Pearson Correlation}} & \multicolumn{2}{c}{-0.90} & \multicolumn{2}{c}{{-0.90}} & \multicolumn{2}{c}{{-0.77}} & \multicolumn{2}{c}{{-0.83}} \\
        \multicolumn{2}{c}{{Spearman Correlation}} & \multicolumn{2}{c}{{-0.95}} & \multicolumn{2}{c}{{-0.98}} & \multicolumn{2}{c}{{-0.99}} & \multicolumn{2}{c}{{-0.90}} \\
        \bottomrule
    \end{tabular}
\end{table}

\subsection{Sensitivity to the Watermark-Induced Proxy Distribution}
\label{appen::pwm_proxy_sensitivity}

We evaluate the sensitivity of RLCracker to the choice of proxy generator used to construct the watermark-induced distribution \(P_{wm}\). We fix the attacker as Qwen3-0.6B, but construct \(P_{wm}\) using either Qwen3-0.6B or Qwen3-8B. As shown in Table~\ref{tab::pwm_proxy_sensitivity}, using Qwen3-8B as the proxy generator leads to a clear performance drop on EWD, SWEET, and PF. This suggests that \(P_{wm}\) is sensitive to proxy choice: a stronger paraphraser such as Qwen3-8B may already substantially rewrite the watermarked text, weakening the local watermark structure that \(P_{wm}\) is intended to capture.

\begin{table}[ht]
\centering
\caption{Sensitivity of RLCracker to the proxy generator used for constructing \(P_{wm}\).}
\label{tab::pwm_proxy_sensitivity}
\small
\begin{tabular}{lcccccc}
\toprule
\multirow{2}{*}{\textbf{\(P_{wm}\) Gen.}} 
& \multicolumn{2}{c}{\textbf{EWD}} 
& \multicolumn{2}{c}{\textbf{SWEET}} 
& \multicolumn{2}{c}{\textbf{PF}} \\
\cmidrule(lr){2-3} \cmidrule(lr){4-5} \cmidrule(lr){6-7}
& \textbf{ESR} & \textbf{P-SP} 
& \textbf{ESR} & \textbf{P-SP} 
& \textbf{ESR} & \textbf{P-SP} \\
\midrule
Qwen3-0.6B & 91.5 & 0.83 & 92.5 & 0.82 & 76.8 & 0.78 \\
Qwen3-8B   & 57.0 & 0.78 & 56.8 & 0.77 & 69.8 & 0.76 \\
\bottomrule
\end{tabular}
\end{table}

\subsection{Impact of Distribution Mismatches on Watermark Removal Attack Success.}
\label{appen::ComparisonSimilarModelFamily}

To investigate the impact of distribution mismatches between the watermark generator and the attacker model on the success rate of watermark removal attacks, we conducted cross-model experiments using the Qwen3 family models: 0.6B, 1.7B, and 4B. For each model, we generated 500-token EWD watermarked texts, which were then utilized to perform watermark removal attacks and to train the RLCracker model.

As shown in the Table~\ref{tab::distributionmismatches}, when the distributions of the watermark generator and attacker model are similar, RLCracker achieves a high ESR, significantly outperforming the other baseline methods. Notably, even when the watermark generator and attacker model are the same, RLCracker can effectively direct the model to generate paraphrased text that diverges from the watermarked distribution. This highlights that RLCracker's effectiveness does not rely on internal output distribution discrepancies between the watermark generator and attacker model.

\begin{table}[ht]
    \centering
    \caption{Limited impact of distribution mismatches on watermark removal attack success}
    \footnotesize
    \label{tab::distributionmismatches}
    \begin{tabular}{llccccccccc}
        \toprule
         \multirow{2}{*}{\diagbox{ \textbf{Att.}}{ \textbf{Gen.}}} &  & \multicolumn{3}{c}{\textbf{Qwen3-0.6B}} & \multicolumn{3}{c}{\textbf{Qwen3-1.7B}} & \multicolumn{3}{c}{\textbf{Qwen3-4B}} \\
        \cmidrule(lr){3-5} \cmidrule(lr){6-8} \cmidrule(lr){9-11}
         & \textbf{Method} & ESR & Rem. & P-SP & ESR & Rem. & P-SP & ESR & Rem. & P-SP \\
        \midrule
        
        \multirow{4}{*}{Qwen3-0.6B} 
        & Base & 12.3 & 13.0 & 0.94 & 21.0 & 21.8 & 0.94 & 28.8 & 30.0 & 0.93 \\ 
        & Sysp. & 17.3 & 18.3 & 0.94 & 48 & 49.5 & 0.92 & 36.5 & 38.5 & 0.92 \\ 
        & Think & 17.3 & 23.5 & 0.91 & 65.0 & 71.5 & 0.87 & 63.8 & 69.5 & 0.87 \\ 
        & RLCracker & 87.0 & 95.8 & 0.81 & 87.3 & 96.0 & 0.82 & 88.3 & 98.0 & 0.81 \\ 
        \midrule
        \multirow{4}{*}{Qwen3-1.7B} 
        & Base & 24.3 & 24.8 & 0.94 & 26.0 & 26.8 & 0.96 & 59.3 & 59.3 & 0.95 \\ 
        & Sysp. & 41.0 & 41.8 & 0.94 & 56.5 & 57.5 & 0.94 & 70.5 & 70.8 & 0.94 \\ 
        & Think & 30.5 & 33.0 & 0.93 & 77.3 & 79.0 & 0.90 & 79.8 & 82.3 & 0.90 \\ 
        & RLCracker & 90.8 & 96.0 & 0.86 & 91.3 & 97.5 & 0.86 & 92.0 & 98.3 & 0.88 \\ 
        \midrule
        \multirow{4}{*}{Qwen3-4B} 
        & Base & 30.8 & 31.3 & 0.96 & 39.0 & 39.5 & 0.97 & 49.8 & 50.0 & 0.96 \\ 
        & Sysp. & 46.5 & 46.8 & 0.95 & 70.5 & 71.5 & 0.95 & 69.8 & 70.5 & 0.95 \\ 
        & Think & 37.5 & 40.0 & 0.93 & 88.5 & 86.8 & 0.91 & 79.3 & 81.8 & 0.92 \\ 
        & RLCracker & 91.5 & 98.3 & 0.87 & 92.3 & 97.8 & 0.87 & 93.3 & 97.5& 0.87\\ 
        \bottomrule
    \end{tabular}
\end{table}

\subsection{Evaluation on Robustly Tuned and Semantic Watermarks}
\label{app:robust_watermarks}

To evaluate whether RLCracker remains effective against watermarking schemes designed or tuned for stronger robustness, we additionally test semantic watermarks, robust KGW variants, and a model-weight watermark. We generate 500-token watermarked datasets using the same protocol as in Section~5 and use Qwen3-4B as the attacker. For GAUSSMark/RLWatermark, we follow the original detection convention and treat samples with $p$-value below $0.05$ as watermarked when computing ESR. 

Overall, as shown in Table~\ref{tab:robust_watermarks}, RLCracker consistently improves ESR over Base and SysP across all robust settings, with ESR above 63.3 in every case. The gains are especially pronounced for robust KGW variants, suggesting that increasing the embedding strength or greenlist rate alone does not eliminate the vulnerability exposed by adaptive distributional attacks. For SemStamp and k-SemStamp, ESR is lower than on several logit-based schemes, indicating that semantic watermarks can be more challenging, but RLCracker still substantially improves over prompting baselines while preserving acceptable semantic similarity.

\begin{table}[h]
\centering
\caption{RLCracker against robustly tuned and semantic watermarking schemes on 500-token texts.}
\label{tab:robust_watermarks}
\begin{tabular}{llccc}
\toprule
Watermark & Method & ESR $\uparrow$ & Rem. $\uparrow$ & P-SP $\uparrow$ \\
\midrule
SemStamp & Base & 11.0 & 62.8 & 0.64 \\
SemStamp & SysP. & 13.8 & 94.3 & 0.51 \\
SemStamp & RLCracker & \textbf{63.3} & 92.5 & 0.71 \\
\midrule
k-SemStamp & Base & 21.8 & 71.25 & 0.70 \\
k-SemStamp & SysP. & 22.0 & 95.3 & 0.62 \\
k-SemStamp & RLCracker & \textbf{75.5} & 97.0 & 0.70 \\
\midrule
KGW, $\gamma=0.5,\delta=8.0$ & Base & 9.5 & 45.0 & 0.69 \\
KGW, $\gamma=0.5,\delta=8.0$ & SysP. & 12.8 & 74.8 & 0.60 \\
KGW, $\gamma=0.5,\delta=8.0$ & RLCracker & \textbf{78.5} & 92.3 & 0.75 \\
\midrule
KGW, $\gamma=0.75,\delta=2.0$ & Base & 41.3 & 93.8 & 0.69 \\
KGW, $\gamma=0.75,\delta=2.0$ & SysP. & 21.0 & 100.0 & 0.57 \\
KGW, $\gamma=0.75,\delta=2.0$ & RLCracker & \textbf{90.5} & 98.5 & 0.81 \\
\midrule
RLWatermark / GAUSSMark & Base & 40.5 & 65.0 & 0.76 \\
RLWatermark / GAUSSMark & SysP. & 59.8 & 86.8 & 0.67 \\
RLWatermark / GAUSSMark & RLCracker & \textbf{87.5} & 92.3 & 0.83 \\
\bottomrule
\end{tabular}
\end{table}

\subsection{Effectiveness Comparison across Adaptive Watermark Removal Attack.}

To further assess the effectiveness of RLCracker, we conducted additional comparisons with two adaptive attacks featuring stronger threat model settings: the DPO attack~\citep{diaa2024optimizing} and the WS attack~\citep{jovanovic2024watermark}. These attacks assume complete knowledge of the watermark detector's details or the full algorithmic details. Specifically, we implemented the DPO attack following~\citep{diaa2024optimizing}, training the Qwen2.5-3B-Instruct model with both 100 preference pairs (DPO(100), matching the data budget used by RLCracker) and 7,000 preference pairs (DPO(7000), as outlined in~\citep{diaa2024optimizing}). We then compared the performance of the DPO attack and RLCracker on both 500-token and 1500-token test sets. For the WS attack, we followed the approach in~\citep{jovanovic2024watermark} and tested its performance using the Qwen2.5-3B-Instruct model on the 500-token KGW and KGW\_self test sets.

As shown in Table~\ref{tab:dpoAttacker1}, RLCracker significantly outperforms both DPO variants across all watermarking schemes, with the performance gap being particularly pronounced on longer (1500-token) sequences. Similarly, as presented in Table~\ref{tab:WSAttack}, RLCracker also outperforms the WS attack, further demonstrating that our improvements are due to the distributional framework rather than adjustments to the model tuning or scale.

\begin{table}[ht]
    \centering
    \caption{Performance of RLCracker and DPO attacker on Qwen2.5-3B-Instruct.}
    \footnotesize
    \label{tab:dpoAttacker1}
    \setlength{\tabcolsep}{6pt}
    \begin{tabular}{llccccccccc}
        \toprule
        \multirow{2}{*}{\textbf{Length}} & \multirow{2}{*}{\textbf{Method}} & \multicolumn{3}{c}{\textbf{Unigram}} & \multicolumn{3}{c}{\textbf{KGW\_self}} & \multicolumn{3}{c}{\textbf{PF}} \\
        \cmidrule(lr){3-5} \cmidrule(lr){6-8} \cmidrule(lr){9-11}
        & & \textbf{ESR} & \textbf{Rem.} & \textbf{P-SP} & \textbf{ESR} & \textbf{Rem.} & \textbf{P-SP} & \textbf{ESR} & \textbf{Rem.} & \textbf{P-SP} \\
        \midrule

        \multirow{3}{*}{500 tokens}
            & DPO(100) & 45.2 & 73.0 & 0.81 & 78.8 & 88.8 & 0.89 & 69.3 & 80.3 & 0.85 \\ 
            & DPO(7000) & 69.3 & 84.5 & 0.71 & 81.0 & 92.5 & 0.82 & 71.8 & 89.5 & 0.79 \\ 
            & RLCracker & 78.5 & 90.2 & 0.81 & 89.8 & 93.5 & 0.86 & 81.8 & 95.2 & 0.84 \\
        \midrule

        \multirow{3}{*}{1500 tokens}
            & DPO(100) & 7.25 & 27.3 & 0.69 & 65.8 & 70.5 & 0.87 & 41.3 & 60.8 & 0.83 \\ 
            & DPO(7000) & 20.8 & 45.3 & 0.67 & 74.0 & 85.3 & 0.84 & 47.0 & 71.5 & 0.75 \\ 
            & RLCracker & 98.5 & 100. & 0.92 & 78.0 & 89.0 & 0.83 & 77.8 & 90.3 & 0.81 \\ 
        \bottomrule
    \end{tabular}
\end{table}

\begin{table}[ht]
    \centering
    \caption{Performance of WS, DPO attacker, and RLCracker on 500 token test set.}
    \footnotesize
    \label{tab:WSAttack}
    \begin{tabular}{lcccccc}
        \toprule
        \multirow{2}{*}{\textbf{Method}} & \multicolumn{3}{c}{\textbf{KGW}} & \multicolumn{3}{c}{\textbf{KGW\_self}} \\
        \cmidrule(lr){2-4} \cmidrule(lr){5-7}
        & \textbf{ESR} & \textbf{Rem.} & \textbf{P-SP} & \textbf{ESR} & \textbf{Rem.} & \textbf{P-SP} \\
        \midrule
        WS & 91.0 & 98.5 & 0.87 & 89.0 & 92.5 & 0.86 \\
        DPO(7000) & 67.3 & 89.0 & 0.73 & 81.0 & 92.5 & 0.82 \\ 
        RLCracker & 97.5 & 99.3 & 0.86 & 89.8 & 93.5 & 0.86 \\ 
        \bottomrule
    \end{tabular}
\end{table}

We further conducted a direct comparison of RLCracker with the LoRA checkpoint provided by ~\cite{diaa2024optimizing}, which was trained using 7,000 preference pairs. As shown in the Table~\ref{tab:dpoAttacker2} and~\ref{tab:dpoAttacker3}, despite being trained on only 7,000 samples and employing smaller base models (e.g., Qwen3-0.6B and Qwen2.5-3B), RLCracker consistently matches or outperforms the DPO model across both 500-token and 1500-token settings for the KGW and KGW\_self watermarks. In the case of the Unigram watermark, where the DPO paraphrasers exhibit considerable underperformance, RLCracker achieves significantly higher ESR and stronger P-SP fidelity, underscoring its robust generalization capability. These results clearly demonstrate that RLCracker not only adapts effectively across various watermark types but also provides comparable or superior performance to the DPO paraphraser, all while operating under significantly lighter training requirements.

\begin{table}[ht]
    \centering
    \caption{Performance of RLCracker and DPO attacker across models.}
    \footnotesize
    \label{tab:dpoAttacker2}
    \setlength{\tabcolsep}{3pt} % 默认是 6pt
    \begin{tabular}{lcccccccccccc}
        \toprule
        \multirow{2}{*}{\textbf{Method}} & \multicolumn{3}{c}{\textbf{KGW (500)}} & \multicolumn{3}{c}{\textbf{KGW (1500)}} & \multicolumn{3}{c}{\textbf{KGW\_self (500)}} & \multicolumn{3}{c}{\textbf{KGW\_self (1500)}}\\
        \cmidrule(lr){2-4} \cmidrule(lr){5-7} \cmidrule(lr){8-10} \cmidrule(lr){11-13}
        & \textbf{ESR} & \textbf{Rem.} & \textbf{P-SP} & \textbf{ESR} & \textbf{Rem.} & \textbf{P-SP} & \textbf{ESR} & \textbf{Rem.} & \textbf{P-SP} & \textbf{ESR} & \textbf{Rem.} & \textbf{P-SP}\\
        \midrule

        DPO(Llama3.1-8B) & 67.5 & 95.3 & 0.71 & 47.5 & 92.5 & 0.76 & 84.5 & 91.3 & 0.86 & 74.0 & 88.3 & 0.84 \\ 
        RLC.(Qwen3-0.6B) & 80.8 & 84.0 & 0.89 & 48.5 & 51.5 & 0.88 & 85.2 & 90.5 & 0.87 & 66.8 & 80.0 & 0.84 \\ 
        RLC.(Qwen2.5-3B) & 97.5 & 99.3 & 0.86 & 58.0 & 67.3 & 0.90 & 89.8 & 93.5 & 0.86 & 78.0 & 89.0 & 0.83 \\ 
        RLC.(Qwen3-8B) & 91.5 & 94.5 & 0.90 & 64.5 & 87.5 & 0.78 & 90.8 & 94.8 & 0.90 & 84.3 & 97.3 & 0.83 \\ 
            
        \bottomrule
    \end{tabular}
\end{table}

\begin{table}[t]
    \centering
    \caption{Performance of RLCracker and DPO attacker across models.}
    \footnotesize
    \label{tab:dpoAttacker3}
    \begin{tabular}{lcccccc}
        \toprule
        \multirow{2}{*}{\textbf{Method}} & \multicolumn{3}{c}{\textbf{Unigram (500)}} & \multicolumn{3}{c}{\textbf{Unigram (1500)}} \\
        \cmidrule(lr){2-4} \cmidrule(lr){5-7}
        & \textbf{ESR} & \textbf{Rem.} & \textbf{P-SP} & \textbf{ESR} & \textbf{Rem.} & \textbf{P-SP} \\
        \midrule
        DPO(Llama3.2-3B) & 58.5 & 100 & 0.70 & 15.5 & 100 & 0.66 \\
        DPO(Qwen2.5-3B) & 61.3 & 85.0 & 0.71 & 16.8 & 96.0 & 0.66 \\ 
        RLCracker(Qwen3-0.6B) & 67.0 & 71.5 & 0.86 & 90.2 & 90.5 & 0.93 \\
        RLCracker(Qwen2.5-3B) & 78.5 & 90.3 & 0.81 & 98.5 & 100. & 0.92 \\ 
        RLCracker(Qwen3-8B) & 80.5 & 88.3 & 0.80 & 81.8 & 88.5 & 0.84 \\ 
        \bottomrule
    \end{tabular}
\end{table}

\subsection{Cross-Watermark Transferability}
\label{appen::cross_watermark_transfer}

In real-world scenarios, an attacker may not know which watermarking scheme was applied to the target text. We first evaluate whether an attack model trained on one watermark can transfer to other watermark families. Specifically, we train Qwen3-4B on EWD and evaluate it against EWD, PF, SWEET, Unigram, and KGW\_self. As shown in Table~\ref{tab::cross_watermark_transfer}, the EWD-trained model achieves strong performance on EWD, but transfers poorly to other watermarking schemes and can even underperform the base model. This suggests that an attacker trained to move away from one watermark-specific distribution does not necessarily move away from other watermark distributions, and may even shift closer to them.

\begin{table}[ht]
\centering
\caption{Cross-watermark transferability of watermark removal.}
\label{tab::cross_watermark_transfer}
\small
\begin{tabular}{lccccc}
\toprule
\textbf{Model} & \textbf{EWD} & \textbf{PF} & \textbf{SWEET} & \textbf{Unigram} & \textbf{KGW\_self} \\
\midrule
BaseModel   & 54.3 & 63.0  & 65.8  & 43.8 & 79.3 \\
EWD-Trained & 93.3 & 36.75 & 41.75 & 13.3 & 41.0 \\
\bottomrule
\end{tabular}
\end{table}

\subsection{Training on Mixed-Watermark Data}
\label{appen::mixed_watermark_training}

We further evaluate whether RLCracker can handle mixed-watermark training data, which better reflects the setting where the applied watermark family is unknown. We construct a 100-example synthetic training set by sampling 25 instances from each of EWD, PF, SWEET, and KGW\_self, and build a separate validation set by sampling 5 instances from each watermark family. We then train Qwen3-4B on this mixed dataset and evaluate it on both 500-token and 1500-token test sets. As shown in Table~\ref{tab::mixed_watermark_training}, mixed-watermark training leads to only a small ESR reduction compared with single-watermark training, with the largest drop below 4\%. These results suggest that RLCracker retains strong watermark removal performance even under mixed-watermark data training.

\begin{table}[ht]
\centering
\caption{Performance of RLCracker trained on single-watermark v.s. mixed-watermark data.}
\label{tab::mixed_watermark_training}
\small
\begin{tabular}{lcccc}
\toprule
\textbf{Method} & \textbf{EWD} & \textbf{SWEET} & \textbf{KGW\_self} & \textbf{PF} \\
\midrule
Base (500)        & 39.3 & 53.3 & 70.0 & 47.8 \\
RLC-Single (500)  & 93.3 & 95.8 & 88.8 & 82.3 \\
RLC-Mixed (500)   & 92.3 & 92.5 & 88.5 & 80.0 \\
\midrule
Base (1500)       & 10.0 & 12.3 & 36.0 & 25.8 \\
RLC-Single (1500) & 64.0 & 65.3 & 82.3 & 64.3 \\
RLC-Mixed (1500)  & 62.0 & 62.5 & 78.8 & 62.5 \\
\bottomrule
\end{tabular}
\end{table}

\subsection{Difference Between Approximated $P_h$ and the True Human-written Distribution.}

In RLCracker, we use a lightweight reference model (e.g., Qwen3-0.6B) to approximate the distribution of human-written text. Since it's not possible to directly obtain the true human-written distribution, we evaluate the effectiveness of this approximation by comparing the detection score distributions of model-generated text and natural text under various watermark detection algorithms. This comparison serves as a proxy to measure how well models of different sizes approximate the human-written distribution.

\begin{figure}[ht]
    \centering
    \includegraphics[width=\textwidth]{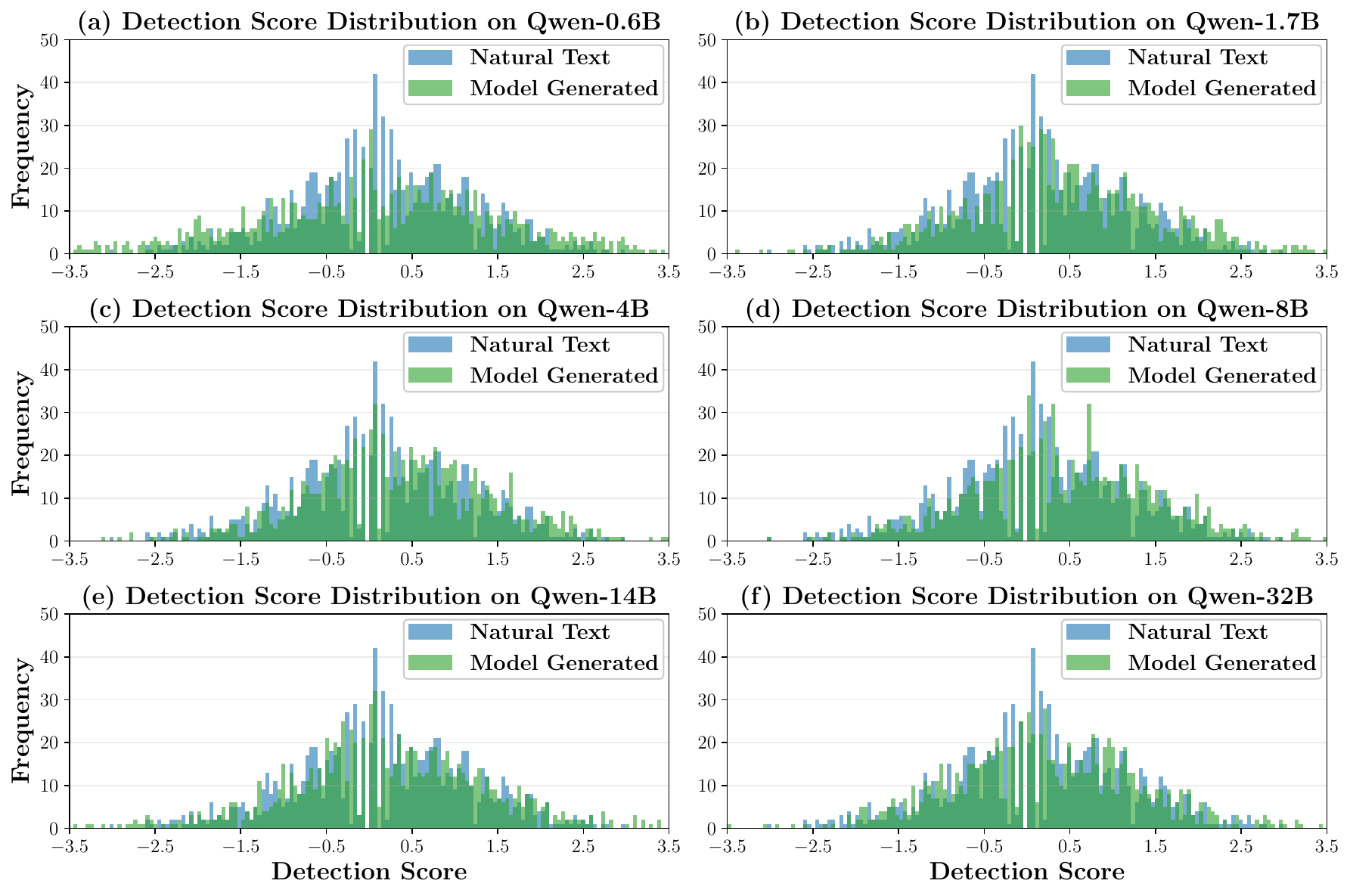}
    \caption{SWEET Detection Score Distribution across Different Model Sizes}
    \label{fig:detectionphSWEET}
    \vspace{-0.3cm}
\end{figure}

\begin{figure}[ht]
    \centering
    \includegraphics[width=\textwidth]{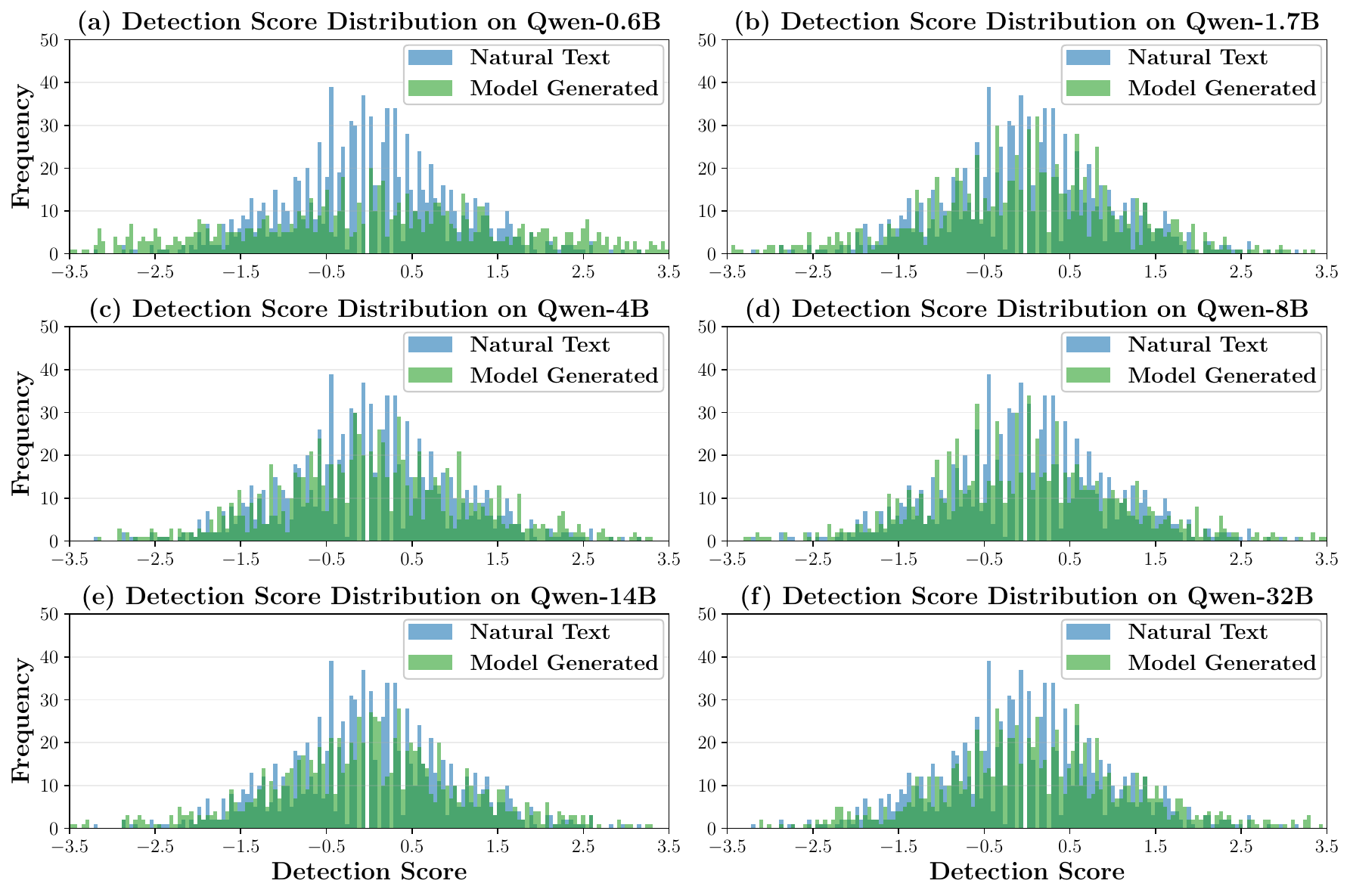}
    \caption{KGW Detection Score Distribution across Different Model Sizes}
    \label{fig:detectionphKGW}
    \vspace{-0.3cm}
\end{figure}

\begin{figure}[ht]
    \centering
    \includegraphics[width=\textwidth]{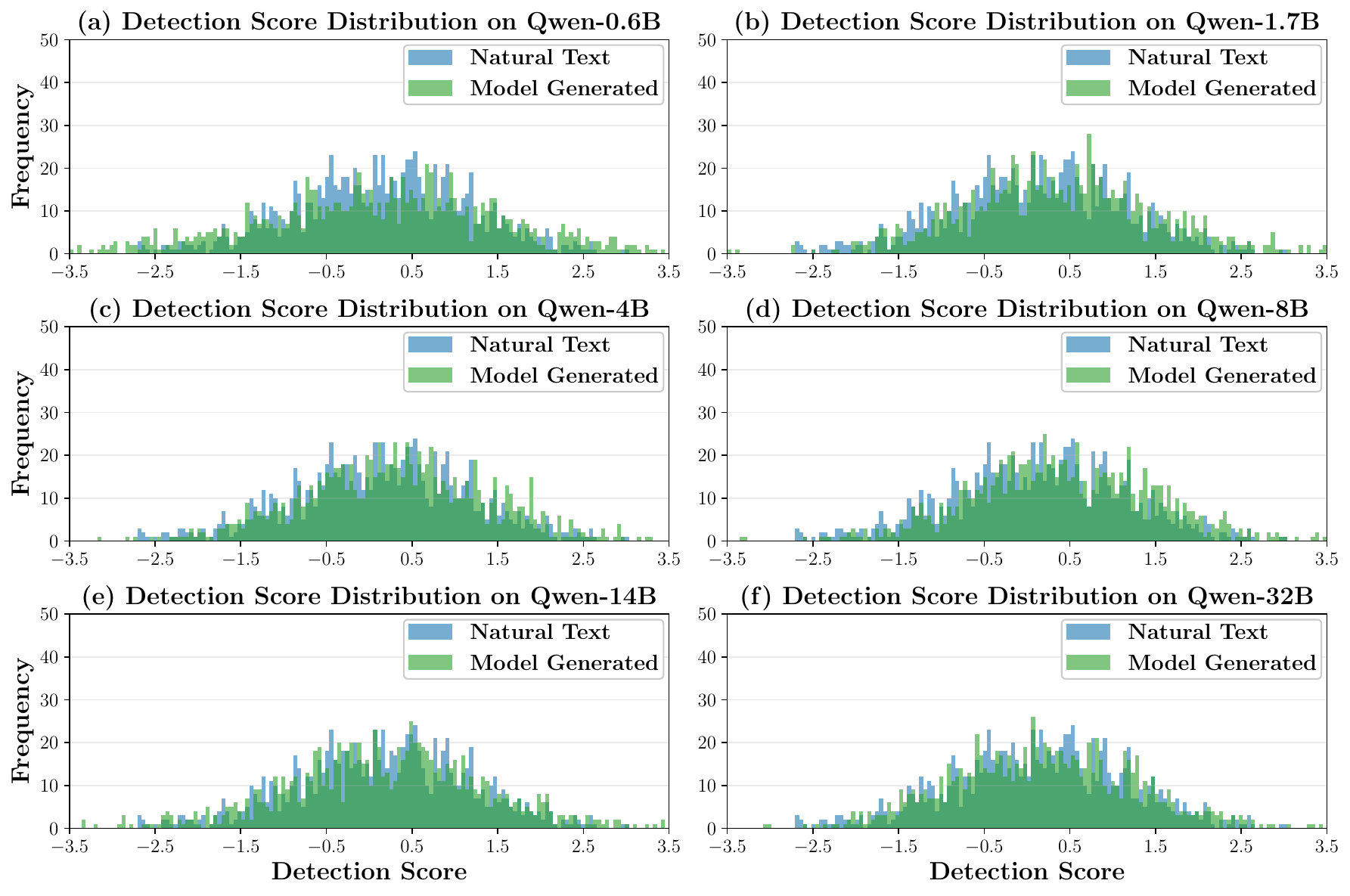}
    \caption{EWD Detection Score Distribution across Different Model Sizes}
    \label{fig:detectionphEWD}
    \vspace{-0.3cm}
\end{figure}

Specifically, we randomly selected 1,000 samples from the C4 dataset, used the model to continue writing, and then calculated the detection scores for both model-generated and natural text using the SWEET, KGW, and EWD watermark detectors. As shown in Figures~\ref{fig:detectionphSWEET},~\ref{fig:detectionphKGW}, and~\ref{fig:detectionphEWD}, we observe that as the model size increases, the detection score distribution of model-generated text shifts closer to that of natural text. This suggests that the model's ability to approximate the human-written distribution ($P_h$) improves as its size grows. We also notice that these shifts are generally subtle, suggesting that even smaller models can effectively approximate the human-written distribution $P_h$.

\begin{table}[ht]
    \centering
    \caption{Rephrased text quality across models and watermarking schemes (500 tokens).}
    \footnotesize
    \label{tab::MainresultDetailtabQuality1}
    \setlength{\tabcolsep}{2.3pt} % 默认是 6pt
    \begin{tabular}{llcccccccccccc}
        \toprule
        \multirow{2}{*}{\textbf{Model}} & \multirow{2}{*}{\textbf{Method}} & \multicolumn{3}{c}{\textbf{EWD}} & \multicolumn{3}{c}{\textbf{SWEET}} & \multicolumn{3}{c}{\textbf{XSIR}} & \multicolumn{3}{c}{\textbf{Unigram}} \\
        \cmidrule(lr){3-5} \cmidrule(lr){6-8} \cmidrule(lr){9-11} \cmidrule(lr){12-14}
        & & ESR & PPL. & GPTS. & ESR & PPL. & GPTS. & ESR & PPL. & GPTS. & ESR & PPL. & GPTS. \\
        \midrule
        
        \multicolumn{2}{c}{Target Watermarked Text}
        & --- & 3.58 & --- & --- & 3.53 & --- & --- & 3.11 & --- & --- & 16.5 & --- \\
        \midrule
        
        \multirow{6}{*}{Qwen3-0.6B} 
            & Base & 13.5 & 3.93 & 8.85 & 21.2 & 3.78 & 8.87 & 29.0 & 3.13 & 8.72 & 35.8 & 16.56 & 8.63 \\ 
        ~ & SysP. & 36.5 & 3.94 & 8.84 & 44.8 & 3.78 & 8.91 & 52.5 & 3.14 & 8.69 & 42.5 & 16.63 & 8.66 \\ 
        ~ & Think & 28.5 & 3.9 & 8.71 & 35.3 & 3.76 & 8.81 & 37.2 & 3.02 & 8.74 & 34.2 & 16.46 & 8.57 \\ 
        ~ & Think+SysP. & 48.2 & 3.91 & 8.82 & 58.8 & 3.77 & 8.84 & 50.7 & 3.06 & 8.73 & 37.5 & 16.49 & 8.59 \\ 
        ~ & SIRA & 0.00 & 5.87 & 4.40 & 0.00 & 5.71 & 4.51 & 0.0 & 4.62 & 8.76 & 0.0 & 4.39 & 4.26 \\ 
        ~ & RLCracker & 91.5 & 3.78 & 8.72 & 92.5 & 3.64 & 8.89 & 86.5 & 2.83 & 4.46 & 67.0 & 16.17 & 8.48 \\ 
        
        \midrule
        \multirow{6}{*}{Qwen3-1.7B} 
            & Base & 34.2 & 3.97 & 8.87 & 42.8 & 3.84 & 8.89 & 43.0 & 3.29 & 8.78 & 34.0 & 16.81 & 8.69 \\ 
        ~ & SysP. & 59.8 & 3.99 & 8.90 & 70.8 & 3.92 & 8.91 & 70.8 & 3.3 & 8.77 & 40.5 & 16.91 & 8.69 \\ 
        ~ & Think & 80.8 & 3.96 & 8.83 & 88.0 & 3.8 & 8.84 & 75.5 & 3.23 & 8.81 & 31.0 & 16.75 & 8.68 \\ 
        ~ & Think+SysP. & 84.0 & 3.96 & 8.85 & 86.5 & 3.82 & 8.85 & 79.8 & 3.26 & 8.78 & 34.8 & 16.8 & 8.69 \\ 
        ~ & SIRA & 0.25 & 5.95 & 4.53 & 0.00 & 5.79 & 4.54 & 0.2 & 5.17 & 4.51 & 0.0 & 4.41 & 4.37 \\ 
        ~ & RLCracker & 91.8 & 3.83 & 8.79 & 91.0 & 3.67 & 8.71 & 86.0 & 3.06 & 8.81 & 73.0 & 16.56 & 8.6 \\ 
        
        \midrule
        \multirow{6}{*}{Qwen3-4B}
            & Base & 39.2 & 4.03 & 8.94 & 53.2 & 4.03 & 8.96 & 50.2 & 3.38 & 8.83 & 39.2 & 17.3 & 8.73 \\ 
        ~ & SysP. & 54.2 & 4.04 & 8.88 & 65.8 & 4.06 & 8.95 & 66.8 & 3.39 & 8.82 & 43.8 & 17.38 & 8.73 \\ 
        ~ & Think & 76.0 & 4.0 & 8.84 & 83.5 & 3.99 & 8.91 & 75.2 & 3.33 & 8.84 & 39.5 & 17.0 & 8.72 \\ 
        ~ & Think+SysP. & 78.8 & 4.03 & 8.93 & 88.0 & 4.02 & 8.92 & 79.2 & 3.34 & 8.84 & 38.8 & 17.29 & 8.72 \\ 
        ~ & SIRA & 0.00 & 5.99 & 4.51 & 0.0 & 5.93 & 4.55 & 0.2 & 5.3 & 8.84 & 0.8 & 4.44 & 4.4 \\ 
        ~ & RLCracker & 93.3 & 4.47 & 8.81 & 95.8 & 4.35 & 8.73 & 85.5 & 3.2 & 4.54 & 73.5 & 16.86 & 8.62 \\ 
        
        \midrule
        \multirow{6}{*}{Qwen3-8B}
            & Base & 72.0 & 4.12 & 8.97 & 70.8 & 4.26 & 9.12 & 77.0 & 3.59 & 8.85 & 40.5 & 18.05 & 8.75 \\ 
        ~ & SysP. & 72.8 & 4.26 & 9.01 & 81.5 & 4.27 & 9.13 & 82.0 & 3.63 & 8.84 & 43.5 & 18.36 & 8.75 \\ 
        ~ & Think & 89.5 & 4.1 & 8.87 & 92.0 & 4.14 & 9.03 & 81.5 & 3.57 & 8.86 & 40.8 & 17.53 & 8.74 \\ 
        ~ & Think+SysP. & 90.2 & 4.12 & 8.91 & 92.0 & 4.22 & 9.06 & 86.8 & 3.58 & 8.86 & 41.0 & 17.74 & 8.74 \\ 
        ~ & SIRA & 0.00 & 6.05 & 4.46 & 0.0 & 6.07 & 4.51 & 0.2 & 5.41 & 8.87 & 0.5 & 4.49 & 4.44 \\ 
        ~ & RLCracker & 94.8 & 3.97 & 8.88 & 96.3 & 3.96 & 8.91 & 90.2 & 3.36 & 4.57 & 80.5 & 17.37 & 8.66 \\ 

        \midrule
        \multirow{4}{*}{\makecell{Qwen2.5-3B\\-Instruct}} 
            & Base & 31.2 & 4.05 & 8.79 & 46.8 & 3.91 & 8.83 & 68.8 & 3.33 & 8.79 & 37.2 & 16.97 & 8.69 \\ 
        ~ & SysP. & 75.5 & 4.07 & 8.82 & 83.0 & 3.98 & 8.86 & 78.5 & 3.38 & 8.81 & 43.5 & 17.12 & 8.75 \\ 
        ~ & SIRA & 0.25 & 6.01 & 4.35 & 0.2 & 5.86 & 4.33 & 0.2 & 5.24 & 4.34 & 0.8 & 4.33 & 4.34 \\ 
        ~ & RLCracker & 93.3 & 3.98 & 8.73 & 95.3 & 3.75 & 8.69 & 89.8 & 3.16 & 8.69 & 78.5 & 16.81 & 8.6 \\ 
        \midrule
        
        \multirow{1}{*}{GPT-4o}
          & ---             & 71.0 & 4.01 & 9.13 & 81.5 & 3.81 & 9.21 & 77.2 & 3.41 & 9.15 & 49.8 & 18.32 & 9.12 \\ 
        \multirow{1}{*}{DIPPER}
          & ---        & 30.3 & 5.93 & 8.76 & 52.8 & 4.06 & 8.77 & 0.0 & 5.21 & 8.79 & 33.3 & 18.92 & 8.79 \\
        \bottomrule
    \end{tabular}
\end{table}

\begin{table}[ht]
    \centering
    \caption{Evasion Success Rate (ESR, \%) across models and watermarking schemes (500 tokens).}
    \footnotesize
    \label{tab::MainresultDetailtab2}
    \setlength{\tabcolsep}{2.3pt} % 默认是 6pt
    \begin{tabular}{llcccccccccccc}
        \toprule
        \multirow{2}{*}{\textbf{Model}} & \multirow{2}{*}{\textbf{Method}} & \multicolumn{3}{c}{\textbf{KGW\_self}} & \multicolumn{3}{c}{\textbf{KGW}} & \multicolumn{3}{c}{\textbf{SIR}} & \multicolumn{3}{c}{\textbf{PF}} \\
        \cmidrule(lr){3-5} \cmidrule(lr){6-8} \cmidrule(lr){9-11} \cmidrule(lr){12-14}
        & & ESR & Rem. & P-SP & ESR & Rem. & P-SP & ESR & Rem. & P-SP & ESR & Rem. & P-SP \\
        \midrule
        
        \multirow{6}{*}{Qwen3-0.6B} 
            & Base           & 35.5 & 39.0 & 0.92 & 60.0 & 90.0 & 0.79 & 92.2 & 98.8 & 0.90 & 14.5 & 18.8 & 0.89 \\
            & SysP.          & 63.2 & 66.8 & 0.90 & 52.2 & 93.5 & 0.74 & 92.5 & 99.2 & 0.89 & 37.2 & 45.0 & 0.87 \\
            & Think          & 66.3 & 69.1 & 0.89 & 45.2 & 93.0 & 0.69 & 89.0 & 98.2 & 0.87 & 27.3 & 36.8 & 0.85 \\
            & Think+SysP.    & 69.2 & 74.8 & 0.87 & 34.5 & 96.5 & 0.64 & 86.0 & 97.2 & 0.85 & 53.8 & 62.0 & 0.85 \\
            & SIRA             & 0.2 & 57.0 & 0.29 & 24.8 & 86.8 & 0.50 & 0.2 & 89.2 & 0.21 & 0.0 & 39.8 & 0.20 \\
            & RLCracker      & 85.2 & 90.5 & 0.87 & 80.8 & 84.0 & 0.89 & 90.5 & 94.2 & 0.91 & 76.8 & 94.8 & 0.78 \\
        
        \midrule
        \multirow{6}{*}{Qwen3-1.7B} 
            & Base           & 56.8 & 59.2 & 0.92 & 62.3 & 90.0 & 0.81 & 94.8 & 99.8 & 0.92 & 36.5 & 41.5 & 0.90 \\
            & SysP.          & 76.2 & 80.5 & 0.90 & 51.2 & 94.5 & 0.73 & 93.8 & 100.0 & 0.89 & 63.2 & 70.0 & 0.88 \\
            & Think          & 86.0 & 96.5 & 0.83 & 28.2 & 98.8 & 0.63 & 77.8 & 98.2 & 0.78 & 71.2 & 86.2 & 0.80 \\
            & Think+SysP.    & 86.0 & 96.8 & 0.83 & 28.2 & 99.5 & 0.63 & 77.0 & 97.8 & 0.77 & 77.5 & 90.5 & 0.80 \\
            & SIRA             & 0.2 & 96.2 & 0.32 & 5.8 & 97.5 & 0.39 & 0.2 & 99.0 & 0.21 & 0.0 & 86.8 & 0.25 \\
            & RLCracker      & 87.5 & 89.2 & 0.88 & 76.8 & 81.0 & 0.91 & 94.5 & 100.0 & 0.91 & 79.5 & 95.8 & 0.80 \\
        
        \midrule
        \multirow{6}{*}{Qwen3-4B}
            & Base           & 70.0 & 73.0 & 0.92 & 72.0 & 89.8 & 0.87 & 52.2 & 53.5 & 0.93 & 47.8 & 52.8 & 0.90 \\
            & SysP.          & 79.2 & 82.8 & 0.91 & 68.8 & 94.8 & 0.82 & 67.5 & 70.2 & 0.91 & 63.0 & 68.5 & 0.90 \\
            & Think          & 85.2 & 92.8 & 0.86 & 51.7 & 97.8 & 0.73 & 79.0 & 90.5 & 0.82 & 71.8 & 83.5 & 0.83 \\
            & Think+SysP.    & 88.2 & 96.5 & 0.85 & 47.5 & 98.2 & 0.72 & 79.5 & 93.2 & 0.80 & 81.2 & 89.2 & 0.84 \\
            & SIRA             & 0.2 & 92.8 & 0.32 & 12.2 & 98.8 & 0.44 & 0.2 & 84.0 & 0.21 & 0.0 & 88.8 & 0.24 \\
            & RLCracker      & 88.8 & 98.5 & 0.84 & 89.0 & 92.2 & 0.90 & 96.8 & 100.0 & 0.92 & 82.2 & 95.0 & 0.82 \\
        
        \midrule
        \multirow{6}{*}{Qwen3-8B}
            & Base           & 81.3 & 89.8 & 0.89 & 74.5 & 91.5 & 0.88 & 74.0 & 76.8 & 0.90 & 68.3 & 75.5 & 0.91 \\
            & SysP.          & 85.5 & 91.5 & 0.90 & 72.0 & 92.8 & 0.85 & 79.8 & 83.2 & 0.88 & 73.0 & 79.5 & 0.90 \\
            & Think          & 83.8 & 94.0 & 0.83 & 53.8 & 95.8 & 0.76 & 76.2 & 92.2 & 0.78 & 71.8 & 82.5 & 0.84 \\
            & Think+SysP.    & 86.2 & 94.5 & 0.82 & 57.5 & 98.0 & 0.75 & 71.0 & 95.2 & 0.76 & 74.0 & 87.8 & 0.81 \\
            & SIRA             & 0.2 & 93.0 & 0.32 & 22.2 & 97.0 & 0.52 & 0.2 & 85.2 & 0.22 & 0.0 & 82.8 & 0.26 \\
            & {RLCracker}   & {{90.8}} & {{94.8}} & {{0.90}} & {{91.5}} & {{94.5}} & {{0.90}} & {{96.3}} & {{99.8}} & {{0.88}} & {{84.3}} & {{91.3}} & {{0.87}}\\

        \midrule
        \multirow{4}{*}{\makecell{Qwen2.5-3B\\-Instruct}} 
            & Base             & 58.8 & 61.3 & 0.92 & 46.0 & 69.5 & 0.86 & 64.0 & 72.8 & 0.88 & 37.0 & 45.0 & 0.90 \\
            & SysP.            & 84.5 & 88.0 & 0.90 & 31.8 & 65.8 & 0.79 & 75.5 & 87.8 & 0.86 & 74.0 & 80.3 & 0.89 \\
            & SIRA             & 0.2 & 92.2 & 0.32 & 2.2 & 99.0 & 0.42 & 0.2 & 90.5 & 0.23 & 0.0 & 78.2 & 0.26 \\
            & RLCracker      & 89.8 & 93.5 & 0.86 & 97.5 & 99.3 & 0.86 & 74.0 & 90.8 & 0.82 & 81.8 & 95.2 & 0.84 \\
            
        \midrule
        \multirow{1}{*}{GPT-4o}
          & ---             & 86.8 & 91.2 & 0.90 & 53.8 & 96.0 & 0.73 & 94.8 & 98.8 & 0.88 & 79.0 & 87.8 & 0.88 \\
        \multirow{1}{*}{DIPPER}
          & ---        &  72.0 & 78.3 & 0.80 & 29.0 & 93.8 & 0.61 & 61.8 & 76.5 & 0.78 & 43.5 & 49.8 & 0.81 \\
        \bottomrule
    \end{tabular}
\end{table}

\begin{table}[ht]
    \centering
    \caption{Rephrased text quality across models and watermarking schemes (500 tokens).}
    \footnotesize
    \label{tab::MainresultDetailtabQuality2}
    \setlength{\tabcolsep}{2.3pt} % 默认是 6pt
    \begin{tabular}{llcccccccccccc}
        \toprule
        \multirow{2}{*}{\textbf{Model}} & \multirow{2}{*}{\textbf{Method}} & \multicolumn{3}{c}{\textbf{KGW\_self}} & \multicolumn{3}{c}{\textbf{KGW}} & \multicolumn{3}{c}{\textbf{SIR}} & \multicolumn{3}{c}{\textbf{PF}} \\
        \cmidrule(lr){3-5} \cmidrule(lr){6-8} \cmidrule(lr){9-11} \cmidrule(lr){12-14}
        & & ESR & PPL. & GPTS. & ESR & PPL. & GPTS. & ESR & PPL. & GPTS. & ESR & PPL. & GPTS. \\
        \midrule
        
        \multicolumn{2}{c}{Target Watermarked Text}
        & --- & 4.36 & --- & --- & 17.5 & --- & --- & 3.24 & --- & --- & 10.6 & --- \\
        \midrule
        
        \multirow{6}{*}{Qwen3-0.6B} 
            & Base & 35.5 & 4.43 & 8.71 & 60.0 & 17.25 & 8.68 & 92.2 & 3.3 & 8.79 & 14.5 & 9.96 & 8.69 \\ 
        ~ & SysP. & 63.2 & 4.43 & 8.73 & 52.2 & 17.32 & 8.7 & 92.5 & 3.35 & 8.79 & 37.2 & 9.97 & 8.7 \\ 
        ~ & Think & 66.3 & 4.39 & 8.7 & 45.2 & 17.15 & 8.66 & 89.0 & 3.26 & 8.78 & 27.3 & 9.67 & 8.68 \\ 
        ~ & Think+SysP. & 69.2 & 4.42 & 8.71 & 34.5 & 17.22 & 8.66 & 86.0 & 3.3 & 8.78 & 53.8 & 9.85 & 8.69 \\ 
        ~ & SIRA & 0.2 & 6.05 & 4.39 & 24.8 & 19.03 & 4.35 & 0.2 & 4.87 & 4.44 & 0.0 & 12.3 & 4.36 \\ 
        ~ & RLCracker & 85.2 & 4.17 & 8.61 & 80.8 & 17.01 & 8.58 & 90.5 & 3.11 & 8.68 & 76.8 & 9.41 & 8.59 \\ 
        
        \midrule
        \multirow{6}{*}{Qwen3-1.7B} 
            & Base & 56.8 & 4.55 & 8.75 & 62.3 & 17.47 & 8.72 & 94.8 & 3.43 & 8.81 & 36.5 & 10.22 & 8.73 \\ 
        ~ & SysP. & 76.2 & 4.57 & 8.76 & 51.2 & 17.54 & 8.73 & 93.8 & 3.52 & 8.82 & 63.2 & 10.23 & 8.73 \\ 
        ~ & Think & 86.0 & 4.49 & 8.75 & 28.2 & 17.41 & 8.71 & 77.8 & 3.38 & 8.81 & 71.2 & 10.11 & 8.72 \\ 
        ~ & Think+SysP. & 86.0 & 4.53 & 8.75 & 28.2 & 17.45 & 8.71 & 77.0 & 3.41 & 8.81 & 77.5 & 10.11 & 8.72 \\ 
        ~ & SIRA & 0.2 & 6.44 & 4.44 & 5.8 & 19.4 & 4.4 & 0.2 & 5.07 & 4.49 & 0.0 & 12.84 & 4.4 \\ 
        ~ & RLCracker & 87.5 & 4.36 & 8.66 & 76.8 & 17.28 & 8.63 & 94.5 & 3.25 & 8.72 & 79.5 & 9.94 & 8.63 \\ 
        
        \midrule
        \multirow{6}{*}{Qwen3-4B}
            & Base & 70.0 & 4.73 & 8.79 & 72.0 & 17.73 & 8.74 & 52.2 & 3.68 & 8.85 & 47.8 & 10.66 & 8.73 \\ 
        ~ & SysP. & 79.2 & 4.8 & 8.8 & 68.8 & 17.74 & 8.76 & 67.5 & 3.69 & 8.85 & 63.0 & 10.67 & 8.79 \\ 
        ~ & Think & 85.2 & 4.7 & 8.78 & 51.7 & 17.66 & 8.73 & 79.0 & 3.62 & 8.84 & 71.8 & 10.46 & 8.63 \\ 
        ~ & Think+SysP. & 88.2 & 4.7 & 8.78 & 47.5 & 17.7 & 8.74 & 79.5 & 3.65 & 8.84 & 81.2 & 10.65 & 8.70 \\ 
        ~ & SIRA & 0.2 & 6.58 & 4.47 & 12.2 & 19.55 & 4.43 & 0.2 & 5.29 & 4.53 & 0.0 & 13.05 & 4.43 \\ 
        ~ & RLCracker & 88.8 & 4.56 & 8.69 & 89.0 & 17.45 & 8.65 & 96.8 & 3.48 & 8.75 & 82.2 & 5.03 & 8.61 \\ 
        
        \midrule
        \multirow{6}{*}{Qwen3-8B}
            & Base & 81.3 & 4.91 & 8.83 & 74.5 & 17.85 & 8.78 & 74.0 & 3.85 & 8.9 & 68.3 & 10.98 & 8.81 \\ 
        ~ & SysP. & 85.5 & 5.06 & 8.86 & 72.0 & 17.91 & 8.83 & 79.8 & 3.98 & 8.9 & 73.0 & 11.16 & 8.82 \\ 
        ~ & Think & 83.8 & 4.87 & 8.81 & 53.8 & 17.81 & 8.78 & 76.2 & 3.76 & 8.87 & 71.8 & 10.81 & 8.78 \\ 
        ~ & Think+SysP. & 86.2 & 4.9 & 8.82 & 57.5 & 17.82 & 8.78 & 71.0 & 3.77 & 8.89 & 74.0 & 10.83 & 8.81 \\ 
        ~ & SIRA & 0.2 & 6.81 & 4.51 & 22.2 & 19.75 & 4.47 & 0.2 & 5.41 & 4.56 & 0.0 & 13.48 & 4.45 \\ 
        ~ & RLCracker & 90.8 & 4.72 & 8.73 & 91.5 & 17.69 & 8.69 & 96.3 & 3.62 & 8.78 & 84.3 & 10.67 & 8.68 \\ 

        \midrule
        \multirow{4}{*}{\makecell{Qwen2.5-3B\\-Instruct}} 
            & Base & 58.8 & 4.53 & 8.79 & 46.0 & 17.51 & 8.77 & 64.0 & 3.52 & 8.87 & 37.0 & 10.52 & 8.76 \\ 
        ~ & SysP. & 84.5 & 4.55 & 8.82 & 31.8 & 17.52 & 8.82 & 75.5 & 3.55 & 8.91 & 74.0 & 10.54 & 8.83 \\ 
        ~ & SIRA & 0.2 & 6.52 & 4.46 & 2.2 & 20.71 & 4.38 & 0.2 & 5.21 & 4.54 & 0.0 & 13.18 & 4.44 \\ 
        ~ & RLCracker & 89.8 & 4.41 & 8.71 & 97.5 & 17.36 & 8.64 & 74.0 & 3.37 & 8.78 & 81.8 & 10.34 & 8.67 \\ 
        \midrule
        
        \multirow{1}{*}{GPT-4o}
          & ---             & 86.8 & 4.36 & 9.15 & 53.8 & 18.23 & 9.14 & 94.8 & 3.28 & 9.16 & 79.0 & 11.23 & 9.06 \\
        \multirow{1}{*}{DIPPER}
          & ---        & 72.0 & 5.09 & 8.69 & 29.0 & 19.93 & 8.69 & 61.8 & 5.07 & 8.78 & 43.5 & 13.85 & 8.79 \\
        \bottomrule
    \end{tabular}
\end{table}

\begin{table}[ht]
    \centering
    \caption{Evasion Success Rate (ESR, \%) across models and watermarking schemes (500 tokens).}
    \footnotesize
    \label{tab::MainresultDetailtab3}
    \begin{tabular}{llcccccc}
        \toprule
        \multirow{2}{*}{\textbf{Model}} & \multirow{2}{*}{\textbf{Method}} & \multicolumn{3}{c}{\textbf{SynthID-Text}} & \multicolumn{3}{c}{\textbf{UPV}} \\
        \cmidrule(lr){3-5} \cmidrule(lr){6-8}
        & & ESR & Rem. & P-SP & ESR & Rem. & P-SP \\
        \midrule
        
        \multirow{6}{*}{Qwen3-0.6B} 
            & Base           & 44.5 & 47.0 & 0.93 & 53.5 & 56.8 & 0.91 \\
            & SysP.          & 73.5 & 76.0 & 0.92 & 76.0 & 79.5 & 0.90 \\
            & Think          & 60.8 & 64.5 & 0.91 & 60.2 & 67.5 & 0.88 \\
            & Think+SysP.    & 81.8 & 85.2 & 0.89 & 79.5 & 84.8 & 0.87 \\
            & SIRA             & 0.0 & 61.0 & 0.19 & 0.2 & 62.0 & 0.30 \\
            & RLCracker      & 96.5 & 99.5 & 0.86 & 91.2 & 97.5 & 0.84 \\
        
        \midrule
        \multirow{6}{*}{Qwen3-1.7B} 
            & Base           & 66.8 & 68.5 & 0.94 & 63.0 & 65.8 & 0.91 \\
            & SysP.          & 90.5 & 92.8 & 0.91 & 87.2 & 90.8 & 0.89 \\
            & Think          & 91.5 & 97.0 & 0.85 & 87.5 & 98.5 & 0.81 \\
            & Think+SysP.    & 92.8 & 99.0 & 0.84 & 90.0 & 99.0 & 0.81 \\
            & SIRA             & 0.2 & 99.8 & 0.23 & 0.0 & 98.5 & 0.32 \\
            & RLCracker      & 95.2 & 98.8 & 0.86 & 90.5 & 97.0 & 0.85 \\
        
        \midrule
        \multirow{6}{*}{Qwen3-4B}
            & Base           & 82.2 & 84.5 & 0.93 & 66.2 & 68.8 & 0.92 \\
            & SysP.          & 93.0 & 94.5 & 0.93 & 83.5 & 85.2 & 0.91 \\
            & Think          & 92.8 & 97.8 & 0.87 & 90.2 & 97.8 & 0.84 \\
            & Think+SysP.    & 95.2 & 99.8 & 0.86 & 91.2 & 97.8 & 0.85 \\
            & SIRA             & 0.2 & 98.8 & 0.21 & 0.2 & 95.8 & 0.32 \\
            & RLCracker      & 96.5 & 99.0 & 0.89 & 94.0 & 99.0 & 0.85 \\
        
        \midrule
        \multirow{6}{*}{Qwen3-8B}
            & Base           & 93.0 & 95.0 & 0.91 & 84.5 & 87.8 & 0.91 \\
            & SysP.          & 97.0 & 99.0 & 0.91 & 90.0 & 93.0 & 0.90 \\
            & Think          & 90.8 & 98.2 & 0.84 & 88.2 & 98.5 & 0.82 \\
            & Think+SysP.    & 93.5 & 99.8 & 0.84 & 91.5 & 99.0 & 0.82 \\
            & SIRA             & 0.2 & 97.2 & 0.22 & 0.0 & 96.2 & 0.34 \\
            & {RLCracker}   & {{98.0}} & {{99.5}} & {{0.91}} & {{95.5}} & {{98.3}} & {{0.89}} \\
            
        \midrule
        \multirow{4}{*}{\makecell{Qwen2.5-3B\\-Instruct}} 
            & Base             & 78.0 & 82.0 & 0.91 & 81.5 & 85.5 & 0.90 \\
            & SysP.            & 94.8 & 99.3 & 0.89 & 92.0 & 97.0 & 0.87 \\
            & SIRA             & 0.2 & 99.0 & 0.23 & 0.2 & 97.5 & 0.33 \\
            & RLCracker      & 93.0 & 98.5 & 0.88 & 90.5 & 98.2 & 0.86 \\
            
        \midrule
        \multirow{1}{*}{GPT-4o}
           & ---             & 96.8 & 99.0 & 0.91 & 92.0 & 96.8 & 0.89 \\
        \multirow{1}{*}{DIPPER}
          & ---        & 90.5 & 96.3 & 0.82 & 81.8 & 89.8 & 0.81 \\
        \bottomrule
    \end{tabular}
\end{table}

\begin{table}[ht]
    \centering
    \caption{Rephrased text quality across models and watermarking schemes (500 tokens).}
    \footnotesize
    \label{tab::MainresultDetailtabQuality3}
    \begin{tabular}{llcccccc}
        \toprule
        \multirow{2}{*}{\textbf{Model}} & \multirow{2}{*}{\textbf{Method}} & \multicolumn{3}{c}{\textbf{SynthID-Text}} & \multicolumn{3}{c}{\textbf{UPV}} \\
        \cmidrule(lr){3-5} \cmidrule(lr){6-8}
        & & ESR & PPL. & GPTS. & ESR & PPL. & GPTS. \\
        \midrule
        
        \multicolumn{2}{c}{Target Watermarked Text}
        & --- & 2.75 & --- & --- & 16.1 & --- \\
        \midrule
        
        \multirow{6}{*}{Qwen3-0.6B} 
            & Base & 44.5 & 2.93 & 8.73 & 53.5 & 16.09 & 8.66 \\ 
        ~ & SysP. & 73.5 & 2.96 & 8.73 & 76 & 16.2 & 8.66 \\ 
        ~ & Think & 60.8 & 2.76 & 8.72 & 60.2 & 16.06 & 8.63 \\ 
        ~ & Think+SysP. & 81.8 & 2.88 & 8.73 & 79.5 & 16.06 & 8.64 \\ 
        ~ & SIRA & 0.0 & 4.01 & 4.37 & 0.2 & 18.33 & 4.32 \\ 
        ~ & RLCracker & 96.5 & 2.6 & 8.62 & 91.2 & 15.78 & 8.54 \\ 
        
        \midrule
        \multirow{6}{*}{Qwen3-1.7B} 
            & Base & 66.8 & 3.05 & 8.76 & 63 & 16.38 & 8.7 \\ 
        ~ & SysP. & 90.5 & 3.06 & 8.76 & 87.2 & 16.44 & 8.7 \\ 
        ~ & Think & 91.5 & 3.0 & 8.75 & 87.5 & 16.34 & 8.69 \\ 
        ~ & Think+SysP. & 92.8 & 3.03 & 8.75 & 90.0 & 16.36 & 8.7 \\ 
        ~ & SIRA & 0.2 & 4.38 & 4.44 & 0.0 & 19.08 & 4.37 \\ 
        ~ & RLCracker & 95.2 & 2.86 & 8.66 & 90.5 & 16.21 & 8.61 \\ 
        
        \midrule
        \multirow{6}{*}{Qwen3-4B}
            & Base & 82.2 & 3.15 & 8.81 & 66.2 & 16.82 & 8.74 \\ 
        ~ & SysP. & 93.0 & 3.21 & 8.82 & 83.5 & 16.89 & 8.74 \\ 
        ~ & Think & 92.8 & 3.1 & 8.77 & 90.2 & 16.64 & 8.72 \\ 
        ~ & Think+SysP. & 95.2 & 3.11 & 8.8 & 91.2 & 16.66 & 8.73 \\ 
        ~ & SIRA & 0.2 & 4.46 & 4.46 & 0.2 & 19.27 & 4.42 \\ 
        ~ & RLCracker & 96.5 & 2.95 & 8.69 & 94.0 & 16.44 & 8.64 \\ 
        
        \midrule
        \multirow{6}{*}{Qwen3-8B}
            & Base & 93.0 & 3.4 & 8.85 & 84.5 & 17.23 & 8.82 \\ 
        ~ & SysP. & 97.0 & 3.83 & 8.9 & 90.0 & 17.25 & 8.84 \\ 
        ~ & Think & 90.8 & 3.35 & 8.84 & 88.2 & 17.03 & 8.78 \\ 
        ~ & Think+SysP. & 93.5 & 3.35 & 8.85 & 91.5 & 17.17 & 8.8 \\ 
        ~ & SIRA & 0.2 & 4.7 & 4.53 & 0.0 & 19.69 & 4.45 \\ 
        ~ & RLCracker & 98.0 & 3.21 & 8.75 & 95.5 & 16.79 & 8.68 \\ 
            
        \midrule
        \multirow{4}{*}{\makecell{Qwen2.5-3B\\-Instruct}} 
            & Base & 78.0 & 3.08 & 8.75 & 81.5 & 16.55 & 8.74 \\ 
        ~ & SysP. & 94.8 & 3.08 & 8.81 & 92.0 & 16.56 & 8.76 \\ 
        ~ & SIRA & 0.2 & 4.34 & 4.43 & 0.2 & 19.35 & 4.32 \\ 
        ~ & RLCracker & 93.0 & 2.92 & 8.66 & 90.5 & 16.39 & 8.62 \\ 
            
        \midrule
        \multirow{1}{*}{GPT-4o}
           & ---             & 96.8 & 2.78 & 9.15 & 92.0 & 16.12 & 9.11 \\ 
        \multirow{1}{*}{DIPPER}
          & ---        & 90.5 & 2.89 & 8.78 & 81.8 & 16.93 & 8.77 \\ 
        \bottomrule
    \end{tabular}
\end{table}

\begin{table}[ht]
    \centering
    \caption{Evasion Success Rate (ESR, \%) across models and watermarking schemes (1500 tokens).}
    \footnotesize
    \label{tab::MainresultDetailtab4}
    \setlength{\tabcolsep}{2.3pt} % 默认是 6pt
    \begin{tabular}{llcccccccccccc}
        \toprule
        \multirow{2}{*}{\textbf{Model}} & \multirow{2}{*}{\textbf{Method}} & \multicolumn{3}{c}{\textbf{EWD}} & \multicolumn{3}{c}{\textbf{SWEET}} & \multicolumn{3}{c}{\textbf{XSIR}} & \multicolumn{3}{c}{\textbf{Unigram}} \\
        \cmidrule(lr){3-5} \cmidrule(lr){6-8} \cmidrule(lr){9-11} \cmidrule(lr){12-14}
        & & ESR & Rem. & P-SP & ESR & Rem. & P-SP & ESR & Rem. & P-SP & ESR & Rem. & P-SP \\
        \midrule
        
        \multirow{6}{*}{Qwen3-0.6B}
            & Base           & 3.5 & 6.2 & 0.81 & 4.2 & 10.2 & 0.79 & 26.0 & 45.5 & 0.71 & 5.0 & 53.0 & 0.71 \\
            & SysP.          & 5.8 & 13.0 & 0.80 & 12.5 & 25.2 & 0.78 & 39.2 & 61.0 & 0.69 & 6.2 & 76.5 & 0.61 \\
            & Think          & 10.8 & 16.5 & 0.81 & 10.2 & 20.0 & 0.80 & 32.0 & 53.8 & 0.73 & 6.0 & 75.5 & 0.59 \\
            & Think+SysP.    & 14.8 & 24.2 & 0.80 & 16.2 & 31.0 & 0.78 & 38.2 & 65.2 & 0.72 & 5.2 & 89.2 & 0.52 \\
            & SIRA             & 0.0 & 38.8 & 0.27 & 0.0 & 48.5 & 0.29 & 0.0 & 0.0 & 0.25 & 0.2 & 62.0 & 0.52 \\
            & RLCracker      & 63.2 & 93.8 & 0.72 & 63.2 & 96.8 & 0.71 & 60.0 & 97.0 & 0.66 & 90.2 & 90.5 & 0.93 \\
        
        \midrule
        \multirow{6}{*}{Qwen3-1.7B}
            & Base           & 4.2 & 8.5 & 0.83 & 6.8 & 14.5 & 0.82 & 30.2 & 49.5 & 0.75 & 5.0 & 60.8 & 0.70 \\
            & SysP.          & 7.5 & 15.5 & 0.82 & 10.5 & 23.0 & 0.82 & 39.5 & 64.8 & 0.74 & 5.8 & 72.5 & 0.65 \\
            & Think          & 42.5 & 71.8 & 0.72 & 48.0 & 80.8 & 0.72 & 53.5 & 91.5 & 0.65 & 4.5 & 96.8 & 0.50 \\
            & Think+SysP.    & 41.5 & 64.5 & 0.74 & 45.5 & 75.8 & 0.73 & 59.0 & 92.5 & 0.68 & 6.2 & 96.2 & 0.53 \\
            & SIRA             & 0.0 & 49.5 & 0.30 & 0.0 & 51.2 & 0.33 & 0.0 & 76.2 & 0.27 & 0.0 & 82.0 & 0.46 \\
            & RLCracker      & 62.7 & 93.5 & 0.72 & 61.3 & 96.8 & 0.71 & 62.2 & 95.2 & 0.67 & 77.8 & 82.5 & 0.86 \\
        
        \midrule
        \multirow{6}{*}{Qwen3-4B}
            & Base           & 10.0 & 19.8 & 0.83 & 12.2 & 27.8 & 0.82 & 38.0 & 83.0 & 0.61 & 4.5 & 43.5 & 0.76 \\
            & SysP.          & 12.5 & 24.8 & 0.83 & 15.0 & 32.5 & 0.82 & 45.0 & 94.0 & 0.61 & 5.5 & 65.0 & 0.69 \\
            & Think          & 40.8 & 61.5 & 0.76 & 45.0 & 71.0 & 0.75 & 62.0 & 90.8 & 0.71 & 4.8 & 79.5 & 0.62 \\
            & Think+SysP.    & 38.2 & 60.5 & 0.76 & 46.2 & 73.0 & 0.75 & 63.2 & 93.2 & 0.70 & 6.5 & 86.8 & 0.60 \\
            & SIRA             & 0.0 & 47.8 & 0.30 & 0.2 & 50.2 & 0.34 & 0.0 & 70.5 & 0.28 & 1.2 & 89.0 & 0.46 \\
            & RLCracker      & 64.8 & 97.2 & 0.72 & 65.2 & 98.2 & 0.72 & 66.3 & 97.2 & 0.72 & 63.7 & 67.2 & 0.89 \\
        
        \midrule
        \multirow{6}{*}{Qwen3-8B}
            & Base           & 33.5 & 51.5 & 0.80 & 30.0 & 59.2 & 0.80 & 48.8 & 74.2 & 0.74 & 5.2 & 61.0 & 0.73 \\
            & SysP.          & 32.0 & 48.8 & 0.80 & 38.8 & 64.5 & 0.83 & 52.5 & 81.0 & 0.72 & 6.8 & 71.2 & 0.69 \\
            & Think          & 54.0 & 79.8 & 0.73 & 51.5 & 73.0 & 0.78 & 57.8 & 86.2 & 0.70 & 7.2 & 76.2 & 0.65 \\
            & Think+SysP.    & 56.2 & 81.0 & 0.74 & 59.8 & 80.5 & 0.74 & 64.5 & 96.8 & 0.68 & 7.2 & 79.0 & 0.66 \\
            & SIRA             & 0.0 & 53.0 & 0.31 & 0.2 & 55.8 & 0.35 & 0.0 & 79.5 & 0.29 & 0.5 & 80.0 & 0.52 \\
            & {RLCracker}   & {{70.0}} & {{90.3}} & {{0.78}} & {{71.3}} & {{95.5}} & {{0.76}} & {{69.3}} & {{96.5}} & {{0.71}} & {{81.8}} & {{88.5}} & {{0.84}}\\

        \midrule
        \multirow{4}{*}{\makecell{Qwen2.5-3B\\-Instruct}} 
            & Base             & 11.5 & 18.8 & 0.83 & 17.0 & 35.5 & 0.83 & 39.2 & 49.8 & 0.85 & 3.5 & 20.5 & 0.85 \\
            & SysP.            & 27.3 & 36.3 & 0.82 & 35.5 & 50.0 & 0.84 & 43.2 & 55.3 & 0.83 & 5.5 &  25.0 & 0.82 \\
            & SIRA             & 0.0 & 33.0 & 0.29 & 0.0 & 50.5 & 0.33 & 0.0 & 84.2 & 0.26 & 0.5 & 88.0 & 0.47 \\
            & RLCracker      & 73.0 & 82.0 & 0.81 & 71.5 & 83.3 & 0.80 & 65.5 & 82.5 & 0.78 & 98.5 &  100.& 0.92 \\

        \midrule
        \multirow{1}{*}{GPT}
          & ---             & 10.2 & 17.2 & 0.86 & 20.5 & 28.5 & 0.88 & 50.2 & 70.5 & 0.78 & 6.8 & 92.8 & 0.58 \\
        \multirow{1}{*}{DIPPER}
          & ---        & 1.15 & 10.0 & 0.62 & 5.75 & 22.3 & 0.66 & 0.0 & 0.0 & 0.60 & 4.50 & 76.5 & 0.50 \\
        \bottomrule
    \end{tabular}
\end{table}

\begin{table}[ht]
    \centering
    \caption{Rephrased text quality across models and watermarking schemes (1500 tokens).}
    \footnotesize
    \label{tab::MainresultDetailtabQuality4}
    \setlength{\tabcolsep}{2.3pt} % 默认是 6pt
    \begin{tabular}{llcccccccccccc}
        \toprule
        \multirow{2}{*}{\textbf{Model}} & \multirow{2}{*}{\textbf{Method}} & \multicolumn{3}{c}{\textbf{EWD}} & \multicolumn{3}{c}{\textbf{SWEET}} & \multicolumn{3}{c}{\textbf{XSIR}} & \multicolumn{3}{c}{\textbf{Unigram}} \\
        \cmidrule(lr){3-5} \cmidrule(lr){6-8} \cmidrule(lr){9-11} \cmidrule(lr){12-14}
        & & ESR & PPL. & GPTS. & ESR & PPL. & GPTS. & ESR & PPL. & GPTS & ESR & PPL. & GPTS. \\
        \midrule
        
        \multicolumn{2}{c}{Target Watermarked Text}
        & --- & 3.76 & --- & --- & 3.61 & --- & --- & 3.81 & --- & --- & 18.1 & --- \\
        \midrule
        
        \multirow{6}{*}{Qwen3-0.6B} 
           & Base & 3.5 & 3.57 & 8.53 & 4.2 & 3.41 & 8.63 & 26.0 & 3.59 & 8.58 & 5.0 & 18.41 & 8.46 \\ 
        ~ & SysP. & 5.8 & 3.58 & 8.54 & 12.5 & 3.41 & 8.63 & 39.2 & 3.6 & 8.6 & 6.2 & 18.44 & 8.47 \\ 
        ~ & Think & 10.8 & 3.55 & 8.51 & 10.2 & 3.36 & 8.61 & 32.0 & 3.54 & 8.57 & 6.0 & 18.34 & 8.45 \\ 
        ~ & Think+SysP. & 14.8 & 3.56 & 8.53 & 16.2 & 3.38 & 8.62 & 38.2 & 3.58 & 8.57 & 5.2 & 18.38 & 8.46 \\ 
        ~ & SIRA & 0.0 & 4.78 & 4.17 & 0.0 & 4.74 & 4.26 & 0.0 & 4.77 & 4.26 & 0.2 & 20.6 & 4.04 \\ 
        ~ & RLCracker & 63.2 & 3.41 & 8.4 & 63.2 & 3.21 & 8.52 & 60.0 & 3.33 & 8.49 & 90.2 & 18.16 & 8.35 \\ 
        
        \midrule
        \multirow{6}{*}{Qwen3-1.7B} 
            & Base & 4.2 & 3.65 & 8.58 & 6.8 & 3.56 & 8.67 & 30.2 & 3.83 & 8.62 & 5.0 & 18.63 & 8.49 \\ 
        ~ & SysP. & 7.5 & 3.68 & 8.58 & 10.5 & 3.57 & 8.67 & 39.5 & 3.87 & 8.63 & 5.8 & 18.64 & 8.49 \\ 
        ~ & Think & 42.5 & 3.62 & 8.55 & 48.0 & 3.5 & 8.65 & 53.5 & 3.76 & 8.62 & 4.5 & 18.61 & 8.49 \\ 
        ~ & Think+SysP. & 41.5 & 3.62 & 8.56 & 45.5 & 3.54 & 8.66 & 59.0 & 3.78 & 8.62 & 6.2 & 18.62 & 8.49 \\ 
        ~ & SIRA & 0.0 & 4.89 & 4.25 & 0.0 & 4.87 & 4.34 & 0.0 & 5.03 & 4.31 & 0.0 & 20.88 & 4.18 \\ 
        ~ & RLCracker & 62.7 & 3.47 & 8.47 & 61.3 & 3.37 & 8.57 & 62.2 & 3.58 & 8.53 & 77.8 & 18.45 & 8.4 \\ 
        
        \midrule
        \multirow{6}{*}{Qwen3-4B}
            & Base & 10.0 & 3.81 & 8.62 & 12.2 & 3.71 & 8.68 & 38 & 3.92 & 8.64 & 4.5 & 18.79 & 8.51 \\ 
        ~ & SysP. & 12.5 & 3.81 & 8.63 & 15.0 & 3.71 & 8.68 & 45.0 & 3.97 & 8.66 & 5.5 & 18.8 & 8.53 \\ 
        ~ & Think & 40.8 & 3.76 & 8.59 & 45.0 & 3.69 & 8.67 & 62.0 & 3.91 & 8.64 & 4.8 & 18.75 & 8.5 \\ 
        ~ & Think+SysP. & 38.2 & 3.8 & 8.62 & 46.2 & 3.7 & 8.67 & 63.2 & 3.91 & 8.64 & 6.5 & 18.78 & 8.51 \\ 
        ~ & SIRA & 0.0 & 5.0 & 4.28 & 0.2 & 5.04 & 4.37 & 0.0 & 5.28 & 4.33 & 1.2 & 21.07 & 4.2 \\ 
        ~ & RLCracker & 64.8 & 3.58 & 8.46 & 65.2 & 3.52 & 8.61 & 66.3 & 3.75 & 8.55 & 63.7 & 18.55 & 8.42 \\ 
        
        \midrule
        \multirow{6}{*}{Qwen3-8B}
            & Base & 33.5 & 3.93 & 8.66 & 30.0 & 3.96 & 8.71 & 48.8 & 4.13 & 8.71 & 5.2 & 19.16 & 8.63 \\ 
        ~ & SysP. & 32.0 & 4.15 & 8.67 & 38.8 & 4.0 & 8.75 & 52.5 & 4.23 & 8.72 & 6.8 & 19.66 & 8.64 \\ 
        ~ & Think & 54.0 & 3.92 & 8.66 & 51.5 & 3.79 & 8.7 & 57.8 & 4.08 & 8.68 & 7.2 & 19.08 & 8.55 \\ 
        ~ & Think+SysP. & 56.2 & 3.92 & 8.66 & 59.8 & 3.91 & 8.71 & 64.5 & 4.1 & 8.69 & 7.2 & 19.1 & 8.57 \\ 
        ~ & SIRA & 0.0 & 5.12 & 4.34 & 0.2 & 5.12 & 4.38 & 0.0 & 5.42 & 4.36 & 0.5 & 21.23 & 4.24 \\ 
        ~ & RLCracker & 70.0 & 3.77 & 8.57 & 71.3 & 3.62 & 8.62 & 69.3 & 3.9 & 8.59 & 81.8 & 18.91 & 8.46 \\ 

        \midrule
        \multirow{4}{*}{\makecell{Qwen2.5-3B\\-Instruct}} 
            & Base & 11.5 & 3.81 & 8.61 & 17.0 & 3.59 & 8.65 & 39.2 & 3.81 & 8.63 & 3.5 & 18.79 & 8.5 \\ 
        ~ & SysP. & 27.3 & 3.83 & 8.63 & 35.5 & 3.64 & 8.78 & 43.2 & 3.85 & 8.68 & 5.5 & 18.8 & 8.54 \\ 
        ~ & SIRA & 0.0 & 4.63 & 4.28 & 0.0 & 4.73 & 4.31 & 0.0 & 4.98 & 4.29 & 0.5 & 20.91 & 4.19 \\ 
        ~ & RLCracker & 73.0 & 3.6 & 8.52 & 71.5 & 3.46 & 8.55 & 65.5 & 3.67 & 8.54 & 98.5 & 18.65 & 8.42 \\
        \midrule
        
        \multirow{1}{*}{GPT-4o}
          & ---             & 10.2 & 3.62 & 8.83 & 20.5 & 3.67 & 8.85 & 50.2 & 3.79 & 8.79 & 6.8 & 18.44 & 8.75 \\ 
        \multirow{1}{*}{DIPPER}
          & ---        & 1.15 & 3.89 & 8.54 & 5.75 & 3.98 & 8.53 & 0.0 & 5.32 & 8.51 & 4.5 & 22.32 & 8.49 \\ 
        \bottomrule
    \end{tabular}
\end{table}

\begin{table}[ht]
    \centering
    \caption{Evasion Success Rate (ESR, \%) across models and watermarking schemes (1500 tokens).}
    \footnotesize
    \label{tab::MainresultDetailtab5}
    \setlength{\tabcolsep}{2.3pt} % 默认是 6pt
    \begin{tabular}{llcccccccccccc}
        \toprule
        \multirow{2}{*}{\textbf{Model}} & \multirow{2}{*}{\textbf{Method}} & \multicolumn{3}{c}{\textbf{KGW\_self}} & \multicolumn{3}{c}{\textbf{KGW}} & \multicolumn{3}{c}{\textbf{SIR}} & \multicolumn{3}{c}{\textbf{PF}} \\
        \cmidrule(lr){3-5} \cmidrule(lr){6-8} \cmidrule(lr){9-11} \cmidrule(lr){12-14}
        & & ESR & Rem. & P-SP & ESR & Rem. & P-SP & ESR & Rem. & P-SP & ESR & Rem. & P-SP \\
        \midrule
        
        \multirow{6}{*}{Qwen3-0.6B}
            & Base           & 17.5 & 23.5 & 0.88 & 6.0 & 33.8 & 0.81 & 73.2 & 98.8 & 0.76 & 11.8 & 17.0 & 0.83 \\
            & SysP.          & 37.0 & 47.2 & 0.87 & 15.0 & 55.8 & 0.77 & 72.1 & 97.5 & 0.76 & 22.0 & 35.0 & 0.81 \\
            & Think          & 33.0 & 42.0 & 0.88 & 15.8 & 50.2 & 0.77 & 73.8 & 98.2 & 0.78 & 22.5 & 37.2 & 0.80 \\
            & Think+SysP.    & 47.8 & 58.5 & 0.86 & 19.8 & 62.7 & 0.76 & 72.2 & 99.8 & 0.76 & 30.0 & 49.8 & 0.79 \\
            & SIRA             & 0.0 & 51.0 & 0.29 & 0.0 & 45.2 & 0.28 & 0.2 & 94.5 & 0.27 & 0.0 & 33.0 & 0.38 \\
            & RLCracker      & 66.8 & 80.0 & 0.84 & 48.5 & 51.5 & 0.88 & 75.8 & 100.0 & 0.75 & 62.0 & 96.5 & 0.72 \\
        
        \midrule
        \multirow{6}{*}{Qwen3-1.7B}
            & Base           & 32.5 & 37.5 & 0.90 & 9.2 & 41.8 & 0.79 & 78.8 & 100.0 & 0.81 & 19.2 & 26.2 & 0.84 \\
            & SysP.          & 41.0 & 49.0 & 0.90 & 16.0 & 55.2 & 0.78 & 78.5 & 100.0 & 0.81 & 30.0 & 41.8 & 0.83 \\
            & Think          & 66.3 & 83.5 & 0.80 & 47.0 & 95.5 & 0.70 & 61.3 & 99.0 & 0.68 & 55.5 & 88.0 & 0.74 \\
            & Think+SysP.    & 68.5 & 73.8 & 0.81 & 52.5 & 97.0 & 0.72 & 64.0 & 98.2 & 0.69 & 60.2 & 92.0 & 0.74 \\
            & SIRA             & 0.0 & 69.5 & 0.31 & 0.0 & 70.0 & 0.30 & 0.2 & 99.8 & 0.29 & 0.0 & 63.2 & 0.43 \\
            & RLCracker      & 68.0 & 78.8 & 0.88 & 60.5 & 74.3 & 0.75 & 80.0 & 99.8 & 0.82 & 61.0 & 92.8 & 0.73 \\
        
        \midrule
        \multirow{6}{*}{Qwen3-4B}
            & Base           & 36.0 & 41.8 & 0.91 & 15.8 & 42.5 & 0.82 & 37.0 & 54.2 & 0.83 & 25.8 & 33.5 & 0.83 \\
            & SysP.          & 40.8 & 47.0 & 0.90 & 19.2 & 57.5 & 0.80 & 45.0 & 63.2 & 0.82 & 34.5 & 45.2 & 0.83 \\
            & Think          & 78.0 & 90.0 & 0.84 & 52.2 & 93.5 & 0.74 & 70.5 & 96.2 & 0.73 & 59.0 & 83.8 & 0.76 \\
            & Think+SysP.    & 79.5 & 93.0 & 0.83 & 54.2 & 96.0 & 0.74 & 64.2 & 90.2 & 0.72 & 63.5 & 91.0 & 0.76 \\
            & SIRA             & 0.0 & 51.0 & 0.29 & 0.0 & 45.2 & 0.28 & 0.2 & 94.5 & 0.27 & 0.0 & 33.0 & 0.38 \\
            & RLCracker      & 82.2 & 97.3 & 0.81 & 57.3 & 70.5 & 0.83 & 65.0 & 96.5 & 0.71 & 64.8 & 96.5 & 0.74 \\
        
        \midrule
        \multirow{6}{*}{Qwen3-8B}
            & Base           & 58.8 & 72.0 & 0.83 & 46.8 & 85.0 & 0.78 & 62.5 & 82.2 & 0.79 & 44.3 & 59.5 & 0.84 \\
            & SysP.          & 64.5 & 75.2 & 0.88 & 51.0 & 90.0 & 0.78 & 64.0 & 84.0 & 0.79 & 50.2 & 68.2 & 0.82 \\
            & Think          & 71.3 & 94.8 & 0.81 & 55.8 & 96.2 & 0.73 & 64.5 & 96.8 & 0.69 & 59.5 & 87.2 & 0.75 \\
            & Think+SysP.    & 78.2 & 97.2 & 0.81 & 58.0 & 98.5 & 0.74 & 62.7 & 96.2 & 0.69 & 63.2 & 93.8 & 0.76 \\
            & SIRA             & 0.0 & 67.2 & 0.32 & 0.0 & 70.2 & 0.31 & 0.0 & 84.0 & 0.30 & 0.2 & 61.5 & 0.44 \\
            & {RLCracker}   & {{84.3}} & {{97.3}} & {{0.83}} & {{64.5}} & {{87.5}} & {{0.78}} & {{82.5}} & {{96.5}} & {{0.78}} & {{73.3}} & {{91.5}} & {{0.79}}\\

        \midrule
        \multirow{4}{*}{\makecell{Qwen2.5-3B\\-Instruct}} 
            & Base             & 38.0 & 47.0 & 0.89 & 20.5 & 40.3 & 0.87 & 56.2 & 66.5 & 0.88 & 15.2 & 28.5 & 0.84 \\
            & SysP.            & 63.7 & 70.0 & 0.87 & 44.8 & 59.8 & 0.84 & 54.2 & 70.0 & 0.84 & 39.5 & 54.8  & 0.84 \\
            & SIRA             & 0.2 & 71.5 & 0.31 & 0.5 & 66.5 & 0.31 & 0.2 & 88.0 & 0.31 & 0.0 & 58.0 & 0.43 \\
            & RLCracker      & 78.0 & 89.0 & 0.83 & 58.0 & 67.3 & 0.90 & 62.0 & 80.3 & 0.81 & 77.8 & 90.3 & 0.81 \\
            
        \midrule
        \multirow{1}{*}{GPT-4o}
          & ---             & 61.8 & 68.2 & 0.90 & 49.2 & 88.5 & 0.79 & 81.8 & 98.5 & 0.82 & 49.3 & 67.0 & 0.80 \\
        \multirow{1}{*}{DIPPER}
          & ---        & 41.5 & 64.0 & 0.72 & 11.5 & 52.8 & 0.65 & 14.8 & 88.9 & 0.49 & 21.0 & 39.0 & 0.69 \\
        \bottomrule
    \end{tabular}
\end{table}

\begin{table}[ht]
    \centering
    \caption{Rephrased text quality across models and watermarking schemes (1500 tokens).}
    \footnotesize
    \label{tab::MainresultDetailtabQuality5}
    \setlength{\tabcolsep}{2.3pt} % 默认是 6pt
    \begin{tabular}{llcccccccccccc}
        \toprule
        \multirow{2}{*}{\textbf{Model}} & \multirow{2}{*}{\textbf{Method}} & \multicolumn{3}{c}{\textbf{KGW\_self}} & \multicolumn{3}{c}{\textbf{KGW}} & \multicolumn{3}{c}{\textbf{SIR}} & \multicolumn{3}{c}{\textbf{PF}} \\
        \cmidrule(lr){3-5} \cmidrule(lr){6-8} \cmidrule(lr){9-11} \cmidrule(lr){12-14}
        & & ESR & PPL. & GPTS. & ESR & PPL. & GPTS. & ESR & PPL. & GPTS. & ESR & PPL. & GPTS. \\
        \midrule
        
        \multicolumn{2}{c}{Target Watermarked Text}
        & --- & 3.57 & --- & --- & 17.9 & --- & --- & 3.8 & --- & --- & 12.8 & --- \\
        \midrule
        
        \multirow{6}{*}{Qwen3-0.6B} 
            & Base & 17.5 & 3.3 & 8.54 & 6.0 & 17.7 & 8.44 & 73.2 & 3.73 & 8.48 & 11.8 & 11.66 & 8.4 \\ 
        ~ & SysP. & 37.0 & 3.34 & 8.55 & 15.0 & 17.73 & 8.45 & 72.1 & 3.73 & 8.49 & 22.0 & 11.68 & 8.4 \\ 
        ~ & Think & 33.0 & 3.29 & 8.53 & 15.8 & 17.38 & 8.42 & 73.8 & 3.67 & 8.45 & 22.5 & 11.38 & 8.36 \\ 
        ~ & Think+SysP. & 47.8 & 3.3 & 8.54 & 19.8 & 17.66 & 8.43 & 72.2 & 3.69 & 8.48 & 30.0 & 11.64 & 8.38 \\ 
        ~ & SIRA & 0.0 & 4.59 & 4.22 & 0.0 & 19.56 & 4.07 & 0.2 & 4.84 & 4.1 & 0.0 & 13.32 & 4.03 \\ 
        ~ & RLCracker & 66.8 & 3.15 & 8.45 & 48.5 & 17.24 & 8.33 & 75.8 & 3.49 & 8.36 & 62.0 & 11.15 & 8.27 \\ 
        
        \midrule
        \multirow{6}{*}{Qwen3-1.7B} 
            & Base & 32.5 & 3.47 & 8.57 & 9.2 & 17.84 & 8.49 & 78.8 & 3.8 & 8.52 & 19.2 & 11.77 & 8.42 \\ 
        ~ & SysP. & 41.0 & 3.48 & 8.57 & 16.0 & 17.85 & 8.5 & 78.5 & 3.8 & 8.52 & 30.0 & 11.8 & 8.44 \\ 
        ~ & Think & 66.3 & 3.44 & 8.57 & 47.0 & 17.8 & 8.49 & 61.3 & 3.77 & 8.51 & 55.5 & 11.75 & 8.41 \\ 
        ~ & Think+SysP. & 68.5 & 3.45 & 8.57 & 52.5 & 17.84 & 8.49 & 64.0 & 3.77 & 8.51 & 60.2 & 11.76 & 8.41 \\ 
        ~ & SIRA & 0.0 & 4.75 & 4.25 & 0.0 & 20.17 & 4.17 & 0.2 & 5.05 & 4.19 & 0.0 & 14.01 & 4.1 \\ 
        ~ & RLCracker & 68.0 & 3.33 & 8.48 & 60.5 & 17.67 & 8.4 & 80.0 & 3.64 & 8.41 & 61 & 11.62 & 8.32 \\ 
        
        \midrule
        \multirow{6}{*}{Qwen3-4B}
            & Base & 36.0 & 3.65 & 8.6 & 15.8 & 18.08 & 8.52 & 37 & 3.83 & 8.55 & 25.8 & 11.99 & 8.46 \\ 
        ~ & SysP. & 40.8 & 3.68 & 8.6 & 19.2 & 18.17 & 8.52 & 45 & 3.85 & 8.57 & 34.5 & 12.02 & 8.46 \\ 
        ~ & Think & 78.0 & 3.62 & 8.58 & 52.2 & 17.98 & 8.52 & 70.5 & 3.82 & 8.55 & 59.0 & 11.91 & 8.45 \\ 
        ~ & Think+SysP. & 79.5 & 3.62 & 8.59 & 54.2 & 18.03 & 8.52 & 64.2 & 3.82 & 8.55 & 63.5 & 11.96 & 8.45 \\ 
        ~ & SIRA & 0.0 & 4.9 & 4.28 & 0.0 & 20.31 & 4.2 & 0.2 & 5.11 & 4.24 & 0.0 & 14.11 & 4.14 \\ 
        ~ & RLCracker & 82.2 & 3.51 & 8.5 & 57.3 & 17.86 & 8.43 & 65.0 & 3.69 & 8.47 & 64.8 & 11.75 & 8.34 \\ 
        
        \midrule
        \multirow{6}{*}{Qwen3-8B}
            & Base & 58.8 & 3.76 & 8.65 & 46.8 & 18.57 & 8.56 & 62.5 & 3.94 & 8.61 & 44.3 & 12.48 & 8.51 \\ 
        ~ & SysP. & 64.5 & 3.77 & 8.66 & 51.0 & 18.64 & 8.57 & 64.0 & 4.06 & 8.61 & 50.2 & 12.49 & 8.56 \\ 
        ~ & Think & 71.3 & 3.74 & 8.63 & 55.8 & 18.48 & 8.55 & 64.5 & 3.91 & 8.59 & 59.5 & 12.2 & 8.49 \\ 
        ~ & Think+SysP. & 78.2 & 3.74 & 8.64 & 58.0 & 18.55 & 8.55 & 62.7 & 3.93 & 8.61 & 63.2 & 12.38 & 8.5 \\ 
        ~ & SIRA & 0.0 & 5.13 & 4.32 & 0.0 & 20.59 & 4.23 & 0.0 & 5.17 & 4.28 & 0.2 & 14.44 & 4.17 \\ 
        ~ & RLCracker & 84.3 & 3.63 & 8.55 & 64.5 & 18.18 & 8.45 & 82.5 & 3.78 & 8.51 & 73.3 & 12.07 & 8.4 \\ 

        \midrule
        \multirow{4}{*}{\makecell{Qwen2.5-3B\\-Instruct}} 
            & Base & 38.0 & 3.56 & 8.56 & 20.5 & 17.9 & 8.48 & 56.2 & 3.83 & 8.57 & 15.2 & 11.83 & 8.47 \\ 
        ~ & SysP. & 63.7 & 3.63 & 8.61 & 44.8 & 17.94 & 8.51 & 54.2 & 3.9 & 8.66 & 39.5 & 11.85 & 8.48 \\ 
        ~ & SIRA & 0.2 & 4.7 & 4.22 & 0.5 & 20.04 & 4.14 & 0.2 & 4.95 & 4.16 & 0.0 & 13.95 & 4.13 \\ 
        ~ & RLCracker & 78.0 & 3.4 & 8.45 & 58 & 17.76 & 8.39 & 62 & 3.67 & 8.44 & 77.8 & 11.7 & 8.37 \\ 
        \midrule
        
        \multirow{1}{*}{GPT-4o}
          & ---             & 61.8 & 3.51 & 8.83 & 49.2 & 17.54 & 8.82 & 81.8 & 3.81 & 8.82 & 49.3 & 11.87 & 8.79 \\ 
        \multirow{1}{*}{DIPPER}
          & ---        & 41.5 & 4.13 & 8.54 & 11.5 & 18.83 & 8.51 & 14.8 & 4.83 & 8.54 & 21 & 12.34 & 8.47 \\ 
        \bottomrule
    \end{tabular}
\end{table}

\begin{table}[ht]
    \centering
    \caption{Evasion Success Rate (ESR, \%) across models and watermarking schemes (1500 tokens).}
    \footnotesize
    \label{tab::MainresultDetailtab6}
    \begin{tabular}{llcccccc}
        \toprule
       \multirow{2}{*}{\textbf{Model}} & \multirow{2}{*}{\textbf{Method}} & \multicolumn{3}{c}{\textbf{SynthID-Text}} & \multicolumn{3}{c}{\textbf{UPV}} \\
        \cmidrule(lr){3-5} \cmidrule(lr){6-8}
        & & ESR & Rem. & P-SP & ESR & Rem. & P-SP \\
        \midrule
        
        \multirow{6}{*}{Qwen3-0.6B}
            & Base           & 38.8 & 48.5 & 0.85 & 39.0 & 47.2 & 0.86 \\
            & SysP.          & 56.5 & 71.5 & 0.83 & 60.0 & 69.5 & 0.84 \\
            & Think          & 51.2 & 65.5 & 0.84 & 47.2 & 61.8 & 0.83 \\
            & Think+SysP.    & 64.8 & 80.5 & 0.83 & 60.5 & 74.8 & 0.81 \\
            & SIRA             & 0.2 & 64.2 & 0.33 & 1.0 & 58.2 & 0.42 \\
            & RLCracker      & 79.8 & 98.8 & 0.80 & 86.5 & 99.2 & 0.79 \\
        
        \midrule
        \multirow{6}{*}{Qwen3-1.7B}
            & Base           & 49.0 & 57.5 & 0.87 & 39.0 & 45.2 & 0.89 \\
            & SysP.          & 60.8 & 74.2 & 0.86 & 48.8 & 56.8 & 0.87 \\
            & Think          & 77.8 & 99.8 & 0.76 & 69.5 & 97.5 & 0.74 \\
            & Think+SysP.    & 79.0 & 99.8 & 0.78 & 80.2 & 98.0 & 0.77 \\
            & SIRA             & 0.2 & 94.5 & 0.36 & 2.0 & 83.5 & 0.45 \\
            & RLCracker      & 78.0 & 96.5 & 0.81 & 79.2 & 93.5 & 0.79 \\
        
        \midrule
        \multirow{6}{*}{Qwen3-4B}
            & Base           & 54.0 & 68.2 & 0.87 & 30.5 & 38.2 & 0.90 \\
            & SysP.          & 70.8 & 83.5 & 0.87 & 41.0 & 49.8 & 0.89 \\
            & Think          & 82.2 & 98.8 & 0.80 & 83.8 & 95.8 & 0.80 \\
            & Think+SysP.    & 83.2 & 99.8 & 0.80 & 84.5 & 96.8 & 0.79 \\
            & SIRA             & 0.2 & 95.5 & 0.36 & 2.0 & 70.2 & 0.47 \\
            & RLCracker      & 80.0 & 98.2 & 0.82 & 83.0 & 96.8 & 0.80 \\
        
        \midrule
        \multirow{6}{*}{Qwen3-8B}
            & Base           & 78.3 & 93.8 & 0.84 & 70.0 & 81.0 & 0.85 \\
            & SysP.          & 82.8 & 97.0 & 0.85 & 78.0 & 87.0 & 0.86 \\
            & Think          & 81.0 & 99.8 & 0.78 & 85.2 & 98.5 & 0.79 \\
            & Think+SysP.    & 81.2 & 100.0 & 0.77 & 89.2 & 99.2 & 0.80 \\
            & SIRA             & 0.2 & 97.5 & 0.37 & 1.2 & 76.2 & 0.48 \\
            & {RLCracker}   & {{85.8}} & {{99.3}} & {{0.84}} & {{87.5}} & {{97.5}} & {{0.81}}\\

        \midrule
        \multirow{4}{*}{\makecell{Qwen2.5-3B\\-Instruct}} 
            & Base             & 81.0 & 89.5 & 0.89 & 80.0 & 89.5 & 0.87 \\
            & SysP.            & 81.2 & 88.8 & 0.86 & 74.8 & 85.8 & 0.83 \\
            & SIRA             & 0.2 & 96.0 & 0.36 & 0.5 & 88.8 & 0.45 \\
            & RLCracker      & 83.3 & 91.0 & 0.82 & 81.5 & 91.5 & 0.85 \\
            
        \midrule
        \multirow{1}{*}{GPT-4o}
          & ---             & 88.2 & 99.5 & 0.89 & 77.8 & 84.5 & 0.88 \\
        \multirow{1}{*}{DIPPER}
          & ---        & 50.2 & 99.5 & 0.66 & 58.0 & 89.8 & 0.72 \\
        \bottomrule
    \end{tabular}
\end{table}

\begin{table}[ht]
    \centering
    \caption{Rephrased text quality across models and watermarking schemes (1500 tokens).}
    \footnotesize
    \label{tab::MainresultDetailtabQuality6}
    \begin{tabular}{llcccccc}
        \toprule
        \multirow{2}{*}{\textbf{Model}} & \multirow{2}{*}{\textbf{Method}} & \multicolumn{3}{c}{\textbf{SynthID-Text}} & \multicolumn{3}{c}{\textbf{UPV}} \\
        \cmidrule(lr){3-5} \cmidrule(lr){6-8}
        & & ESR & PPL. & GPTS. & ESR & PPL. & GPTS. \\
        \midrule
        
        \multicolumn{2}{c}{Target Watermarked Text}
        & --- & 2.85 & --- & --- & 21.7 & --- \\
        \midrule
        
        \multirow{6}{*}{Qwen3-0.6B} 
            & Base & 38.8 & 2.86 & 8.49 & 39.0 & 21.5 & 8.43 \\ 
        ~ & SysP. & 56.5 & 2.87 & 8.49 & 60.0 & 21.57 & 8.44 \\ 
        ~ & Think & 51.2 & 2.81 & 8.47 & 47.2 & 21.42 & 8.41 \\ 
        ~ & Think+SysP. & 64.8 & 2.82 & 8.48 & 60.5 & 21.46 & 8.41 \\ 
        ~ & SIRA & 0.25 & 3.43 & 4.09 & 1.0 & 23.78 & 4.06 \\ 
        ~ & RLCracker & 79.8 & 2.7 & 8.37 & 86.5 & 21.13 & 8.31 \\ 
        
        \midrule
        \multirow{6}{*}{Qwen3-1.7B} 
            & Base & 49.0 & 2.93 & 8.54 & 39.0 & 21.93 & 8.46 \\ 
        ~ & SysP. & 60.8 & 2.94 & 8.55 & 48.8 & 21.95 & 8.47 \\ 
        ~ & Think & 77.8 & 2.91 & 8.53 & 69.5 & 21.79 & 8.45 \\ 
        ~ & Think+SysP. & 79.0 & 2.92 & 8.53 & 80.2 & 21.89 & 8.46 \\ 
        ~ & SIRA & 0.25 & 3.59 & 4.21 & 2.0 & 24.39 & 4.14 \\ 
        ~ & RLCracker & 78.0 & 2.78 & 8.44 & 79.2 & 21.58 & 8.37 \\ 
        
        \midrule
        \multirow{6}{*}{Qwen3-4B}
            & Base & 54.0 & 2.98 & 8.57 & 30.5 & 22.16 & 8.48 \\ 
        ~ & SysP. & 70.8 & 3.01 & 8.58 & 41.0 & 22.2 & 8.49 \\ 
        ~ & Think & 82.2 & 2.98 & 8.56 & 83.8 & 21.99 & 8.48 \\ 
        ~ & Think+SysP. & 83.2 & 2.98 & 8.57 & 84.5 & 22.02 & 8.48 \\ 
        ~ & SIRA & 0.2 & 3.67 & 4.26 & 2.0 & 24.67 & 4.18 \\ 
        ~ & RLCracker & 80.0 & 2.86 & 8.48 & 83 & 21.84 & 8.4 \\ 
        
        \midrule
        \multirow{6}{*}{Qwen3-8B}
            & Base & 78.3 & 3.12 & 8.63 & 70.0 & 23.03 & 8.5 \\ 
        ~ & SysP. & 82.8 & 3.12 & 8.63 & 78.0 & 23.11 & 8.55 \\ 
        ~ & Think & 81.0 & 3.06 & 8.59 & 85.2 & 22.38 & 8.5 \\ 
        ~ & Think+SysP. & 81.2 & 3.08 & 8.6 & 89.2 & 22.67 & 8.5 \\ 
        ~ & SIRA & 0.2 & 3.72 & 4.28 & 1.2 & 24.94 & 4.19 \\ 
        ~ & RLCracker & 85.8 & 2.91 & 8.51 & 87.5 & 22.17 & 8.41 \\ 
            
        \midrule
        \multirow{4}{*}{\makecell{Qwen2.5-3B\\-Instruct}} 
            & Base & 81.0 & 2.96 & 8.57 & 80.0 & 21.71 & 8.49 \\ 
        ~ & SysP. & 81.2 & 2.97 & 8.61 & 74.8 & 21.72 & 8.49 \\ 
        ~ & SIRA & 0.2 & 3.72 & 4.15 & 0.5 & 24.4 & 4.09 \\ 
        ~ & RLCracker & 83.3 & 2.83 & 8.49 & 81.5 & 21.57 & 8.4 \\ 
            
        \midrule
        \multirow{1}{*}{GPT-4o}
           & ---             & 88.2 & 2.63 & 8.81 & 77.8 & 21.14 & 8.78 \\ 
        \multirow{1}{*}{DIPPER}
          & ---        & 50.2 & 2.93 & 8.51 & 58.0 & 22.52 & 8.46 \\ 
        \bottomrule
    \end{tabular}
\end{table}